\documentclass[authoryear,preprint,review,12pt]{elsarticle}

\usepackage{amssymb}

\usepackage{siunitx}
\usepackage{booktabs}
\usepackage{rotating}
\usepackage{multirow}
\usepackage{subcaption}
\usepackage{float}
\usepackage{tabularx}
\usepackage{geometry}
\usepackage{lscape}
\usepackage{url}
\usepackage{color}
\usepackage[table]{xcolor}
\usepackage{textgreek}
\usepackage{setspace}

\begin{document}

\begin{frontmatter}



\title{Spectral reflectance analysis of type 3 carbonaceous chondrites and search for their asteroidal parent bodies}


\author[label1]{J. Eschrig \corref{correspondingauthor}}
\cortext[correspondingauthor]{Corresponding author.}
\ead[label2]{jolantha.eschrig@univ-grenoble-alpes.fr}
\author[label1]{L. Bonal}
\author[label1]{P. Beck}
\author[label1]{T. J. Prestgard}

\address[label1]{Univ. Grenoble Alpes, IPAG, F-38000 Grenoble, France.}

\begin{abstract}
Non-differentiated asteroids are among the most primitive objects in our Solar System, having escaped intense heating mechanisms. To help us understand the information contained in reflectance spectra measured on asteroids, we analyzed meteorites in the laboratory. We present an in-depth analysis of a large set of reflectance spectra: 23 CV3, 15 CO3, 4 CR2 and 31 Unequilibrated Ordinary Chondrites (UOCs). Each of the samples has a well characterized thermal history . Variations in the reflectance spectra are observed between and within each chondrite group.
UOCs systematically exhibit deeper absorption features, distinguishing them from carbonaceous chondrites. The CR2 samples presented in this study are easily distinguished from type 3 chondrites by exhibiting the \SI{1}{\micro\meter} band at lower wavelengths. CV and CO chondrites exhibit comparable mineralogical compositions and can, therefore, not be distinguished solely based on their spectral features. In the case of CV chondrites, the \SI{1}{\micro\meter} band depth increases with increasing metamorphic grade, while among CO chondrites, the spectra exhibit increasing visual slopes. \\
By comparing the chondrite spectra with the spectra of various end member asteroids we are able to suggest several possible genetic links to the studied chondrites. The method in this work is supported by observed match between UOC and S-type asteroid spectra.
A further link is found between CV/CO chondrites and Eos and L-type asteroids. Finally, CK chondrite spectral features match with Eos and K-type asteroids. Lastly, we underline the potential of the \SI{3}{\micro\meter} band to constrain asteroid-meteorite links.
\end{abstract}


\end{frontmatter}



\newcommand{\AR}{AR$_\textrm{Hyd/Si-O}$ }
\newcommand{\Area}{IBD$_\textrm{Hyd}$ }
\newcommand{\x}{(*) }

\section{Introduction}
\label{Sec:Introduction}
While some asteroids are clearly differentiated (e.g., Vesta), several asteroid types appear to have escaped differentiation (e.g., \cite{Vernazza2017}).
 This makes them primitive objects and as such, they are important for the understanding of the formation and evolution of the Solar System. 
Strides towards understanding their physical and chemical characteristics have been made in the last decades, leading to the discovery of a great variety of asteroid types (e.g. \cite{Bus1999}, \cite{Zellner1985} and later \cite{DEMEO}). Recently, \cite{Greenwood2020} estimated the number of parent asteroids needed to represent our meteorite collection on Earth between 95 and 148.\\
There are several techniques constraining the physical and chemical properties and/or the composition of asteroids. Among them, reflectance spectroscopy is one of the most widely applied methods (\cite{Burbine02}, \cite{Reddy2015}). 
For one, it is a remote characterization method that allows for the analysis of asteroid surfaces from the ground or a spacecraft. As a ground based method it can, therefore, constitutes a low cost technique in comparison to, e.g. space missions that return samples. Most importantly however, the possibility of applying this technique in laboratory on Earth enables a direct comparison with a series of samples.  
While spectral measurements can reveal a lot of information, the reflectance spectra acquired for asteroids in the 0.45 - \SI{2.45}{\micro\meter} region show only two main absorption features, which can be very faint in some cases \citep{DEMEO}. To improve the interpretation of asteroid spectra, it can be useful to do some measurements on well known samples. Several in depth studies of reflectance spectroscopy on meteorites have been led in the past (e.g. \cite{Cloutis_CR}, \cite{Cloutis_CO}, \cite{Cloutis2012}, \cite{Gaffey1976}, \cite{Vernazza2014}).\\
Based on these previous studies, chondritic meteorites have been found to display three main absorption bands in their spectra at \SI{1}{\micro\meter}, \SI{2}{\micro\meter} and \SI{3}{\micro\meter}. 
The origin of these features are attributed to anhydrous and hydrous silicates whose exact compositions will vary with the considered chondrite group as well as the secondary processes they experienced on their asteroidal parent bodies. Consequently, the position and depth of the absorption bands are modified.\\
The chondrite groups considered in this work are carbonaceous chondrites (CV, CO, CR) and Unequilibrated Ordinary Chondrites (UOC). The main conclusions from past works are the following:

\begin{itemize}
\item \textbf{CV chondrites \citep{Cloutis2012}}: Based on the 11 CV chondrites measured in this previous work, the position and morphology of the \SI{1}{\micro\meter} absorption band is consistent with the presence of ferrous olivine. The \SI{2}{\micro\meter} region shows signs of being affected by Fe-bearing pyroxene, but can also exhibit contributions due to spinel. The spectral slopes in the \SI{1}{\micro\meter} and \SI{2}{\micro\meter} region become increasingly positive or red sloped with increasing metamorphic grade. While \cite{Cloutis2012} did expect to see an increase in reflectance, bluer spectral slopes and better defined mafic silicate bands with increasing thermal metamorphism, no correlation between the spectral features and the metamorphic grade could be observed with the considered samples.
\item \textbf{CO chondrites \citep{Cloutis_CO}}: This study included 16 CO chondrites. The major silicate contributing to the \SI{1}{\micro\meter} and \SI{2}{\micro\meter} absorption features is olivine. The abundance of low-Ca pyroxene is expected to be relatively low with possible contributions due to spinel. The position and band depth of the \SI{1}{\micro\meter} absorption feature are both correlated with the metamorphic grade. The overall reflectance in the \SI{700}{\nano\meter} region increases with higher thermal metamorphism. This peak reflectance appears to correlate with the \SI{1}{\micro\meter} and \SI{2}{\micro\meter} band depths as well as with the overall spectral slope.
\item \textbf{CR chondrites \citep{Cloutis_CR}}: This study was based in 14 CR chondrites. Generally, CR chondrites exhibit only weak \SI{1}{\micro\meter} and \SI{2}{\micro\meter} absorption features due to the low amount of Fe-rich silicates present in the samples. The \SI{1}{\micro\meter} band is related to the presence of low-Fe olivine, low-Fe pyroxene and  phyllosilicates. In case of terrestrial weathering, a contribution at \SI{900}{\nano\meter} due to Fe oxyhydroxides can be present. Low-Fe pyroxene shows a second absorption feature at \SI{2}{\micro\meter}. The spectra show a red overall spectral slope. The \SI{1}{\micro\meter} absorption band depth is correlated with the weathering degree of the samples. With increasing aqueous alteration the amount of phyllosilicates increases, which is reflected in a more pronounced absorption band at \SI{900}{\nano\meter}.
\item \textbf{UOCs \citep{Gaffey1976} and \cite{Vernazza2014}}: Ordinary chondrites (OC) have been well studied in the past showing reddish slopes, a broad peak reflectance in the visible wavelength range (near \SI{700}{\nano\meter}) which is dependent on the metamorphic grade as well as olivine and pyroxene absorption features in the \SI{1}{\micro\meter} and \SI{2}{\micro\meter} region. \cite{Gaffey1976} found feldspar to be linked to various near-IR absorption bands located at \SI{900}{\nano\meter} and \SI{970}{\nano\meter}, near \SI{1900}{\nano\meter}, as well as an inflection in the spectra near \SI{1300}{\nano\meter}. However, the presence of feldspar explaining these absorption features has since been disproved (e.g. \cite{Crown1987}). These observations are greatly based on Equilibrated ordinary chondites (EOCs). UOCs have recently been studied in comparison to EOCs by \cite{Vernazza2014}, which analyzed a set of 53 UOCs. Generally, UOCs exhibit stronger absorption features than carbonaceous chondrites.
\end{itemize}
The present work presents several novelties: we conduct the analysis of more than 70 chondrites (CV, CO, CR and UOCs) and nearly all chondrites we consider have a well constraint post-accretion history. Their metamorphic grade was previously determined by \cite{Bonal2006}, \cite{Bonal2007}, \cite{Bonal2016} and \cite{Quirico2009}. The aqueous alteration history of CV chondrites was assessed in \cite{Bonal_TGA}. This allows us to analyze the consequences of secondary processes on the reflectance spectra of meteorites and, thus, of asteroids. 
We contribute to the analysis of UOC spectra which, until recently \citep{Vernazza2014} have received much less attention EOCs. However, this will not be the main focus of the present paper.
By taking advantage of a new spectro-radio goniometer available at the Institute de Plan\'etologie et d'Astrophysique Grenoble (IPAG) \citep{SHADOWS} we enable measurement under vacuum and at higher temperatures than ambient. This allows us to eliminate terrestrial water contamination.
We introduce a set of spectral properties, including a range of absorption band depths and positions as well as spectral slopes, following earlier works by \cite{Cloutis2012}. 
Our objectives are to (i) deepen our understanding of reflectance spectral features, (ii) understand if some of these features are controlled by secondary processes such as thermal metamorphism, and (iii) establish some genetic link between asteroids and carbonaceous chondrites, that are not firmly identified for type 3 chondrites.

\section{Sample lists and Experimental procedure}
\subsection{Sample list}
\label{Sec:MeasuredSamples}
In this paper 4 CR \citep{Eschrig_CR}, 23 CV \citep{Eschrig_CV}, 31 UOCs \citep{Eschrig_UOC} and 15 CO \citep{Eschrig_CO}  were measured. All data are available online in the GhoSST/SSHADE spectral database. With the exception of the four type 2 CR chondrites, all samples considered in this work are type 3 chondrites. 
CV chondrites can generally be sub-divided into oxidized (CV$_\textrm{Ox}$) and reduced (CV$_\textrm{Red}$) based on their metal abundance as well as their Ni in sulfides and Ni in metal content. According to this classification CV$_\textrm{Ox}$ can be further sub-divided into CV$_\textrm{OxA}$ and CV$_\textrm{OxB}$ with CV$_\textrm{OxA}$ exhibiting a higher metal abundance and lower
Ni in sulfide content than CV$_\textrm{OxB}$.
The samples considered in this work were previously classified by \cite{Bonal_TGA}.
6 of the 23 CV chondrites are classified as CV$_\textrm{Red}$, 9 as CV$_\textrm{OxA}$ and 8 as CV$_\textrm{OxB}$. A list of the considered samples as well as their classification is given in Tables \ref{Tab:ReflRes_Carb} and \ref{Tab:ReflRes2_Carb}. 
UOCs are sub-divided into three classes based on their total iron abundance. H types exhibit a total average metal abundance of 8.4 vol.$\%$ while for L types the value reaches 4.1 vol.$\%$ and for LL types it is 2.0 vol.$\%$ \citep{ClasMet}. 10 of the 31 UOCs considered in this work are classified as H, 15 as L, and 6 as LL as indicated in Tables \ref{Tab:ReflRes_Ord} and \ref{Tab:ReflRes2_Ord}. All the meteorites characterized in the present work are remaining samples that had been initially allocated for previous studies (\cite{Bonal2016} for CV, CO and UOCs) by the NASA Meteorite Working Group (Johnson Space Center, Houston, USA). Most of the samples presented in this work have a well constraint thermal history. The petrologic type (PT) previously determined by \cite{Bonal2006}, \cite{Bonal2007}, \cite{Bonal2016} and \cite{Quirico2009} through Raman spectroscopy is given for each sample in Tables \ref{Tab:ReflRes_Carb} and \ref{Tab:ReflRes_Ord} as well. However, all plotting was done using the FWHM$_\textrm{D}$ values determined from the Raman spectra, since this value allows for a more precise distinction of the metamorphic grades. For more explanation we refer to \cite{Bonal2016}.\\
Additionally to the chondrites that were measured in the present work, a set of 17 asteroidal end member spectra (\cite{DEMEO}, Database: \url{http://smass.mit.edu/busdemeoclass.html}), 24 Eos family member spectra \citep{EOS} and 9 CK chondrite spectra (RELAB Database (\url{http://www.planetary.brown.edu/relab/})) were included. This was done to improve our understanding of asteroid spectra and investigate possible links between asteroid and chondrite types. 
The asteroidal end member spectra range from \SI{450}{\nano\meter} to \SI{2450}{\nano\meter}. The 24 Eos family spectra range from \SI{850}{\nano\meter} to \SI{2485}{\nano\meter}. The 9 CK chondrite spectra range from \SI{300}{\nano\meter} to \SI{2600}{\nano\meter}. 
All literature spectra were treated in the same way as the chondrite spectra measured in this work (see Section \ref{Sec:ExpProc}). 
Since the spectral ranges of the literature spectra are smaller than those of our experimental data, not all spectral features could be compared between the data sets. Indeed, none of the literature spectra include the hydration band in the \SI{3}{\micro\meter} range. Moreover, C, Cg, Cgh, Ch, B, T, X, Xc, Xe, Xk and D-type end member spectra exhibit only a faint or no \SI{2}{\micro\meter} absorption band which is why the band depth and position are not considered. The same is true for CK chondrite spectra, however, their \SI{2}{\micro\meter} absorption band features are still considered in later comparisons while keeping the larger uncertainties in mind.

\definecolor{grund}{gray}{.7} 

\begin{singlespace}
\begin{table}[]
\centering
\caption{Spectral parameters determined from the reflectance spectra of carbonaceous chondrites (CO, CV and CR). The Integrated Band Depth (\Area), and position of the hydration band (Hyd. Band Pos.) in the \SI{3}{\micro\meter} range are shown for each sample. Furthermore, the spectral slope in the \SI{1}{\micro\meter} and \SI{2}{\micro\meter} range as well as the Visual slope are listed. For each spectral value, the average and the standard deviation are given. For sample marked by \x the sapphire window correction was done using the measuring program (see Section \ref{Sec:ExpProc}). For sample for which the measurements were done at ambient temperatures the \Area value is highlighted in grey. The petrologic type (PT) as determined by \cite{Bonal2006}, \cite{Bonal2007} and \cite{Bonal2016} are given as well. If no value is given, the determination of the PT was unsuccessful in \cite{Bonal2016}.}
\label{Tab:ReflRes_Carb}
\scalebox{0.7}{
\begin{tabular}{lllcllllr}
\toprule
 &    &			
 \multicolumn{1}{p{2cm}}{\centering \Area (\%)} & 
 \multicolumn{1}{p{2cm}}{\centering Hyd. Band Pos. (\si{\nano\meter})} &				\multicolumn{1}{p{2cm}}{\centering \SI{1}{\micro\meter} slope ($10^{-7}$~\si{\per\nano\meter})} & 				\multicolumn{1}{p{2cm}}{\centering \SI{2}{\micro\meter} slope ($10^{-7}$~\si{\per\nano\meter})} & 	\multicolumn{1}{p{2cm}}{\centering Visual slope ($10^{-5}$~\si{\per\nano\meter})} &  	\multicolumn{1}{p{2cm}}{\centering PT} &
 	 \\ \midrule
 	\multirow{9}{*}{\begin{turn}{90} \textbf{CV$_\textrm{OxA}$ chondrites} \end{turn}}  & ALH 81003  & 5.9 $\pm$ 2.7  & 2960 & -18.3$\pm$14.8 &      -13.1$\pm$12.3 &       9.5$\pm$0.3 & $>$3.6\\
        & Allende    & 6.9  $\pm$ 1.8  & 2940 & 106.0$\pm$7.5 &       25.4$\pm$14.6 &       6.5$\pm$0.4 & $>$3.6\\
        & Axtell     & 19.9 $\pm$ 1.2  & 3000 & 123.0$\pm$2.5 &       -68.4$\pm$6.7 &      30.9$\pm$0.8 & $>$3.6\\
        & GRA 06101  & 5.1  $\pm$ 1.4  & 2940 & 118.0$\pm$3.1 &       10.0$\pm$14.3 &      16.4$\pm$1.8 & $>$3.6\\
        & LAP 02206  & 1.6  $\pm$ 1.21 & 2900 & 76.1$\pm$2.1 &         1.0$\pm$8.0 &      13.1$\pm$0.9 & $>$3.7\\
        & MIL 07002  & 10.8 $\pm$ 1.5  & 2940 & 158.0$\pm$4.7 &        9.0$\pm$31.9 &      22.6$\pm$0.3 & -\\
        & MIL 07671  & 7.4  $\pm$ 1.3  & 2900 & 97.1$\pm$2.8 &      -17.8$\pm$12.3 &      19.5$\pm$0.6 & 3.1\\
        & MIL 091010 & 9.8  $\pm$ 1.6  & 2900 & 72.1$\pm$2.7 &      -12.2$\pm$10.0 &      17.9$\pm$1.1 & $>$3.6\\
        & QUE 94688  & 19.6 $\pm$ 1.6  & 2920 & -5.6$\pm$3.0 &       -99.7$\pm$9.4 &      26.5$\pm$1.8& $>$3.6\\ \midrule
 &   \textbf{average}   &  9.7 $\pm$	2.1 &	2933 &	80.7$\pm$19.5 &  -18.4$\pm$13.5 &  18.1$\pm$2.6 &
  \\ \midrule
\multirow{8}{*}{\begin{turn}{90} \textbf{CV$_\textrm{OxB}$ chondrites} \end{turn}} & ALH 85006  & 13.7 $\pm$ 2.9  & 2960 & -29.3$\pm$11.2 &    12.1$\pm$16.7 &      9.5$\pm$0.2 & 3.6\\
        & Grosnaja   & 10.1  $\pm$ 1.8  & 2960 & -68.7$\pm$2.0 &    -31.7$\pm$6.1 &      4.6$\pm$0.5& $\sim$3.6\\
        & Kaba       & 11.8 $\pm$ 2.0  & 2980 & -12.9$\pm$3.2 &     31.0$\pm$6.9 &      6.7$\pm$0.1& 3.1\\
        & LAR 06317  & 16.3 $\pm$ 2.5  & 2980 & 1.8$\pm$6.8 &     21.3$\pm$9.5 &     11.1$\pm$0.2 & 3.4-3.6\\
        & MCY 05219  & 5.0  $\pm$ 3.0  & 2960 & -88.6$\pm$4.6 &   -12.2$\pm$16.7 &      4.0$\pm$0.2 & $>$3.6\\
        & MET 00761  & 14.2 $\pm$ 3.1  & 2960 & -88.8$\pm$7.1 &   -16.8$\pm$26.4 &     21.6$\pm$2.5 & 3.6\\
        & MET 01074 \x  & 1.4  $\pm$ 0.6  & 2800 & -40.8$\pm$0.8 &     11.6$\pm$2.9 &      3.2$\pm$0.5 & 3.6\\
        & Mokoia  \x   & 1.9  $\pm$ 2.1 & 2900 & 36.0$\pm$2.4 &    21.3$\pm$11.3 &      5.8$\pm$0.6 & $\sim$3.6\\ \midrule
 &    \textbf{average}          & 9.3 $\pm$	2.0  &	2967 &	-36.4$\pm$15.7 &  4.6$\pm$7.8 &  8.3$\pm$2.1 &
  \\ \midrule
\multirow{6}{*}{\begin{turn}{90} \textbf{CV$_\textrm{Red}$} \end{turn}}
 & Efremovka & 12.5 $\pm$ 1.7 & 2960 & -11.7$\pm$7.4 &     -22.2$\pm$5.2 &      20.6$\pm$0.6 & 3.1-3.4 \\
 & GRO 95652 & 13.7 $\pm$ 2.3 & 2960 & -19.3$\pm$3.2 &     -9.0$\pm$12.4 &      19.2$\pm$0.9 & 3.6 \\
 & MIL 07277 & 6.3  $\pm$ 1.3 & 2940 & -27.8$\pm$2.9 &     -5.6$\pm$14.7 &      17.1$\pm$0.7& 3.4-3.6 \\
 & RBT 04302 & 12.1 $\pm$ 1.7 & 2920 & -39.2$\pm$3.5 &      -8.9$\pm$5.8 &      20.0$\pm$1.1 & 3.1-3.4 \\
 & Leoville \x  & 8.9 $\pm$ 3.7 & 2880 & 26.5$\pm$7.1 &     30.3$\pm$14.8 &       9.2$\pm$0.4 & 3.1-3.4\\
 & Vigarano \x  & 6.2  $\pm$ 3.3 & 2900 & -18.2$\pm$12.0 &     -9.2$\pm$20.6 &      14.2$\pm$0.2 & 3.1-3.4\\ \midrule
 &    \textbf{average}       & 9.9 $\pm$ 1.3 & 2927 & -15.0$\pm$9.1 &  -4.1$\pm$7.3 &  16.7$\pm$1.8 & \\ \midrule
\multirow{14}{*}{\begin{turn}{90} \textbf{CO chondrites} \end{turn}}
 & ALHA 77003  & 6.3  $\pm$ 1.5  & 2980 & 18.6 $\pm$3.1 &      -14.6 $\pm$13.1 &      25.0 $\pm$0.6 & $>$3.6\\
        & DOM 08006  & 12.1 $\pm$ 2.3  & 2960 & 34.1 $\pm$2.5 &        38.4 $\pm$9.2 &      10.0 $\pm$0.3& 3.0\\
        & MIL 05024  & 9.2  $\pm$ 2.1  & 2900 & -11.9 $\pm$2.8 &        32.7 $\pm$9.6 &      10.8 $\pm$0.4 & 3.1\\
        & MIL 07193  & 11.0 $\pm$ 2.2  & 2900 & -31.1 $\pm$2.3 &         8.2 $\pm$6.6 &      11.0 $\pm$0.5 & 3.1\\
        & ALH 83108 \x & \cellcolor{lightgray}{7.8 $\pm$ 0.3}  & 2960 & -8.7 $\pm$3.5 &      -106.0 $\pm$3.1 &      36.7 $\pm$2.0  & -\\
        & EET 92126 \x & \cellcolor{lightgray}{11.2 $\pm$ 0.3}  & 2960 & -13.5 $\pm$0.9 &      -102.0 $\pm$2.0 &      51.1 $\pm$1.2 & -\\
        & MET 00737 \x & \cellcolor{lightgray}{11.4 $\pm$ 0.3} & 2920 & -10.4 $\pm$1.2 &       -21.7 $\pm$2.4 &      41.5 $\pm$1.3  & -\\
        & MIL 05104 \x & \cellcolor{lightgray}{12.4 $\pm$ 0.4}  & 2900 & 24.1 $\pm$1.2 &        65.0 $\pm$2.0 &      19.5 $\pm$0.8 & 3.1\\
        & MIL 07709 \x & \cellcolor{lightgray}{13.3  $\pm$ 0.3}  & 2940 & 90.0 $\pm$1.0 &       -28.0 $\pm$2.1 &      22.7 $\pm$4.1 & 3.7\\
        & Moss  \x     & \cellcolor{lightgray}{5.6 $\pm$ 0.4}  & 2880 & 95.5 $\pm$1.6 &        11.4 $\pm$3.1 &      15.3 $\pm$1.8 & 3.6\\
        & Kainsaz \x   & \cellcolor{lightgray}{6.1 $\pm$ 0.5}  & 2960 & 80.2 $\pm$0.9 &        54.2 $\pm$1.9 &      16.8 $\pm$0.2  & 3.6\\
        & LAP 031117 \x & \cellcolor{lightgray}{15.7 $\pm$ 1.0}  & 2940 & 12.3 $\pm$0.7 &        26.3 $\pm$1.9 &      10.0 $\pm$0.4 & 3.05\\
        & QUE 97416 \x  & \cellcolor{lightgray}{16.5 $\pm$ 0.3}  & 2960 & 51.9 $\pm$1.2 &       -62.5 $\pm$3.0 &      52.4 $\pm$1.8 & -\\
        & ALH 85003 \x  & \cellcolor{lightgray}{9.8  $\pm$ 0.3}  & 2960 & 20.2 $\pm$1.1 &       -84.2 $\pm$3.2 &      37.2 $\pm$2.2 & -\\
        & DOM 03238 \x  & \cellcolor{lightgray}{9.8 $\pm$ 0.8}  & 2960 & -8.5 $\pm$1.2 &        31.0 $\pm$3.3 &      13.2 $\pm$0.5 & 3.1\\ \midrule
 &    \textbf{average}        & 9.7 $\pm$ 1.3  & 2935 & 22.9 $\pm$10.4 &  -10.1 $\pm$14.5 &  24.9 $\pm$3.9 & 
 \\ \midrule
\multirow{4}{*}{\begin{turn}{90} \textbf{CR} \end{turn}}
  & EET 92042  & 17.2 $\pm$ 1.0  & 2920 & 142.0$\pm$2.2 &      78.9$\pm$12.3 &      25.3$\pm$0.8 \\
        & GRA 95229  & 20.9 $\pm$ 1.0  & 2960 & 127.0$\pm$3.0 &       42.9$\pm$9.9 &      31.8$\pm$1.4 \\
        & LAP 04720  & 17.4 $\pm$ 1.5  & 2940 & 55.1$\pm$2.6 &       43.6$\pm$7.4 &      24.6$\pm$1.5 \\
        & MIL 090657 & 20.9 $\pm$ 1.4  & 2900 & 51.4$\pm$5.8 &        2.5$\pm$8.3 &      21.8$\pm$1.2  \\ \midrule
&    \textbf{average}         & 19.1 $\pm$ 1.0  & 2930 & 93.9$\pm$23.7 &  42.0$\pm$15.6 &  25.9$\pm$2.1  \\
\bottomrule
\end{tabular}}
\end{table}

\begin{table}[]
\centering
\caption{Spectral parameters determined from the reflectance spectra of carbonaceous chondrites (CO, CV and CR). Listed are the peak reflectance value in the \SI{700}{\nano\meter} range as well as the band depths and positions of the \SI{1}{\micro\meter} and \SI{2}{\micro\meter} bands. For each spectral value, the average and the standard error are given. For sample marked by \x the sapphire window correction was done using the measuring program (see Section \ref{Sec:ExpProc}). Samples which are provided with highlighted \SI{2}{\micro\meter} band depth and position values exhibit very faint spectral features, therefore, making the exact determination of these spectral parameters difficult.}
\label{Tab:ReflRes2_Carb}
\scalebox{0.7}{
\begin{tabular}{lllllcc}
\toprule
 &    & 
 \multicolumn{1}{p{2cm}}{\centering \SI{700}{\nano\meter} peak (\%)} &
 	\multicolumn{1}{p{2cm}}{\centering\SI{1}{\micro\meter} band depth \\ (\%)} & 
 	\multicolumn{1}{p{2cm}}{\centering \SI{2}{\micro\meter} band depth \\ (\%)} & 
 	\multicolumn{1}{p{2cm}}{\centering \SI{1}{\micro\meter} \\ band \\ pos. (\si{\nano\meter})} & 
 	\multicolumn{1}{p{2cm}}{\centering \SI{2}{\micro\meter} \\ band \\ pos. (\si{\nano\meter})} \\ \midrule
 	\multirow{9}{*}{\begin{turn}{90} \textbf{CV$_\textrm{OxA}$ chondrites} \end{turn}} & ALH 81003 &      13.2$\pm$0.1 &           5.3$\pm$1.8 &           5.7$\pm$0.9 &             1080 &             2040 \\
  & Allende &       9.8$\pm$ $<$ 0.1 &           4.4$\pm$0.9 &           4.3$\pm$0.8 &             1060 &             1960 \\
  &  Axtell &      10.8$\pm$ $<$ 0.1 &           6.9$\pm$0.4 &           7.4$\pm$0.5 &             1060 &             1960 \\
  &GRA 06101 &      11.1$\pm$ $<$ 0.1 &           4.0$\pm$0.5 &           3.3$\pm$1.0 &             1040 &             2100 \\
  &LAP 02206 &       9.8$\pm$ $<$ 0.1 &           5.8$\pm$0.5 &           4.1$\pm$0.7 &             1080 &             1940 \\
  &MIL 07002 &      14.4$\pm$ $<$ 0.1 &           5.9$\pm$0.5 &           4.4$\pm$0.9 &             1080 &             1960 \\
  &MIL 07671 &      10.5$\pm$ $<$ 0.1 &           6.2$\pm$0.6 &           4.3$\pm$0.9 &             1060 &             1960 \\
 &MIL 091010 &      10.1$\pm$ $<$ 0.1 &           5.1$\pm$0.6 &           4.0$\pm$0.7 &             1040 &             1960 \\
 & QUE 94688 &      10.8$\pm$ $<$ 0.1 &           7.2$\pm$0.6 &           6.9$\pm$0.7 &             1060 &             1960 \\ \midrule
 &    \textbf{average}        & 11.2$\pm$0.5 &       5.6$\pm$0.4 &       4.9$\pm$0.5 &       1062 &      1982
\\ \midrule
\multirow{8}{*}{\begin{turn}{90} \textbf{CV$_\textrm{OxB}$ chondrites} \end{turn}} & ALH 85006 &      9.4$\pm$0.1 &           3.9$\pm$1.4 &           4.9$\pm$1.3 &             1040 &             1960 \\
 & Grosnaja &      8.6$\pm$ $<$0.1 &           3.7$\pm$0.5 & \cellcolor{lightgray}{3.5 $\pm$ 0.6} & 1060 & \cellcolor{lightgray}{2380} \\
 &    Kaba &      7.3$\pm$ $<$0.1 &           2.6$\pm$0.7 &           2.8$\pm$0.8 &             1100 &             1980 \\
 &LAR 06317 &      9.3$\pm$ $<$0.1 &           3.8$\pm$1.0 &           4.4$\pm$0.7 &             1060 &             1960 \\
 &MCY 05219 &      8.7$\pm$ $<$0.1 &           2.9$\pm$0.9 &           3.6$\pm$1.2 &             1060 &             1960 \\
 &MET 00761 &     11.1$\pm$ $<$0.1 &           2.8$\pm$1.0 &           3.6$\pm$1.3 &             1060 &             1920 \\
 &  Mokoia &      6.8$\pm$ $<$0.1 &           4.4$\pm$0.8 &           2.6$\pm$1.0 &             1060 &             1920 \\
 &MET 01074 &      7.6$\pm$ $<$0.1 &           2.3$\pm$0.2 &           2.1$\pm$0.4 &             1060 &             1920 \\ \midrule
 &    \textbf{average}        & 8.6$\pm$0.5 &       3.3$\pm$0.3 &       3.0$\pm$0.5 &       1062 &     1946 
\\ \midrule
\multirow{6}{*}{\begin{turn}{90} \textbf{CV$_\textrm{Red}$} \end{turn}}
 &Efremovka &     13.1$\pm$ $<$0.1 &           2.3$\pm$0.7 &           2.5$\pm$0.4 &             1040 &             1960 \\
 & GRO 95652 &      7.1$\pm$ $<$0.1 &           4.3$\pm$0.9  & \cellcolor{lightgray}{4.3 $\pm$ 1.0} & 1020 & \cellcolor{lightgray}{2380} \\
 & MIL 07277 &     10.2$\pm$ $<$0.1 &           3.3$\pm$0.7 &           3.1$\pm$1.1 &             1040 &             1960 \\
 & RBT 04302 &      8.1$\pm$ $<$0.1 &           3.0$\pm$0.8 &           3.9$\pm$0.8 &             1040 &             1860 \\
 & Leoville \x &      6.7$\pm$ $<$0.1 &           1.4$\pm$3.2 &           1.2$\pm$1.7 &             1380 &             1860 \\
 & Vigarano \x &      7.5$\pm$ $<$0.1 &           2.6$\pm$2.0 &           2.7$\pm$2.5 &             1380 &             1900 \\ \midrule
 & \textbf{average}  & 8.8$\pm$1.0 &       2.8$\pm$0.4 &       2.7$\pm$0.4 &      1150 &      1908 
\\ \midrule
\multirow{16}{*}{\begin{turn}{90} \textbf{CO chondrites} \end{turn}}
 &ALHA 77003 &      10.5 $\pm$ $<$0.1 &           7.3 $\pm$0.6 &           5.3 $\pm$1.0 &             1080 &             1960 \\
 & DOM 08006 &       5.0 $\pm$ $<$0.1 &           2.8 $\pm$0.4 &  \cellcolor{lightgray}{3.4 $\pm$ 1.1} & 980  & \cellcolor{lightgray}{2460} \\
 & MIL 05024 &       5.5 $\pm$ $<$0.1 &           4.6 $\pm$0.9 & \cellcolor{lightgray}{2.5 $\pm$ 1.0} & 1120 & \cellcolor{lightgray}{2400} \\
 &MIL 07193 &       5.2 $\pm$ $<$0.1 &           3.2 $\pm$1.1 & \cellcolor{lightgray}{2.8 $\pm$ 1.2} & 1000 & \cellcolor{lightgray}{2300} \\
 &ALH 83108 \x &      19.2 $\pm$ $<$0.1 &           9.9 $\pm$0.2 &           6.4 $\pm$0.1 &             1060 &             1860 \\
 &EET 092126 \x &      17.3 $\pm$ $<$0.1 &           7.5 $\pm$0.1 &           5.0 $\pm$0.1 &             1060 &             1900 \\
 & MET 00737 \x &      16.1 $\pm$ $<$0.1 &           8.7 $\pm$0.1 &           6.5 $\pm$0.2 &             1060 &             1900 \\
  &MIL 05104 \x&       9.5 $\pm$ $<$0.1 &           3.4 $\pm$0.2 &           3.0 $\pm$0.2 &             1400 &             1840 \\
 &MIL 07709 \x&      11.6 $\pm$ $<$0.1 &           5.2 $\pm$0.2 &           3.1 $\pm$0.2 &             1060 &             1940 \\
  &    Moss \x&      13.2 $\pm$ $<$0.1 &           4.9 $\pm$0.2 &           3.4 $\pm$0.2 &             1060 &             1840 \\
  & Kainsaz \x&      10.1 $\pm$ $<$0.1 &           2.2 $\pm$ $<$0.1 &           2.2 $\pm$0.3 &              940 &             1840 \\
 &LAP 031117 \x&       4.5 $\pm$ $<$0.1 &           3.4 $\pm$0.3 &           1.4 $\pm$0.5 &             1380 &             1840 \\
 & QUE 97416 \x&      15.9 $\pm$ $<$0.1 &           5.2 $\pm$0.2 &           4.4 $\pm$0.2 &             1060 &             1840 \\
 & ALH 85003 \x&      16.8 $\pm$ $<$0.1 &           6.2 $\pm$0.2 &           4.8 $\pm$0.2 &             1080 &             2100 \\
 & DOM 03238 \x&       6.6 $\pm$ $<$0.1 &           2.4 $\pm$0.3 &           1.2 $\pm$0.3 &             1400 &             1840 \\ \midrule
 &   \textbf{average}          & 11.1 $\pm$1.3 &       5.1 $\pm$0.6 &       3.8 $\pm$0.5 &      1116 &      1885 \\ \midrule
\multirow{4}{*}{\begin{turn}{90} \textbf{CR} \end{turn}}
 & EET 92042 &      9.7$\pm$ $<$0.1 &           2.9$\pm$0.1 & \cellcolor{lightgray}{3.2 $\pm$ 0.7} & 920 & \cellcolor{lightgray}{2400} \\
 & GRA 95229 &     11.4$\pm$ $<$0.1 &           4.5$\pm$0.1 &           1.9$\pm$0.5 &              920 &             1820  \\
 & LAP 04720 &      9.9$\pm$ $<$0.1 &           5.0$\pm$0.1 & \cellcolor{lightgray}{4.0 $\pm$ 0.6} & 920 & \cellcolor{lightgray}{2400} \\
 & MIL 090657 &      8.1$\pm$ $<$0.1 &           5.4$\pm$0.2 &           1.7$\pm$0.8 &              920 &             1940 \\ \midrule
 &    \textbf{average}        & 9.8$\pm$0.7 &       4.4$\pm$0.5 &       1.8$\pm$0.1 &        920 &      1880 \\ \bottomrule
\end{tabular}}
\end{table}

\begin{table}[]
\centering
\caption{Spectral parameters determined from the reflectance spectra of UOCs. The Integrated Band Depth (\Area), and position of the hydration band (Hyd. Band Pos.) in the \SI{3}{\micro\meter} range are shown for each sample. Furthermore, the spectral slope in the \SI{1}{\micro\meter} and \SI{2}{\micro\meter} range as well as the Visual slope are listed. For each spectral value, the average and the standard deviation are given. For sample marked by \x the sapphire window correction was done using the measuring program (see Section \ref{Sec:ExpProc}). For sample for which the measurements were done at ambient temperatures the \Area value is highlighted in grey. The petrologic type (PT) as determined by \cite{Bonal2006}, \cite{Bonal2007} and \cite{Bonal2016} are given as well. If no value is given, the determination of the PT was unsuccessful in \cite{Bonal2016}. The ``M'' for ALHA 78119 stands for highly structured carbonaceous matter for which no reference sample was given to determine the PT \citep{Bonal2016}.}
\label{Tab:ReflRes_Ord}
\scalebox{0.8}{
\begin{tabular}{lllcllllr}
\toprule
 &    &			
 \multicolumn{1}{p{2cm}}{\centering \Area (\%)} & 
 \multicolumn{1}{p{2cm}}{\centering Hyd. Band Pos (\si{\nano\meter})} &				\multicolumn{1}{p{2cm}}{\centering \SI{1}{\micro\meter} slope ($10^{-7}$~\si{\per\nano\meter})} & 				\multicolumn{1}{p{2cm}}{\centering \SI{2}{\micro\meter} slope ($10^{-7}$~\si{\per\nano\meter})} & 	\multicolumn{1}{p{2cm}}{\centering Visual slope ($10^{-5}$~\si{\per\nano\meter})} & \multicolumn{1}{p{2cm}}{\centering PT} &
 	 \\ \midrule
\multirow{10}{*}{\begin{turn}{90} \textbf{H} \end{turn}}
   & BTN 00302 & 0.5  $\pm$ 1.0 & 2900 & 58.2  $\pm$ 2.3  & 52.2  $\pm$ 3.9  & 16.6 $\pm$ 1.8 & 3.1-3.4\\
        & EET 83248 & 17.2 $\pm$ 0.9  & 2900 & 69.6  $\pm$ 1.5  & 3.9   $\pm$ 3.7  & 30.7 $\pm$ 3.6 & $>$3.6\\
        & RBT 04251 \x & 10.4 $\pm$ 0.2  & 2900 & 129.1 $\pm$ 1.1  & 51.9  $\pm$ 1.6  & 42.6 $\pm$ 3.5 & 3.4\\
        & MCY 05218 \x & 16.3 $\pm$ 0.4  & 2920 & 129.0 $\pm$ 1.0  & 59.3  $\pm$ 1.6  & 35.2 $\pm$ 3.6 & 3.05-3.1\\
        & WIS 91627 \x & 15.8 $\pm$ 0.2  & 2900 & 108.6 $\pm$ 0.8  & -9.9  $\pm$ 2.6  & 36.9 $\pm$ 4.1 & $>$3.6\\ 
& DOM 08468 \x & 5.5 $\pm$ 4.5 & 2920 & 44.3  $\pm$ 6.5  & 16.3 $\pm$ 38.0 & 25.5 $\pm$ 0.8 & 3.6\\
        & LAR 04382 \x & 13.4 $\pm$ 2.2 & 2900 & 122.3 $\pm$ 7.8  & 18.3 $\pm$ 24.5 & 40.5 $\pm$ 2.2 & 3.1-3.4\\
        & MAC 88174 \x & 5.4 $\pm$ 2.4 & 2920 & 77.9  $\pm$ 4.2  & 74.5 $\pm$ 22.1 & 30.0 $\pm$ 0.6 & $>$3.6\\
        & MET 00506 \x & 22.2 $\pm$ 0.5 & 2900 & 168.6 $\pm$ 1.5  & 96.3 $\pm$ 4.6  & 11.3 $\pm$ 2.7 & 3.1\\
        & WSG 95300 \x & 8.8 $\pm$ 2.8 & 2920 & 99.6  $\pm$ 2.6  & 53.7 $\pm$ 22.9 & 30.2 $\pm$ 1.3 & 3.4\\ \midrule
 &  \textbf{average}  & 11.6 $\pm$ 2.1 & 2908 & 100.9 $\pm$ 12.1 & 44.5 $\pm$ 11.2 & 29.9 $\pm$ 3.2 & \\ \midrule
\multirow{15}{*}{\begin{turn}{90} \textbf{L} \end{turn}}
		& EET 90066 & 17.7 $\pm$ 1.3  & 2900 & 139.9 $\pm$ 2.1  & 25.5  $\pm$ 6.2  & 30.2 $\pm$ 3.5 & 3.1\\
        & EET 90628 & 10.9 $\pm$ 0.9  & 2920 & 110.4 $\pm$ 3.0  & 44.9  $\pm$ 6.4  & 38.2 $\pm$ 2.5 & 3.0\\
        & GRO 06054 \x & 12.8 $\pm$ 0.9  & 2900 & 98.4  $\pm$ 2.1  & 34.0  $\pm$ 5.7  & 35.0 $\pm$ 3.0 & 3.05\\
        & LEW 87248 \x & 4.8  $\pm$ 0.3  & 2900 & 43.3  $\pm$ 0.7  & 71.3  $\pm$ 2.2  & 29.3 $\pm$ 2.4 & 3.0\\
        & MET 00489 \x & 10.3 $\pm$ 0.3  & 2900 & 127.6 $\pm$ 0.6  & 32.4  $\pm$ 1.2  & 47.2 $\pm$ 2.7 & 3.05-3.1\\
        & LEW 87284 \x & 4.9  $\pm$ 0.2  & 2900 & 83.4  $\pm$ 0.8  & 89.6  $\pm$ 3.1  & 39.1 $\pm$ 1.3 & 3.1-3.4\\
        & ALH 83008 \x & \cellcolor{lightgray}{20.7 $\pm$ 0.4}  & 2980 & 89.7  $\pm$ 0.8  & -0.1  $\pm$ 2.3  & 28.5 $\pm$ 3.1 & $>$3.6\\
        & ALH 84086 \x & \cellcolor{lightgray}{7.5  $\pm$ 0.2}  & 2960 & 180.9 $\pm$ 2.5  & 126.9 $\pm$ 3.2  & 44.7 $\pm$ 1.9 & -\\
        & ALH 84120 \x & \cellcolor{lightgray}{6.7  $\pm$ 0.2}  & 3000 & 83.6  $\pm$ 3.3  & 123.8 $\pm$ 3.0  & 41.7 $\pm$ 1.8 & -\\ 
        & DOM 03287 \x & 13.2 $\pm$ 2.6 & 2900 & 13.7  $\pm$ 8.8  & 13.7 $\pm$ 18.4 & 33.6 $\pm$ 1.7 & 3.6\\
        & EET 87735 \x & 12.7 $\pm$ 2.9 & 2920 & 92.8  $\pm$ 2.3  & 15.9 $\pm$ 6.4  & 19.8 $\pm$ 2.4 & 3.05-3.1\\
        & LEW 88617 \x & 11.9 $\pm$ 2.4 & 2900 & 137.5 $\pm$ 4.6  & 49.1 $\pm$ 14.9 & 34.1 $\pm$ 2.3 & 3.6\\
        & LEW 88632 \x & 18.4 $\pm$ 1.5 & 2900 & 155.4 $\pm$ 9.4  & 10.3 $\pm$ 11.0 & 40.7 $\pm$ 3.2 & 3.4\\
        & MIL 05050 \x & 13.9 $\pm$ 1.9 & 2800 & 224.4 $\pm$ 10.3 & 81.7 $\pm$ 11.0 & 31.6 $\pm$ 1.0 & 3.1\\
        & MIL 05076 \x & 18.1 $\pm$ 2.0 & 2920 & 113.1 $\pm$ 2.8  & -2.3 $\pm$ 8.7  & 25.7 $\pm$ 1.8 & 3.4\\ \midrule
 &   \textbf{average}       & 12.3 $\pm$ 1.3 & 2913 & 112.8 $\pm$ 13.5 & 51.9 $\pm$ 11.3 & 34.6 $\pm$ 1.9 & \\ \midrule
\multirow{6}{*}{\begin{turn}{90} \textbf{LL} \end{turn}}
   & ALHA 76004 & 10.5 $\pm$ 1.0  & 2900 & -8.1  $\pm$ 3.0  & -4.7  $\pm$ 5.2  & 30.0 $\pm$ 1.2 & 3.1-3.4\\
        & TIL 82408 \x & 14.9 $\pm$ 0.3  & 2920 & 180.7 $\pm$ 0.6  & 128.7 $\pm$ 2.8  & 34.1 $\pm$ 1.6 & 3.05-3.1\\
        & EET 96188 \x & 12.4 $\pm$ 0.3  & 2900 & 137.1 $\pm$ 0.5  & 69.1  $\pm$ 2.3  & 44.7 $\pm$ 4.0 & 3.1-3.4\\
        & LAR 06279 \x & 18.4 $\pm$ 0.2  & 2940 & 218.2 $\pm$ 2.9  & 51.2  $\pm$ 2.3  & 49.2 $\pm$ 5.1 & 3.05-3.1\\ 
    & ALHA 78119 \x & 12.7  $\pm$ 2.4 & 2940 & 76.6  $\pm$ 4.3  & 33.8 $\pm$ 16.5 & 33.8 $\pm$ 1.8 & M\\
        & LAR 06469 \x & 14.3  $\pm$ 0.8 & 2900 & -0.3  $\pm$ 1.1  & 39.3 $\pm$ 6.7  & 27.9 $\pm$ 1.9 & $>$3.6\\ \midrule
 &    \textbf{average} & 13.9 $\pm$ 1.1 & 2917 & 100.7 $\pm$ 38.4 & 59.6 $\pm$ 18.0 & 36.6 $\pm$ 3.5 &
 \\ \bottomrule
\end{tabular}}
\end{table}

\begin{table}[]
\centering
\caption{Spectral parameters determined from the reflectance spectra of UOCs. Listed are the peak reflectance value in the \SI{700}{\nano\meter} range as well as the band depths and positions of the \SI{1}{\micro\meter} and \SI{2}{\micro\meter} band. For each spectral value, the average and the standard error are given. For sample marked by \x the sapphire window correction was done using the measuring program (see Section \ref{Sec:ExpProc}).}
\label{Tab:ReflRes2_Ord}
\scalebox{0.8}{
\begin{tabular}{lllllcc}
\toprule
 &    & 
 \multicolumn{1}{p{2cm}}{\centering \SI{700}{\nano\meter} peak (\%)} &
 	\multicolumn{1}{p{2cm}}{\centering\SI{1}{\micro\meter} band depth (\%)} & 
 	\multicolumn{1}{p{2cm}}{\centering \SI{2}{\micro\meter} band depth (\%)} & 
 	\multicolumn{1}{p{2cm}}{\centering \SI{1}{\micro\meter} \\ band \\ pos. (\si{\nano\meter})} & 
 	\multicolumn{1}{p{2cm}}{\centering \SI{2}{\micro\meter} \\ band \\ pos. (\si{\nano\meter})} \\ \midrule
\multirow{10}{*}{\begin{turn}{90} \textbf{H} \end{turn}}
          & BTN 00302 & 14.1 $\pm$ 0.1 & 11.5 $\pm$ 0.1 & 5.5  $\pm$ 0.2 & 940  & 1920 \\
        & EET 83248 & 11.7 $\pm$ $<$ 0.1  & 11.8 $\pm$ 0.1 & 6.5  $\pm$ 0.3 & 940  & 1960 \\
        & RBT 04251 \x & 16.3 $\pm$ $<$ 0.1  & 13.6 $\pm$ $<$ 0.1 & 5.7  $\pm$ 0.1 & 940  & 1840 \\
        & MCY 05218 \x & 12.4 $\pm$ $<$ 0.1 & 10.2 $\pm$ 0.1 & 4.0  $\pm$ 0.1 & 940  & 1860 \\
        & WIS 91627 \x & 15.6 $\pm$ $<$ 0.1 & 15.9 $\pm$ $<$ 0.1 & 8.5  $\pm$ 0.1 & 940  & 1840 \\ 
 & DOM 08468 \x & 11.4 $\pm$ $<$ 0.1 & 12.4 $\pm$ 0.1 & 11.8 $\pm$ 0.1 & 960  & 1920 \\
        & LAR 04382 \x & 14.8 $\pm$ $<$ 0.1 & 17.1 $\pm$ 0.1 & 15.8 $\pm$ 0.1 & 940  & 1960 \\
        & MAC 88174 \x & 13.9 $\pm$ $<$ 0.1 & 12.8 $\pm$ 0.1 & 14.5 $\pm$ $<$ 0.1 & 940  & 1940 \\
        & MET 00506 \x & 11.9 $\pm$ $<$ 0.1 & 14.0 $\pm$ 0.5 & 13.3 $\pm$ $<$ 0.1 & 1000 & 2400 \\
        & WSG 95300 \x & 11.1 $\pm$ $<$ 0.1 & 11.5 $\pm$ 1.5 & 12.0 $\pm$ $<$ 0.1 & 1000 & 1940 \\ \midrule
 &  \textbf{average}& 13.3 $\pm$ 0.6 & 13.1 $\pm$ 0.7 & 9.8  $\pm$ 1.3 & 954  & 1958  \\ \midrule
\multirow{9}{*}{\begin{turn}{90} \textbf{L} \end{turn}}  
           & EET 90066 & 10.3 $\pm$ $<$ 0.1 & 11.4 $\pm$ 0.4 & 5.5  $\pm$ 0.4 & 1000 & 1920 \\
        & EET 90628 & 13.5 $\pm$ $<$ 0.1  & 15.1 $\pm$ 0.1 & 7.5  $\pm$ 0.2 & 960  & 1840 \\
        & GRO 06054 \x & 11.7 $\pm$ $<$ 0.1  & 12.6 $\pm$ 0.4 & 5.5  $\pm$ 0.3 & 1000 & 1880 \\
        & LEW 87248 \x & 14.2 $\pm$ $<$ 0.1 & 11.5 $\pm$ $<$ 0.1 & 5.1  $\pm$ 0.1 & 940  & 1840 \\
        & MET 00489 \x & 17.3 $\pm$ $<$ 0.1 & 18.8 $\pm$ $<$ 0.1 & 8.3  $\pm$ 0.1 & 940  & 1940 \\
        & LEW 87284 \x & 15.5 $\pm$ $<$ 0.1 & 12.3 $\pm$ $<$ 0.1 & 5.3  $\pm$ 0.1 & 920  & 1840 \\
        & ALH 83008 \x & 12.6 $\pm$ $<$ 0.1 & 12.5 $\pm$ $<$ 0.1 & 5.0  $\pm$ 0.1 & 940  & 1860 \\
        & ALH 84086 \x & 20.1 $\pm$ $<$ 0.1  & 19.5 $\pm$ 0.1 & 10.5 $\pm$ 0.1 & 940  & 1980 \\
        & ALH 84120 \x & 20.3 $\pm$ $<$ 0.1  & 14.1 $\pm$ 0.1 & 7.5  $\pm$ 0.1 & 920  & 1980 \\ 	
 		& DOM 03287 \x & 12.9 $\pm$ $<$ 0.1 & 13.0 $\pm$ 0.1 & 13.1 $\pm$ 0.1 & 940  & 1940 \\
        & EET 87735\x  & 9.4  $\pm$ $<$ 0.1 & 12.2 $\pm$ $<$ 0.1 & 10.1 $\pm$ $<$ 0.1 & 940  & 1940 \\
        & LEW 88617 \x & 13.3 $\pm$ $<$ 0.1 & 14.6 $\pm$ 0.1 & 14.5 $\pm$ $<$ 0.1 & 960  & 1960 \\
        & LEW 88632 \x & 14.2 $\pm$ $<$ 0.1 & 15.8 $\pm$ 0.3 & 15.4 $\pm$ 0.1 & 1020 & 1960 \\
        & MIL 05050 \x & 11.3 $\pm$ $<$ 0.1 & 9.4  $\pm$ 1.5 & 13.2 $\pm$ 0.1 & 1000 & 1940 \\
        & MIL 05076 \x & 10.7 $\pm$ $<$ 0.1 & 11.1 $\pm$ 0.1 & 11.7 $\pm$ $<$ 0.1 & 940  & 1920 \\ \midrule
 &     \textbf{average}     & 13.8 $\pm$ 0.9 & 13.6 $\pm$ 0.7 & 9.3  $\pm$ 0.9 & 957  & 1916 \\ \midrule
    \multirow{6}{*}{\begin{turn}{90} \textbf{LL} \end{turn}} 
  & ALHA 76004 & 13.9 $\pm$ $<$ 0.1  & 15.5 $\pm$ 0.1 & 8.0  $\pm$ 0.2 & 940  & 1920 \\
      & TIL 82408 \x & 12.5 $\pm$ $<$ 0.1 & 12.1 $\pm$ $<$ 0.1 & 5.4  $\pm$ 0.1 & 940  & 1840 \\
        & EET 96188 \x & 16.2 $\pm$ $<$ 0.1 & 14.9 $\pm$ $<$ 0.1 & 6.2  $\pm$ 0.0 & 940  & 1840 \\
        & LAR 06279 \x & 17.7 $\pm$ $<$ 0.1  & 15.3 $\pm$ 0.1 & 7.3  $\pm$ 0.1 & 940  & 1840 \\ 
 		& ALHA 78119 \x & 14.5 $\pm$ $<$ 0.1 & 16.0 $\pm$ 0.1 & 15.1 $\pm$ $<$ 0.1 & 940 & 1940 \\
        & LAR 06469 \x & 13.2 $\pm$ $<$ 0.1 & 11.2 $\pm$ $<$ 0.1 & 13.2 $\pm$ $<$ 0.1 & 940 & 1960 \\ \midrule
 & \textbf{average} & 14.7 $\pm$ 0.8 & 14.2 $\pm$ 0.8 & 9.3  $\pm$ 1.6 & 940 & 1893
 \\ \bottomrule
\end{tabular}}
\end{table}

\end{singlespace}

\subsection{Experimental procedure}
\label{Sec:ExpProc}
The reflectance spectroscopy was measured using the SHADOWS instrument \citep{SHADOWS}, a spectro-radio goniometer available at IPAG (France). 
The chondrites were all carefully ground using a mortar and pestle until they resembled a fine powder with sub-millimeter grain sizes. 
In previous works by \cite{Garenne2016} the volume weighted average grain size of hand ground chondrites was estimated to be between \SI{100}{\micro\meter} and \SI{200}{\micro\meter}. However, no sieving was performed to determine the exact grain size of the powders. The samples we are looking at are dark and we are interested in a heterogeneous powder containing a continuum of grain sizes as would be expected on asteroid surfaces.
To ensure the representativity of all chondrite components, approximately \SI{100}{\milli\gram} of sample were ground and \SI{50}{\milli\gram} were used to fill the sample holder. To further ensure the comparability between different chondrites the measurements were done on flat surfaces. Thus, a spatula was set onto the edge of the sample holder and run over the top, smoothing and flattening the surface of the powdered sample.\\
The spectra were obtained in the \SI{340}{\nano\meter} to \SI{4200}{\nano\meter} region and were measured under vacuum ($P < \SI{1e-4}{\milli\bar}$) and at \SI{80}{\celsius}. These temperature and pressure conditions ensured the desorption of most terrestrial water contamination, therefore, leaving the hydration feature mostly controlled by chondritic hydration. It needs to be noted here that a few samples, which are highlighted in grey in Tables \ref{Tab:ReflRes_Carb} and \ref{Tab:ReflRes_Ord}, were measured under vaccum but at ambient temperature.
Each sample was measured with a spectral resolution of \SI{20}{\nano\meter} and a measuring geometry of $i = \SI{0}{\degree}$, $e = \SI{30}{\degree}$. This geometry was chosen as a standard for good comparability between measurements of different laboratories. 
To normalize the spectra to the Lambertian surface, two reference spectra were measured prior to the samples. In the wavelength range between \SI{340}{\nano\meter} and \SI{2100}{\nano\meter} a spectralon$^\textrm{TM}$ standard was used. For longer wavelengths between \SI{2100}{\nano\meter} and \SI{4200}{\nano\meter} an infragold$^\textrm{TM}$ standard was used.\\
The environmental cell containing the sample holder was closed by a sapphire window. To correct for the reflection on this window, the sample was first measured without the window, then with the window and finally with the window under vacuum and at \SI{80}{\celsius}. By comparing the spectra measured with and without the window a correction factor could be calculated that allowed for the correction of the values measured under vacuum and at \SI{80}{\celsius}. During the course of the measurements done in this work the correction of the window reflectance was integrated into the program used for the measurements. This allowed for direct measurements under vacuum and at \SI{80}{\celsius}. Although no difference could be observed between manually and non-manually corrected samples, Tables \ref{Tab:ReflRes_Carb} to \ref{Tab:ReflRes2_Ord} indicate which method was applied to each sample.

\subsection{Analytical procedure}
\label{Ch:Analytical}
\begin{figure*}[h]
\centering
\includegraphics[width=\textwidth]{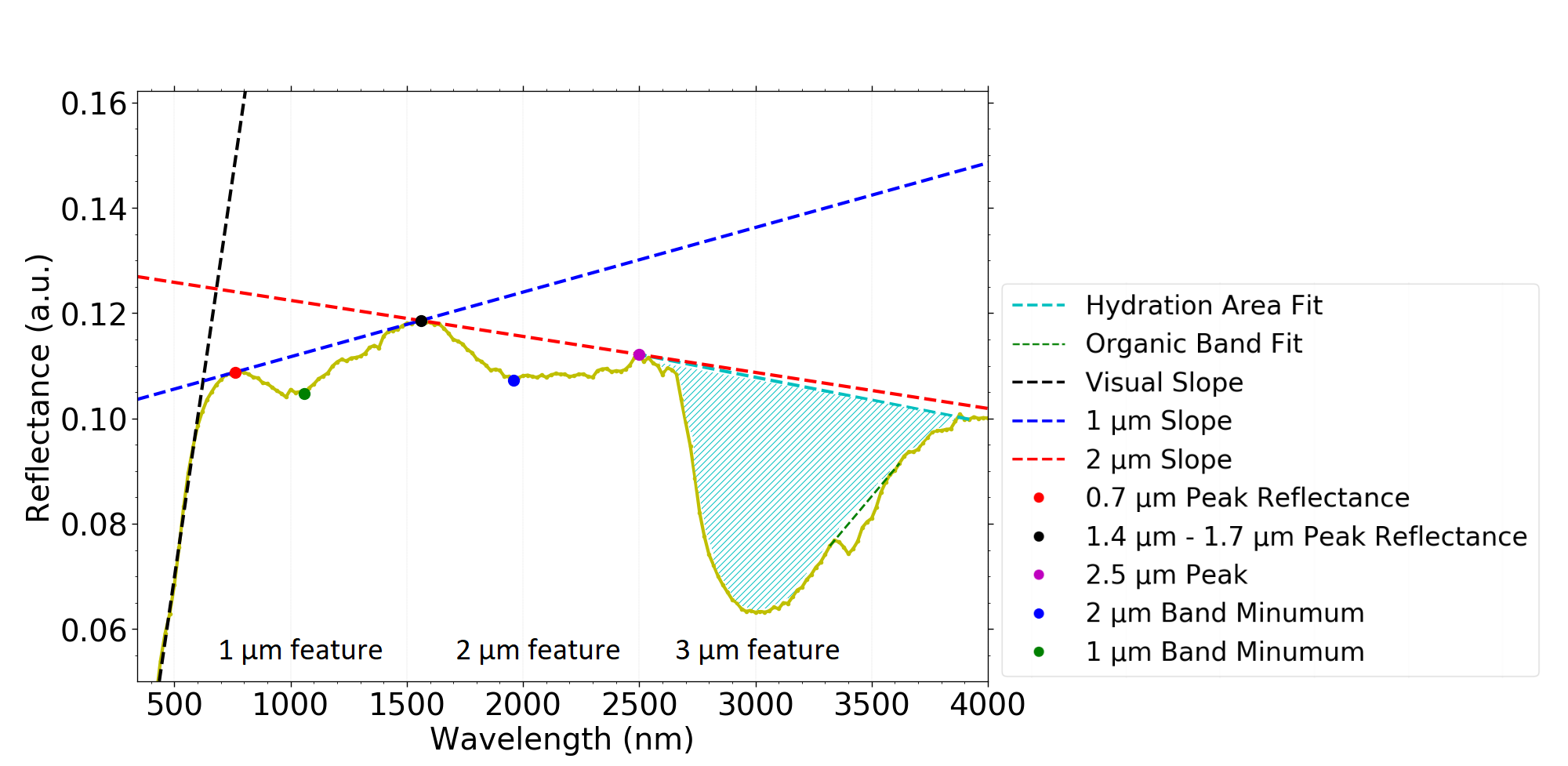}
\caption{Reflectance spectrum of the CV chondrite Axtell indicating the different spectral values which are systematically determined for each sample. The different slopes considered are the visual slope in black, the \SI{1}{\micro\meter} region spectral slope in blue and the \SI{2}{\micro\meter} region spectral slope in red. The cyan line indicates the linear fit to the hydration area. The green dotted line shows the linear fit to the organic bands. The peak reflectance in the visible wavelength range is shown as a red dot, the peak reflectance between \SI{1400}{\nano\meter} and \SI{1700}{\nano\meter} is indicated by a black dot and the peak reflectance between \SI{2300}{\nano\meter} and \SI{2600}{\nano\meter} is shown as a purple dot. The \SI{1}{\micro\meter} and \SI{2}{\micro\meter} band minima are shown as green and blue dots, respectively.}
\label{fig:SpectralValues}
\end{figure*}

The obtained reflectance spectra are analyzed following previous works by \cite{Cloutis2012}. A set of spectral features including absorption band depths and positions as well as spectral slopes and peak reflectance values are introduced. A summary of all spectral features that are determined from each individual spectrum is shown in Figure \ref{fig:SpectralValues}.\\
The \SI{1}{\micro\meter} and \SI{2}{\micro\meter} absorption bands are fit with linear baselines which are then used to determine the depth of the band as calculated by \cite{Spectroscopy}. Furthermore, the positions of the bands are determined at the minimum of reflectance. Besides the two absorption features at \SI{1}{\micro\meter} and \SI{2}{\micro\meter}, an absorption band attributed to hydration (OH and H$_2$O) can be seen at \SI{3}{\micro\meter}.
To characterize it, the Integrated Band Depth (IBD$_\textrm{Hyd}$) is calculated. For this, the organic bands (if present) located in the \SI{3500}{\nano\meter} range are excluded by linearly fitting this region. Subsequently, the IBD$_\textrm{Hyd}$ of the hydration band is calculated according to \cite{MillikenMustard} between \SI{2550}{\nano\meter} and \SI{4000}{\nano\meter}. 
The peak reflectance value in the visible wavelength range is determined between \SI{500}{\nano\meter} and \SI{760}{\nano\meter}. The visual slope is determined by calculating the steepest slope in the \SI{340}{\nano\meter} to \SI{520}{\nano\meter} range and linearly fitting the points around this area. The overall spectral slope is defined as the inclination of the entire spectrum. Additionally, the spectral slope in the \SI{1}{\micro\meter} and \SI{2}{\micro\meter} range are considered by determining the slope of the \SI{1}{\micro\meter} and \SI{2}{\micro\meter} baselines fits.\\
It should be noted that some spectra display weak absorption features, leading to the determination of the band positions occasionally being ambiguous. 
This is the case of CV samples ALH 85006 and GRO 95652, CO samples DOM 08006, MIL 05024, MIL 07193, LAP 031117 and DOM 03238 as well as CR samples EET 92042 and LAP 04720 (Tables \ref{Tab:ReflRes_Carb} and \ref{Tab:ReflRes2_Carb} as well as Fig. \ref{Fig:Ox_Features}, \ref{Fig:CO_Features} and \ref{Fig:CR_Features}).

\section{Results}
\label{Sec:MeasuringResults}
\begin{figure*}
\centering
\begin{subfigure}[c]{0.49\textwidth}
\includegraphics[width=\textwidth]{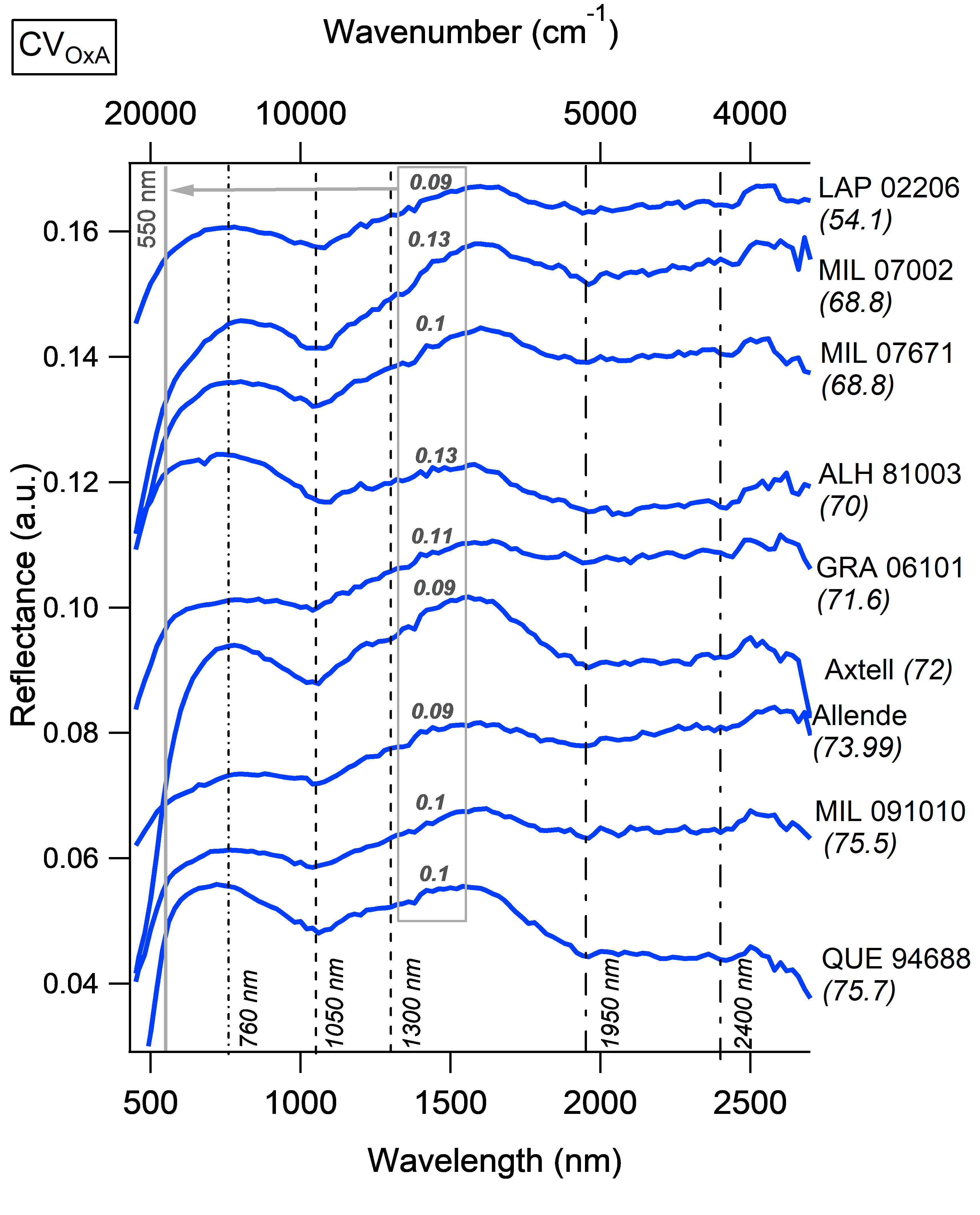}
\subcaption{}
\label{Fig:OXA_Feature}
\end{subfigure}
\begin{subfigure}[c]{0.49\textwidth}
\includegraphics[width=\textwidth]{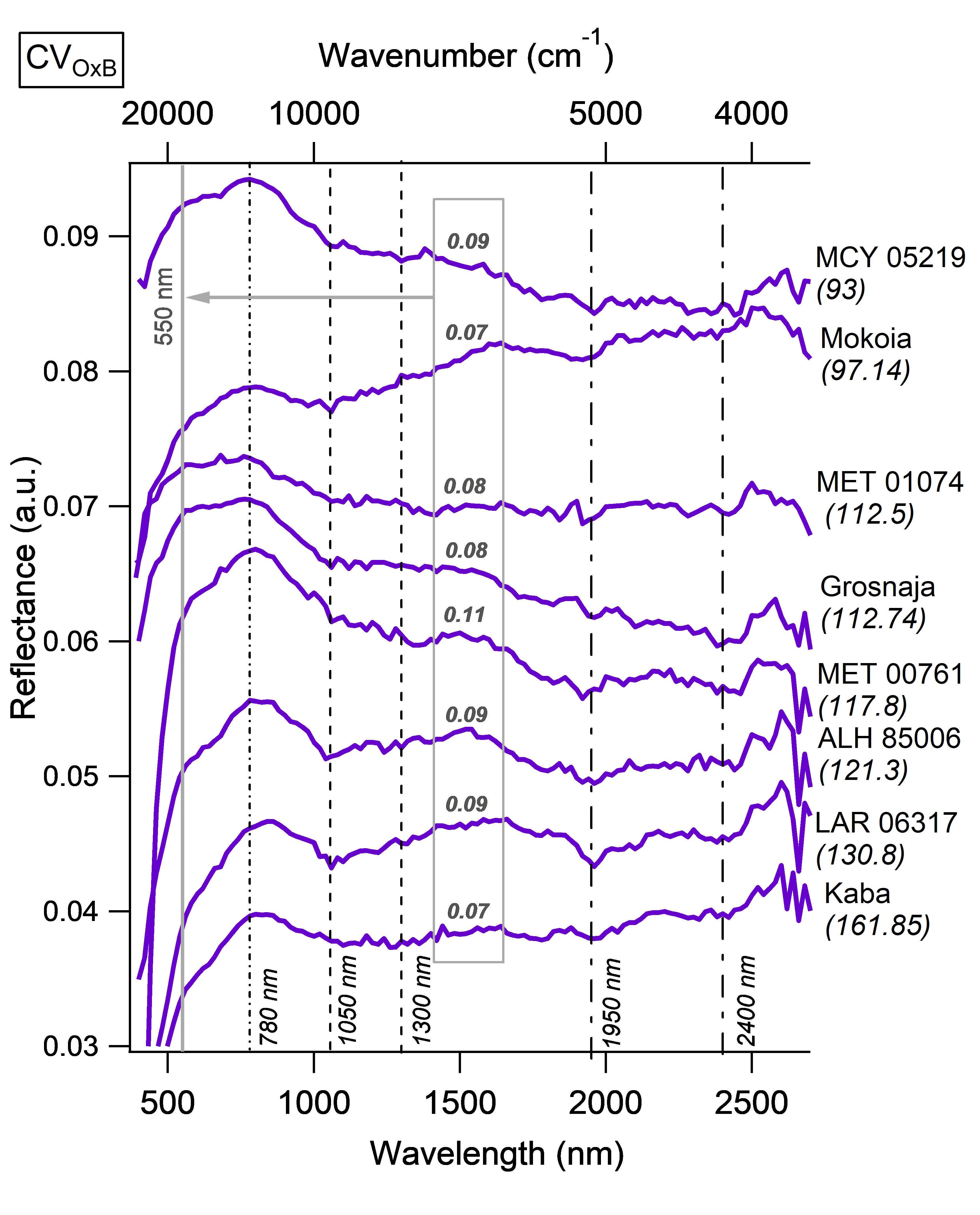}
\subcaption{}
\label{Fig:OXB_Feature}
\end{subfigure}
\begin{subfigure}[c]{0.49\textwidth}
\includegraphics[width=\textwidth]{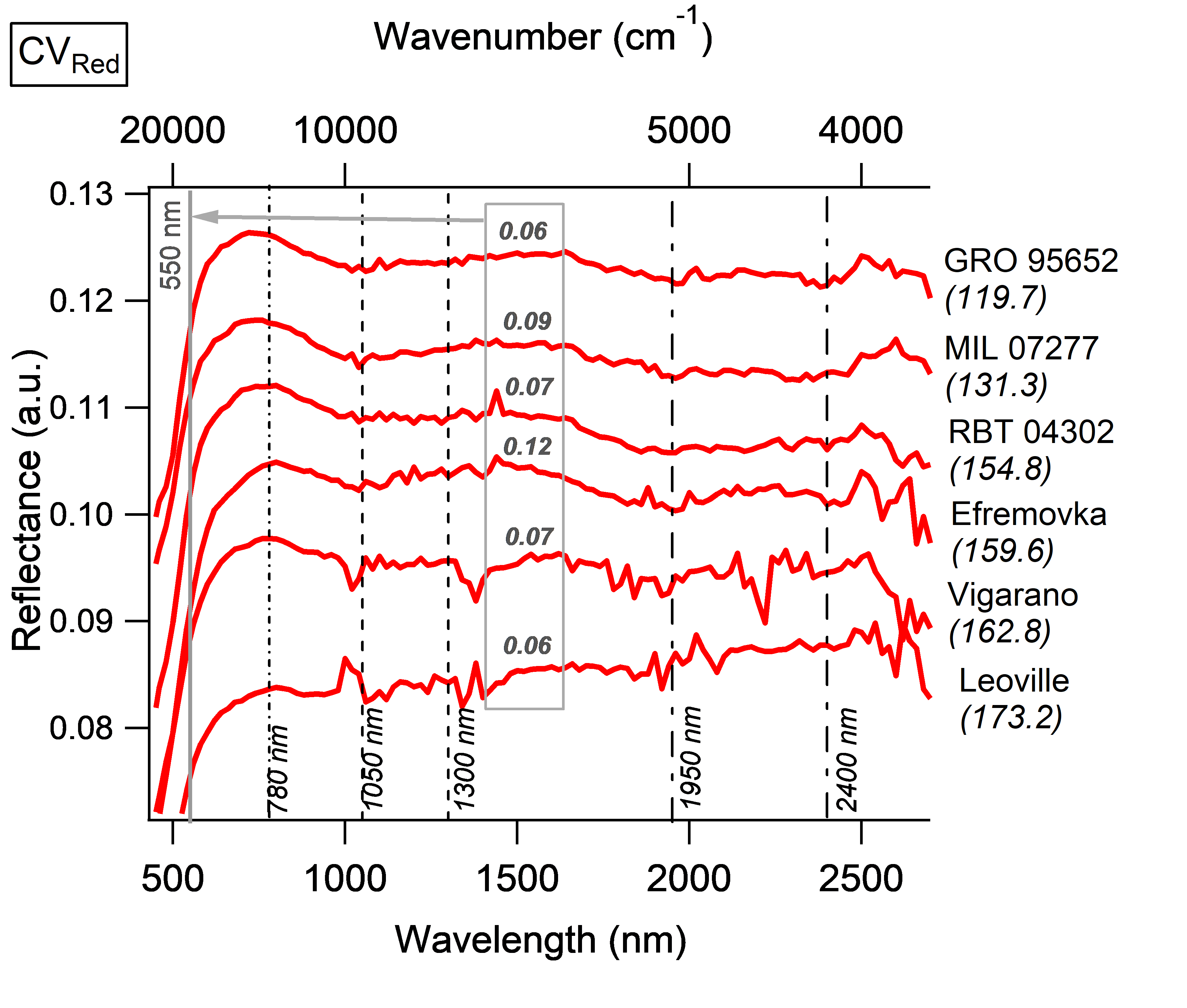}
\subcaption{}
\label{Fig:Red_Features}
\end{subfigure}
\caption{Reflectance spectra obtained (a) for CV$_\textrm{OxA}$ (b) for CV$_\textrm{OxB}$ and (c) for CV$_\textrm{Red}$. Spectra are shown between \SI{400}{\nano\meter} and \SI{2600}{\nano\meter} and are plotted with an vertical offset, for better visibility. The reflectance value at \SI{550}{\nano\meter} for each spectrum is given in the gray box. They are sorted by metamorphic grade with metamorphic grades increasing from bottom to top as indicated by the FWHM$_\textrm{D}$ (cm$^{-1}$) values (\cite{Bonal2016}) given in parenthesis behind each sample name. The dotted lines indicate the position of the peak reflectance values in the visible wavelength range. The dashed lines indicate the position of the olivine absorption features in the \SI{1}{\micro\meter} spectral region. The dash-dotted line indicates the spinel or pyroxene absorption features in the \SI{2}{\micro\meter} region.}
\label{Fig:Ox_Features}
\end{figure*}

\begin{figure*}[h]
\centering
\includegraphics[width=0.5\textwidth]{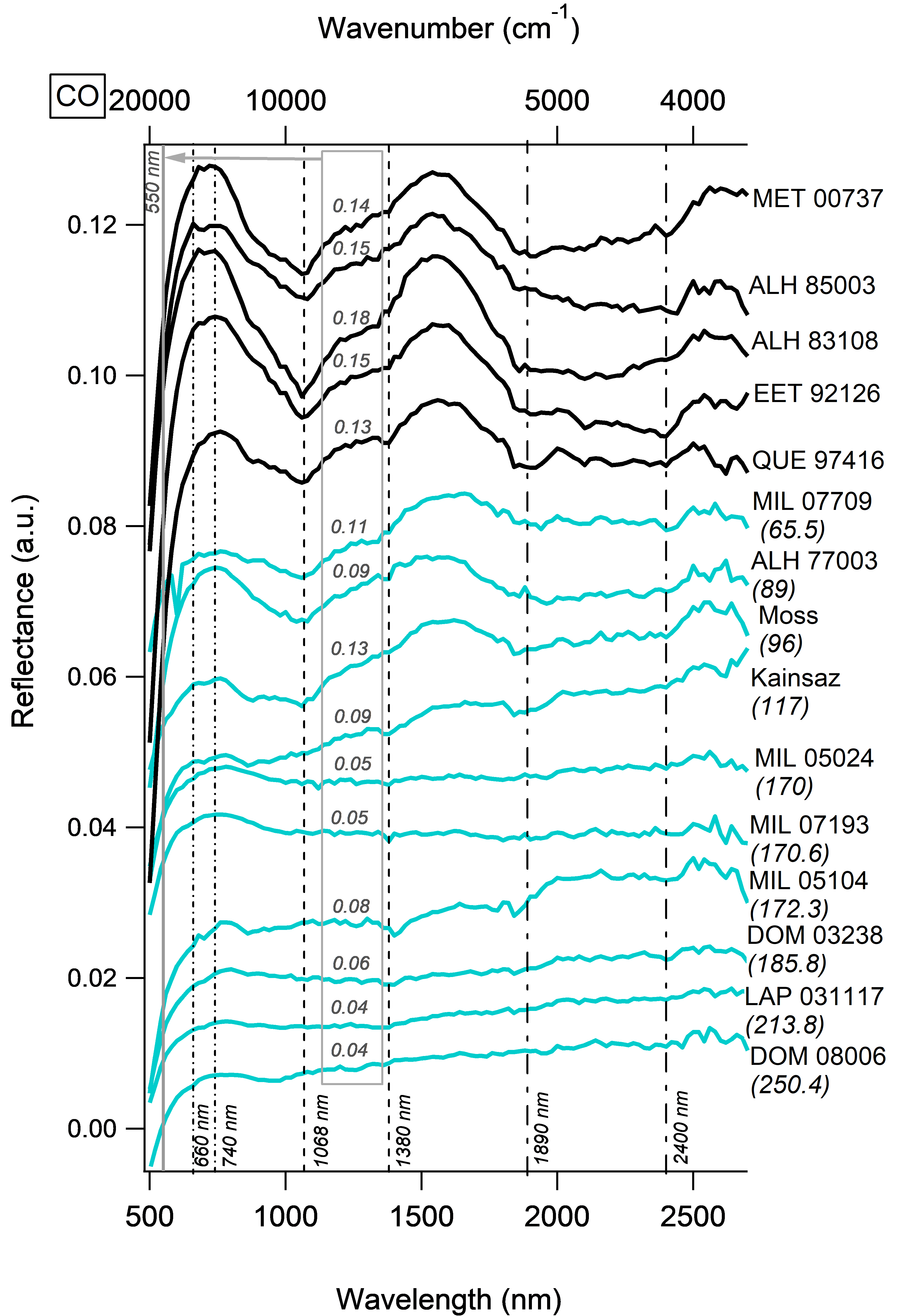}
\caption{Reflectance spectra obtained for COs. Spectra are shown between \SI{500}{\nano\meter} and \SI{2700}{\nano\meter} and are plotted with an vertical offset, for better visibility. The reflectance value at \SI{550}{\nano\meter} for each spectrum is given in the gray box. They are sorted by metamorphic grade with metamorphic grades increasing from bottom to top as indicated by the FWHM$_\textrm{D}$ (cm$^{-1}$) values \citep{Bonal2016} given in parenthesis behind each sample name. The dotted lines indicate the position of the peak reflectance values in the visible wavelength range. The dashed lines indicate the position of the olivine absorption features in the \SI{1}{\micro\meter} spectral region. The dash-dotted line indicates the spinel or pyroxene absorption features in the \SI{2}{\micro\meter} region.\\
Five of the 15 samples measured were not provided with FWHM$_\textrm{D}$ values by \cite{Bonal2016}. They are plotted in black. 
}
\label{Fig:CO_Features}
\end{figure*}

\begin{figure*}
\centering
\begin{subfigure}[c]{0.49\textwidth}
\includegraphics[width=\textwidth]{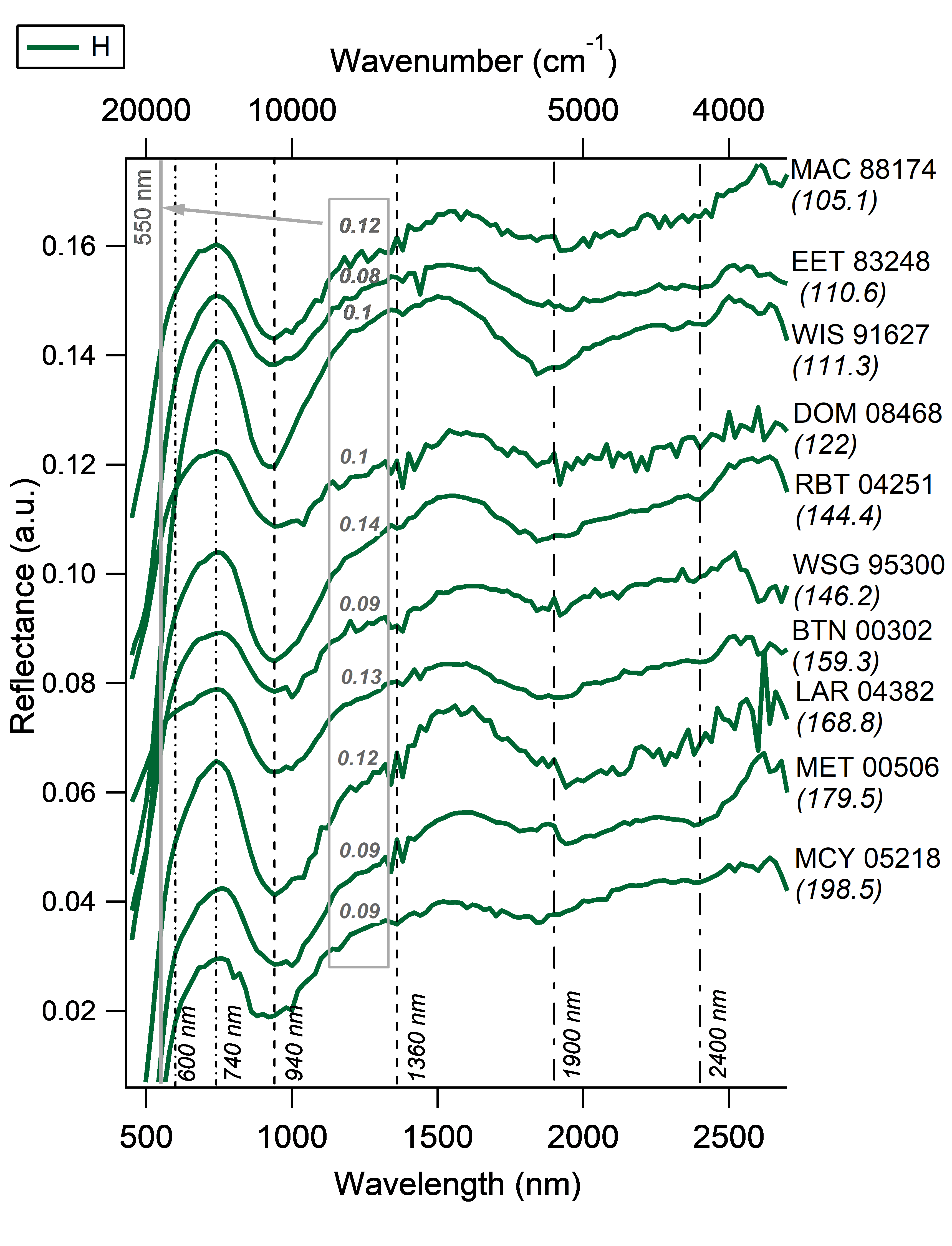}
\subcaption{}
\label{Fig:H_LL_Feature}
\end{subfigure}
\begin{subfigure}[c]{0.49\textwidth}
\includegraphics[width=\textwidth]{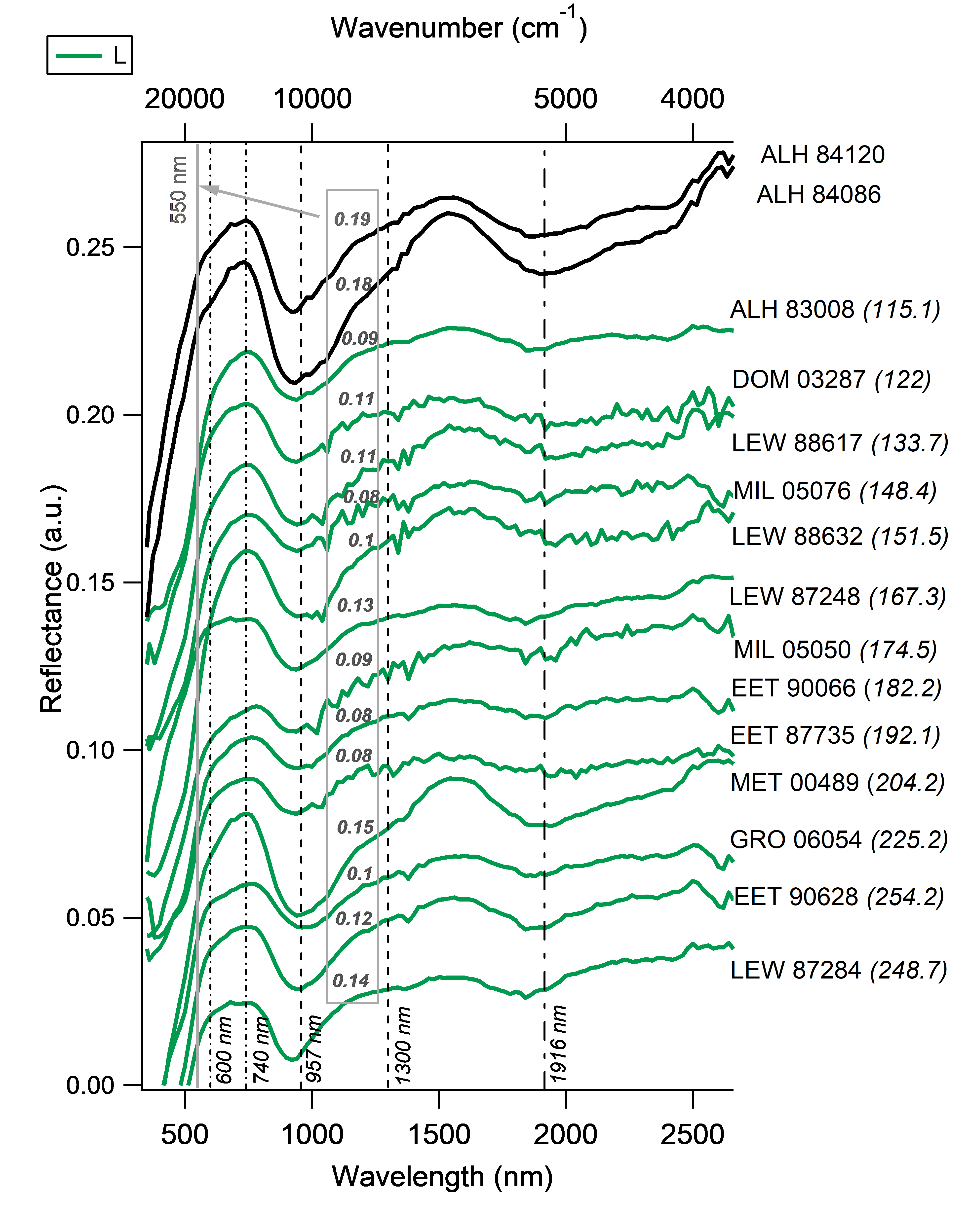}
\subcaption{}
\label{Fig:L_Feature}
\end{subfigure}
\begin{subfigure}[c]{0.49\textwidth}
\includegraphics[width=\textwidth]{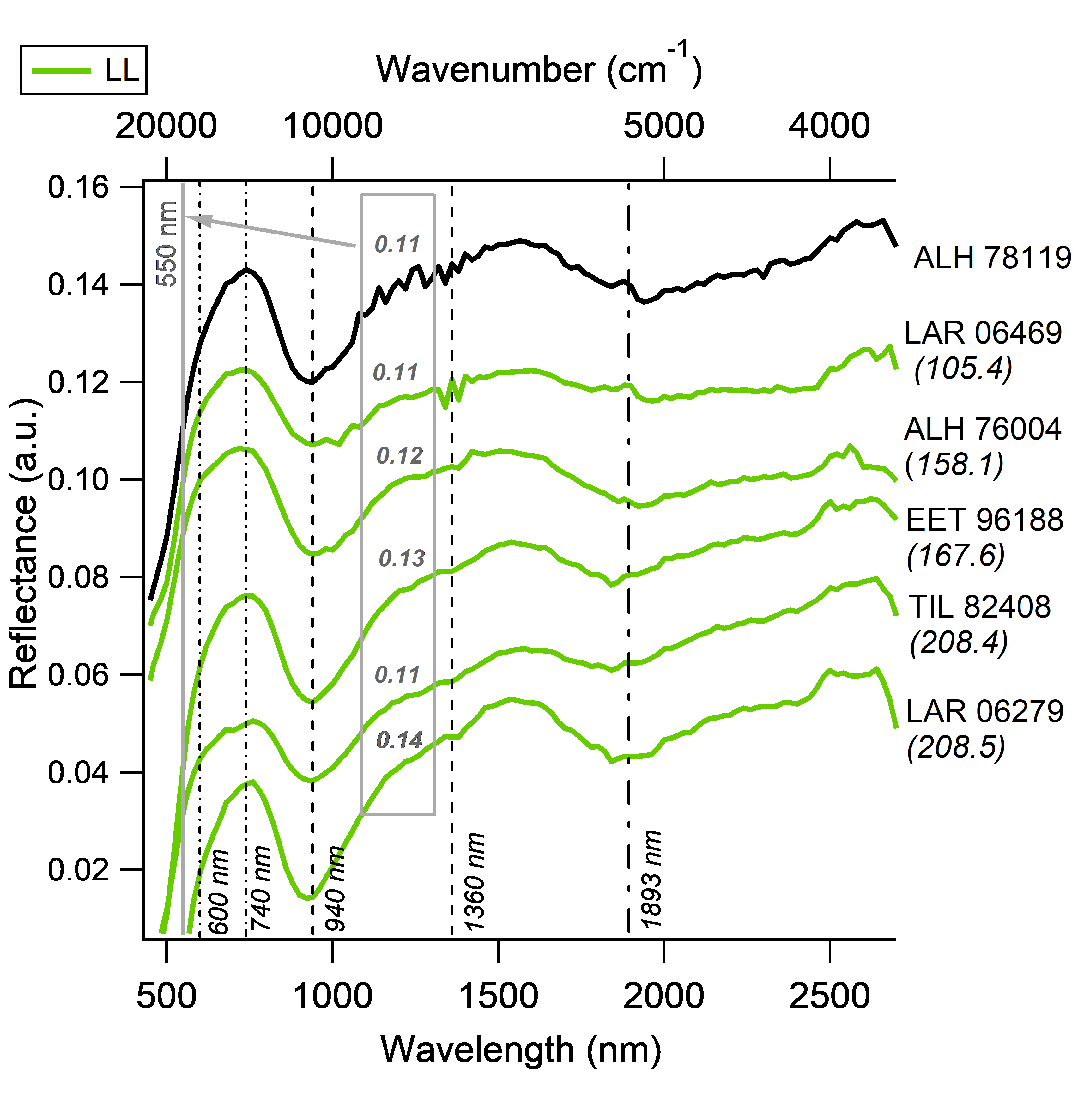}
\subcaption{}
\label{Fig:L_Feature}
\end{subfigure}
\caption{Reflectance spectra obtained for (a) H (b) L and (c) LL UOCs. Spectra are shown between \SI{350}{\nano\meter} and \SI{2660}{\nano\meter} and are plotted with an vertical offset, for better visibility. The reflectance value at \SI{550}{\nano\meter} for each spectrum is given in the gray box. They are sorted by metamorphic grade with metamorphic grades increasing from bottom to top as indicated by the FWHM$_\textrm{D}$ (cm$^{-1}$) values \citep{Bonal2016} given in parenthesis behind each sample name. The dotted lines indicate the position of the peak reflectance values in the visible wavelength range. The dashed lines indicate the position of the olivine absorption features in the \SI{1}{\micro\meter} spectral region. The dash-dotted line indicate pyroxene absorption feature in the \SI{2}{\micro\meter} region.
Two of the L type UOC measured were not provided with FWHM$_\textrm{D}$ values by \cite{Bonal2016}. They are plotted in black. 
}
\label{Fig:Ord_Features}
\end{figure*}

\begin{figure*}[h]
\centering
\includegraphics[width=0.5\textwidth]{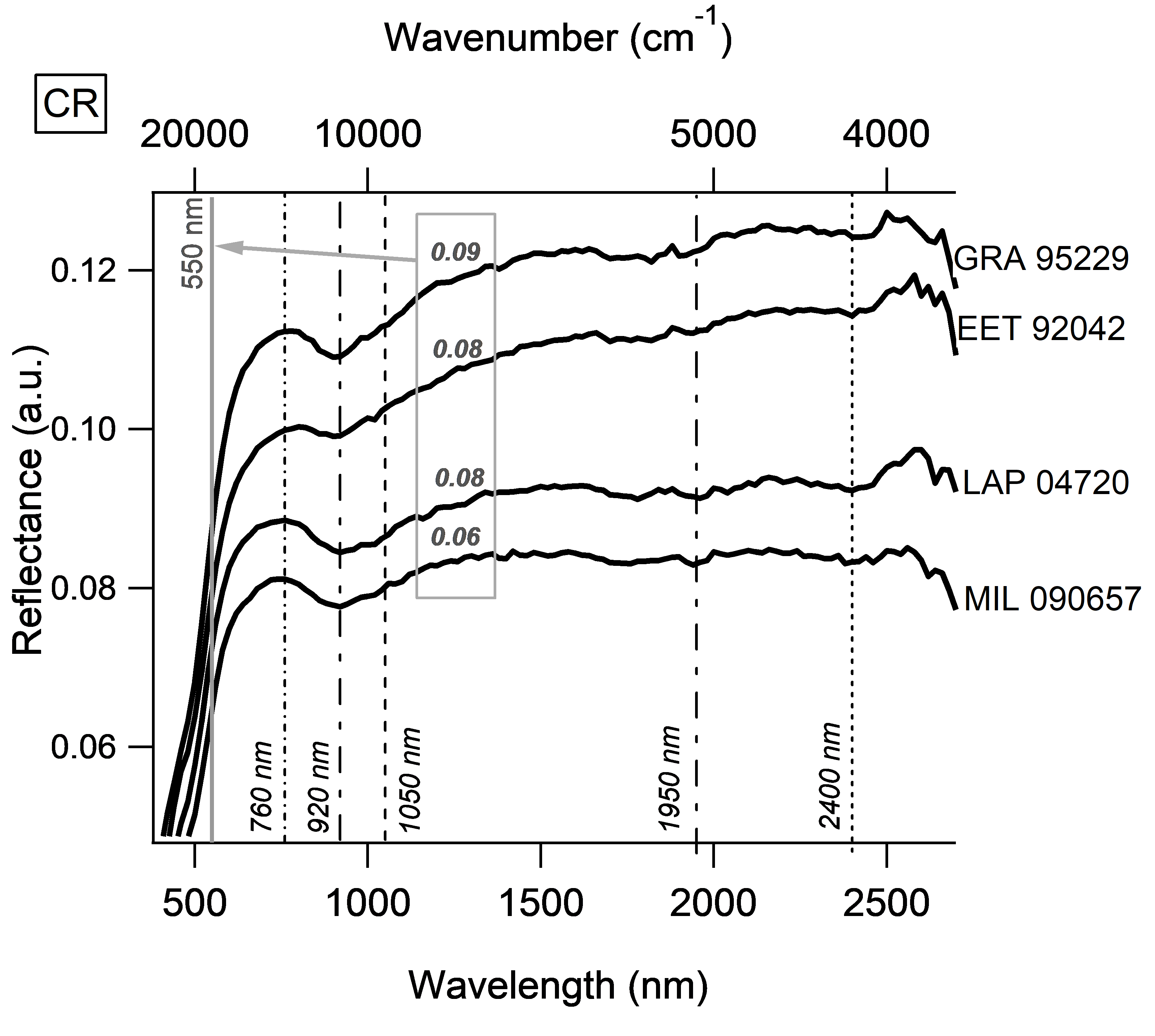}
\caption{Reflectance spectra obtained for CRs. Spectra are shown between \SI{360}{\nano\meter} and \SI{2600}{\nano\meter} and are plotted with an vertical offset, for better visibility. The reflectance value at \SI{550}{\nano\meter} for each spectrum is given in the gray box. The dotted lines indicate the position of the peak reflectance values in the visible wavelength range. The dashed lines indicate the position of the olivine absorption features in the \SI{1}{\micro\meter} spectral region. The dash-dotted line indicate pyroxene absorption feature in the \SI{1}{\micro\meter} and \SI{2}{\micro\meter} region.}
\label{Fig:CR_Features}
\end{figure*}

The reflectance spectra obtained for the 23 CVs, 15 COs, 4 CRs and 31 UOCs show variable spectral features. Significant spectral variations are visible between chondrite groups as well as within a given chondrite group (see Figs. \ref{Fig:Ox_Features} to \ref{Fig:Ord_Hyd}). 
To allow for the quantification of these variations, the spectral features introduced in Section \ref{Ch:Analytical} are determined. The resulting parameters are listed in Tables \ref{Tab:ReflRes_Carb}, \ref{Tab:ReflRes2_Carb} (for carbonaceous chondrites) and \ref{Tab:ReflRes_Ord}, \ref{Tab:ReflRes2_Ord} (for UOCs). Furthermore, average spectral parameters are given to ease the investigation of spectral differences between chondrite groups.\\
The spectra of CVs Leoville and Vigarano as well as the UOCs ALH 78119, DOM 03287, DOM 08468, EET 87735, LAR 04382, LAR 06469, LEW 88617, LEW 88632, MAC 88174, MET 00506, MIL 05050, MIL 05076 and WSG 95300 exhibit a much lower Signal to Noise Ratio (SNR) (Figs. \ref{Fig:Ox_Features} and \ref{Fig:Ord_Features}) than the other samples. This results from the aging of the IR detector. To treat these samples, a gaussian smoothing of the spectra was performed before the determination of the spectral features. It needs to be kept in mind, that especially for samples Leoville and Vigarano which exhibit faint absorption features, the determined band depths and positions have a higher uncertainty.\\
Five of the CO chondrites measured in this work (MET 00737, ALH 85003, ALH 83108, EET 92126 and QUE 97416) and two of the UOCs (ALH 84120 and ALH 84086) could not be previously characterized by Raman spectroscopy \citep{Bonal2016}. They are plotted in black in Figure \ref{Fig:CO_Features} and \ref{Fig:Ord_Features}.
For the five COs the intense absorption features as well as the trend of the spectral features becoming more pronounced with increasing metamorphic grade (Fig. \ref{Fig:CO_Features}) point towards all five of these samples being highly metamorphosed. This is consistent with one possible hypothesis to explain the lack of successful Raman acquisitions by \cite{Bonal2016}. \\
Since the main focus of this work is type 3 chondrites, only four CR2 chondrites were measured. This does not allow for statistically significant statements, however, the results are included for comparison. 
In the following sections, each spectral feature is described for the different chondrite groups.

\subsection{Overall characterization of the reflectance spectra}
\subsubsection{700 nm region peak reflectance}
The reflectance spectra measured for CV chondrites show slight variations between samples of different sub-classifications (Fig. \ref{Fig:Ox_Features} and Tables \ref{Tab:ReflRes_Carb} and \ref{Tab:ReflRes2_Carb}). Indeed, CV$_\textrm{OxA}$ samples exhibit the most pronounced absorption features and generally show a higher reflectance than CV$_\textrm{OxB}$ and CV$_\textrm{Red}$. On average, the peak reflectance values in the visible wavelength range reach $11.2 \pm 0.5\%$ for CV$_\textrm{OxA}$, $8.6 \pm 0.5\%$ for CV$_\textrm{OxB}$ and $8.8 \pm 1.0\%$ for CV$_\textrm{Red}$ (Table \ref{Tab:ReflRes2_Carb}). The positions of these features vary between \SI{720}{\nano\meter} and \SI{760}{\nano\meter} for CV$_\textrm{OxA}$ and are less variable for CV$_\textrm{OxB}$, which are located at \SI{780}{\nano\meter}. For CV$_\textrm{Red}$ this feature is located at approximately \SI{780}{\nano\meter} and shifts to lower wavelengths with increasing metamorphic grade (dotted lines in Fig. \ref{Fig:Ox_Features}). The peak reflectance of GRO 95652, the most metamorphosed CV$_\textrm{Red}$, is located at \SI{720}{\nano\meter}. Interestingly, a second feature is visible at \SI{560}{\nano\meter} (dotted lines in Fig. \ref{Fig:OXB_Feature}) for CV$_\textrm{OxB}$. This feature is not visible for CV$_\textrm{OxA}$ and CV$_\textrm{Red}$.\\
The largest spectral variability within a chondrite group is seen for CO chondrites. As shown in Figure \ref{Fig:CO_Features}, spectral features become more pronounced with increasing metamorphic grade.
All samples exhibit a peak reflectance in the visible wavelength range with values ranging from $4.5~\pm <0.1\%$ for LAP 031117 to $19.2~\pm <0.1\%$ for ALH 83108 (Table \ref{Tab:ReflRes_Carb}). On average, the peak reflectance for CO chondrites reaches $11.1 \pm 1.3\%$. This value is comparable to the average peak reflectance value determined for CV$_\textrm{OxA}$ (Table \ref{Tab:ReflRes_Carb} and Fig. \ref{Fig:CO_Features}). 
The position of the peak reflectance in the visible wavelength range is located close to \SI{740}{\nano\meter}. 
A second, weaker feature closer to \SI{660}{\nano\meter} can be seen for samples LAP 031117, DOM 03238, MIL 05104, Kainsaz, Moss, EET 92126, ALH 83108, ALH 85003 and MET 00737. \\
All UOCs exhibit reflectance spectra with distinct absorption features generally much more pronounced than for carbonaceous chondrites (Fig. \ref{Fig:Ord_Features}).
The peak reflectance in the visible wavelength range is located at \SI{740}{\nano\meter} and exhibits average values of $13.3 \pm 0.6 \%$ for H, $13.8 \pm 0.9 \%$ for L and $14.7 \pm 0.8 \%$ for LL. For some samples a shoulder at around \SI{600}{\nano\meter} is seen, especially for samples classified as L-types, such as GRO 06054, EET 90066 and EET 90628 (Fig. \ref{Fig:L_Feature}).\\
The four considered CR chondrites show rather featureless spectra with a red overall spectral slope (Fig. \ref{Fig:CR_Features}). 
The peak reflectance value in the visible wavelength range is located at \SI{760}{\nano\meter} and has an average value of $(9.8 \pm 0.7)\%$ (Table \ref{Tab:ReflRes_Carb} and Fig. \ref{Fig:CO_Features}).

\subsubsection{1000 nm region absorption feature}
\label{Sec:1000nmAbsorptionFeature}
For CV chondrites, the \SI{1}{\micro\meter} absorption band depth decreases from CV$_\textrm{OxA}$ to CV$_\textrm{OxB}$ to CV$_\textrm{Red}$. On average the band depth values reach $5.6 \pm 0.4\%$, $3.3 \pm 0.3\%$ and $2.8 \pm 0.4\%$, respectively. 
The main minimum of the \SI{1}{\micro\meter} band is located at \SI{1050}{\nano\meter}. A slightly less pronounced feature around \SI{1300}{\nano\meter} can be observed for CV$_\textrm{OxA}$ but is less visible for CV$_\textrm{OxB}$ and CV$_\textrm{Red}$ (dashed lines in Fig. \ref{Fig:Ox_Features}). 
The positions of these features are associated with ferrous olivine present in the samples \citep{Cloutis2012}.
The tilt of the spectra in the \SI{1}{\micro\meter} region is red to neutrally sloped for CV$_\textrm{OxA}$ (average value of $80.7 \pm 19.5~\si{\per\nano\meter}$) while it is neutrally to blue sloped for CV$_\textrm{OxB}$ and CV$_\textrm{Red}$ (average values of $-36.4 \pm 15.7~\si{\per\nano\meter}$ and $-15.0 \pm 9.1~\si{\per\nano\meter}$, respectively). 
There are some exceptions to these overall observations. Indeed, the CV$_\textrm{OxA}$ ALH 81003 and QUE 94688 are neutrally to blue sloped. LAR 06317 (CV$_\textrm{OxB}$) is neutrally sloped and Mokoia (CV$_\textrm{OxB}$) is redder slope than the other CV$_\textrm{OxB}$.\\
For CO chondrites the \SI{1}{\micro\meter} absorption band is pronounced for high metamorphic grades (average absorption band depth of $5.1 \pm 0.6\%$). On the other hand, CO chondrites with low metamorphic grades such as DOM 08006 (3.0) and LAP 031117 (3.05) exhibit virtually no spectral features (Fig. \ref{Fig:CO_Features}). The main minimum is also located at \SI{1050}{\nano\meter} and a second feature at \SI{1300}{\nano\meter} is observed. These features indicate the presence of olivine.  
The slope of the \SI{1}{\micro\meter} region varies between red, blue and neutrally sloped samples (Table \ref{Tab:ReflRes_Carb}, Fig. \ref{Fig:CO_Features}).\\ 
For UOCs the \SI{1}{\micro\meter} region has a narrow, deep absorption band. The average \SI{1}{\micro\meter} band depths are $13.3 \pm 0.6~\%$, $13.6 \pm 0.9~\%$ and $14.2 \pm 0.8~\%$ for H, L and LL, respectively. The \SI{1}{\micro\meter} absorption band is thus more than twice as deep as those observed for carbonaceous chondrites.
A slight shift in the average position of the \SI{1}{\micro\meter} feature can be observed between H chondrites (\SI{954}{\nano\meter}), L chondrites (\SI{957}{\nano\meter}) and LL chondrites (\SI{940}{\nano\meter}) (Table \ref{Tab:ReflRes2_Ord}).
A fainter feature around \SI{1300}{\nano\meter} is seen as well. These two features are due to the presence of calcium-rich Fe-bearing pyroxene, which exhibits absorption features around \SI{940}{\nano\meter}, and olivine which has a side band at \SI{1300}{\nano\meter} \citep{Cloutis2012}. Generally, the \SI{1}{\micro\meter} band area is red sloped with the exception of LL samples ALH 76004 and LAR 06469. These two samples exhibit neutrally to blue \SI{1}{\micro\meter} slopes.\\
For CR chondrites, the \SI{1}{\micro\meter} region shows a narrow absorption band located at \SI{920}{\nano\meter}. This feature is most likely due to Fe oxy-hydroxides, but can also be associated with low-Fe pyroxene, olivine or Fe- bearing phyllosilicates \citep{Cloutis_CR}. Since the CR chondrites measured in this work experienced strong aqueous alteration it is expected to see hydrated minerals such as oxy-hydroxides as well as phyllosilicates like serpentine, saponite and chlorite. A slight indent at approximately \SI{1050}{\nano\meter} points towards the presence of low-Fe olivine. The average \SI{1}{\micro\meter} band depth of $4.4 \pm 0.5 \%$ is comparable to that of CO chondrites (Table \ref{Tab:ReflRes2_Carb}). The \SI{1}{\micro\meter} region is red sloped for all samples (average slope of $93.9 \pm 23.7~\si{\per\nano\meter}$).

\subsubsection{2000 nm region absorption feature}
For carbonaceous chondrites, the \SI{2}{\micro\meter} region is dominated by contributions of pyroxene and spinel \citep{Cloutis2012}. At \SI{2400}{\nano\meter} a faint absoption feature observed in previous works by \cite{Cloutis2012} was attributed to the presence of fassaite.\\
For the CV chondrites analyzed in this work, the \SI{2}{\micro\meter} absorption band shows a minimum at \SI{1950}{\nano\meter} (dash-dotted lines Fig. \ref{Fig:Ox_Features}). Grosnaja (CV$_\textrm{OxB}$) and GRO 95652 (CV$_\textrm{Red}$) exhibit a \SI{2}{\micro\meter} position at much longer wavelengths. This is due to the very shallow \SI{2}{\micro\meter} absorption feature seen for these chondrites which makes the band position determination imprecise (Fig. \ref{Fig:Ox_Features}).
The average \SI{2}{\micro\meter} band depths are $4.9 \pm 0.5~\%$, $3.0 \pm 0.5~\%$ and $2.7 \pm 0.4~\%$ for CV$_\textrm{OxA}$, CV$_\textrm{OxB}$ and CV$_\textrm{Red}$, respectively (Table \ref{Tab:ReflRes_Carb}). Therefore, the \SI{1}{\micro\meter} and \SI{2}{\micro\meter} bands show similar depths for the CV chondrites considered here (see Section \ref{Sec:1000nmAbsorptionFeature}).
For CV$_\textrm{OxA}$ the tilt of the spectra in the \SI{2}{\micro\meter} region is variable: from blue sloped for QUE 94688 ($-99.7 \pm 9.4~\si{\per\nano\meter}$), to neutrally sloped for LAP 02206 ($1.0 \pm 8.0~\si{\per\nano\meter}$), and red sloped for Allende ($25.4 \pm 14.6~\si{\per\nano\meter}$) (Table \ref{Tab:ReflRes_Carb} and Fig. \ref{Fig:OXA_Feature}). On average CV$_\textrm{OxA}$ are blue sloped ($-18.4 \pm 13.5~\si{\per\nano\meter}$). The tilt in the \SI{2}{\micro\meter} range of CV$_\textrm{OxB}$ and CV$_\textrm{Red}$ samples is also variable. On average CV$_\textrm{OxB}$ and CV$_\textrm{Red}$ are neutrally sloped ($4.6 \pm 7.8~\si{\per\nano\meter}$ and $-4.1 \pm 7.3~\si{\per\nano\meter}$, respectively).\\
For CO chondrites, the \SI{2}{\micro\meter} region shows a feature at \SI{1850}{\nano\meter} pointing towards the presence of low-Ca pyroxene but spinel could contribute as well \citep{Cloutis_CO}. A feature at \SI{2400}{\nano\meter} is also observed. The \SI{2}{\micro\meter} band is slightly less deep than the \SI{1}{\micro\meter} band showing an average depth of $3.8 \pm 0.5~\%$. This band depth is comparable to the other carbonaceous chondrite groups. The spectral slope in the \SI{2}{\micro\meter} range is variable as well, ranging from red to blue sloped samples (Table \ref{Tab:ReflRes_Carb}). On average the slope is $-10.1 \pm 14.5~\si{\per\nano\meter}$\\
For UOCs, the \SI{2}{\micro\meter} feature generally is much broader and shallower than the \SI{1}{\micro\meter} feature. It is located around \SI{1900}{\nano\meter} (Fig. \ref{Fig:Ord_Features}) and shows an average depth of $9.8 \pm 1.3~\%$, $9.3 \pm 0.9~\%$ and $9.3 \pm 1.6~\%$ for H, L and LL types, respectively (Table \ref{Tab:ReflRes2_Ord}). A faint feature at \SI{2400}{\nano\meter} is seen. The slope in the \SI{2}{\micro\meter} range is predominantly red sloped for all UOC types. The exceptions are L chondrites ALH 83008 ($-0.1 \pm 2.3~\si{\per\nano\meter}$) and MIL 05076 ($-2.3\pm 8.7~\si{\per\nano\meter}$), LL chondrite ALH 76004 ($-4.7 \pm 5.2~\si{\per\nano\meter}$), and H chondrites EET 83248 ($3.9 \pm 3.7~\si{\per\nano\meter}$) and WIS 91627 ($-9.9 \pm 2.6~\si{\per\nano\meter}$) which are neutral to blue sloped (Table \ref{Tab:ReflRes_Ord} and Fig. \ref{Fig:Ord_Features}). On average the slopes are $44.5 \pm 11.2~\si{\per\nano\meter}$, $51.9 \pm 11.3~\si{\per\nano\meter}$ and $59.6 \pm 18~\si{\per\nano\meter}$) for H, L and LL types, respectively.\\
The \SI{2}{\micro\meter} region of CR chondrites is rather broad and shows several smaller absorption features. At \SI{1950}{\nano\meter} a feature due to low-Fe pyroxene is observed in all spectra. An additional feature is seen at approximately \SI{2440}{\nano\meter} (Fig. \ref{Fig:CR_Features}). There are several contributions of phyllosilicates between \SI{2310}{\nano\meter} and \SI{2380}{\nano\meter} such as serpentine, saponite and chlorite, which could possibly explain this absorption band. The \SI{2}{\micro\meter} band is generally less pronounced than the \SI{1}{\micro\meter} band with an average band depth of $1.8 \pm 0.1~\%$ (Table \ref{Tab:ReflRes2_Carb}). The spectral tilt in this region is red sloped (average slope of $42.0 \pm 15.6~\si{\per\nano\meter}$).

\subsubsection{Overall spectral slope and visual slope}
For CV chondrites, the overall spectral slope is variable between sub-groups.
Generally, the overall spectral slopes of CV$_\textrm{OxA}$ are red to neutrally sloped. In contrast the overall spectral slope of CV$_\textrm{OxB}$ are blue sloped and those of CV$_\textrm{Red}$ are neutrally to red sloped. Some exceptions to these trends can be observed.
The spectra of ALH 81003 (CV$_\textrm{OxA}$) and QUE 94688 (CV$_\textrm{OxA}$) exhibit a blue slope (Fig. \ref{Fig:Ox_Features}). The spectrum of Mokoia is characterized by a strong red overall spectral slope (Fig. \ref{Fig:OXB_Feature}). This could be related to terrestrial weathering (Section \ref{Sec:Slopes}).
While the terrestrial weathering degree of the hand-specimen of ALH 81003 and QUE 94688 is moderate (\cite{Grossman1998} and \cite{Grossman1994}, respectively), a whitish/greyish color could be noted during the preparation of our allocated samples. This coloration is usually a sign of terrestrial weathering.  
The visual slope in the \SI{360}{\nano\meter} to \SI{600}{\nano\meter} region reaches an average value of $(18.1 \pm 2.6)\cdot 10^{-5}$\si{\per\nano\meter} for CV$_\textrm{OxA}$, $(8.3 \pm 2.1)\cdot 10^{-5}$\si{\per\nano\meter} for CV$_\textrm{OxB}$ and $(16.7 \pm 1.8)\cdot 10^{-5}$\si{\per\nano\meter} for CV$_\textrm{Red}$  (Table \ref{Tab:ReflRes_Carb}). The steepness of the visual slope decreases in the order CV$_\textrm{OxA}$ to CV$_\textrm{Red}$ to CV$_\textrm{OxB}$.\\
For CO chondrites, the overall spectra are generally neutrally sloped with the exception of the samples Moss, Kainsaz and MIL 05104 which show red slopes (Fig. \ref{Fig:CO_Features}).
On average, the visual slope reaches $(24.9 \pm 3.9) \cdot 10^{-5}$\si{\per\nano\meter}. This value is comparable to the average visual slope determined for CV$_\textrm{OxA}$ chondrites (Table \ref{Tab:ReflRes_Carb}).\\
All UOCs have a red overall spectral slope (Fig. \ref{Fig:Ord_Features}). The visual slope reaches average values of $(29.9 \pm 3.2) \cdot 10^{-5}$\si{\nano\meter} for H, $(34.6 \pm 1.9) \cdot 10^{-5}$\si{\nano\meter} for L and $(36.6 \pm 3.5) \cdot 10^{-5}$\si{\nano\meter} for LL (Table \ref{Tab:ReflRes_Ord}). Thus, UOCs exhibit much steeper visual slopes than carbonaceous chondrites (Table \ref{Tab:ReflRes_Carb} vs. Table \ref{Tab:ReflRes_Ord}).\\
Lastly, CR chondrites have an red overall spectral slope (Fig. \ref{Fig:CR_Features}). The visual slope of the samples has an average value of $(25.9 \pm 2.1)\cdot 10^{-5}$\si{\per\nano\meter} and, thus, is comparable to the average value for carbonaceous chondrites.

\begin{figure*}
\centering
\begin{subfigure}[c]{0.49\textwidth}
\includegraphics[width=\textwidth]{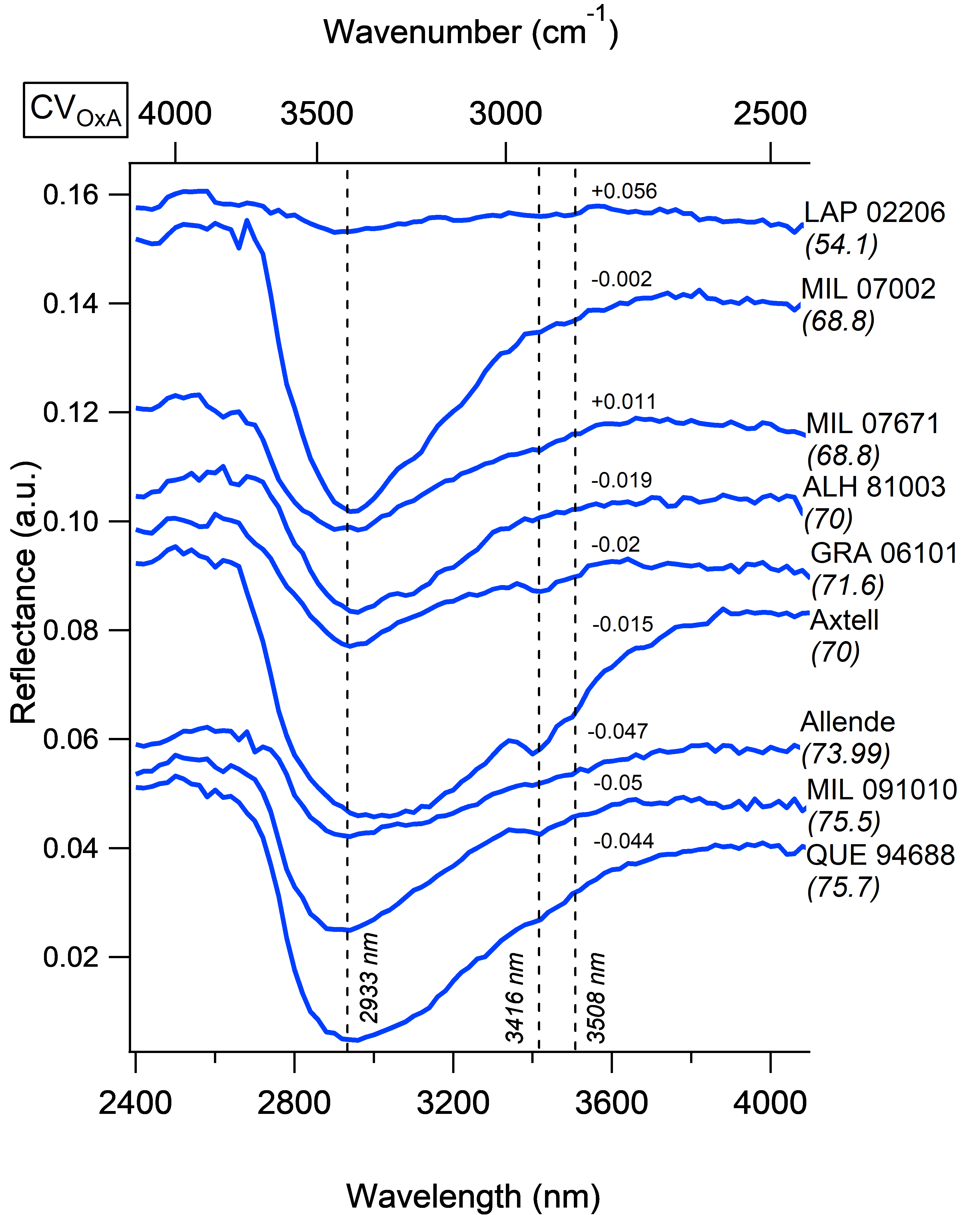}
\subcaption{}
\label{Fig:OxA_Hyd}
\end{subfigure}
\begin{subfigure}[c]{0.49\textwidth}
\includegraphics[width=\textwidth]{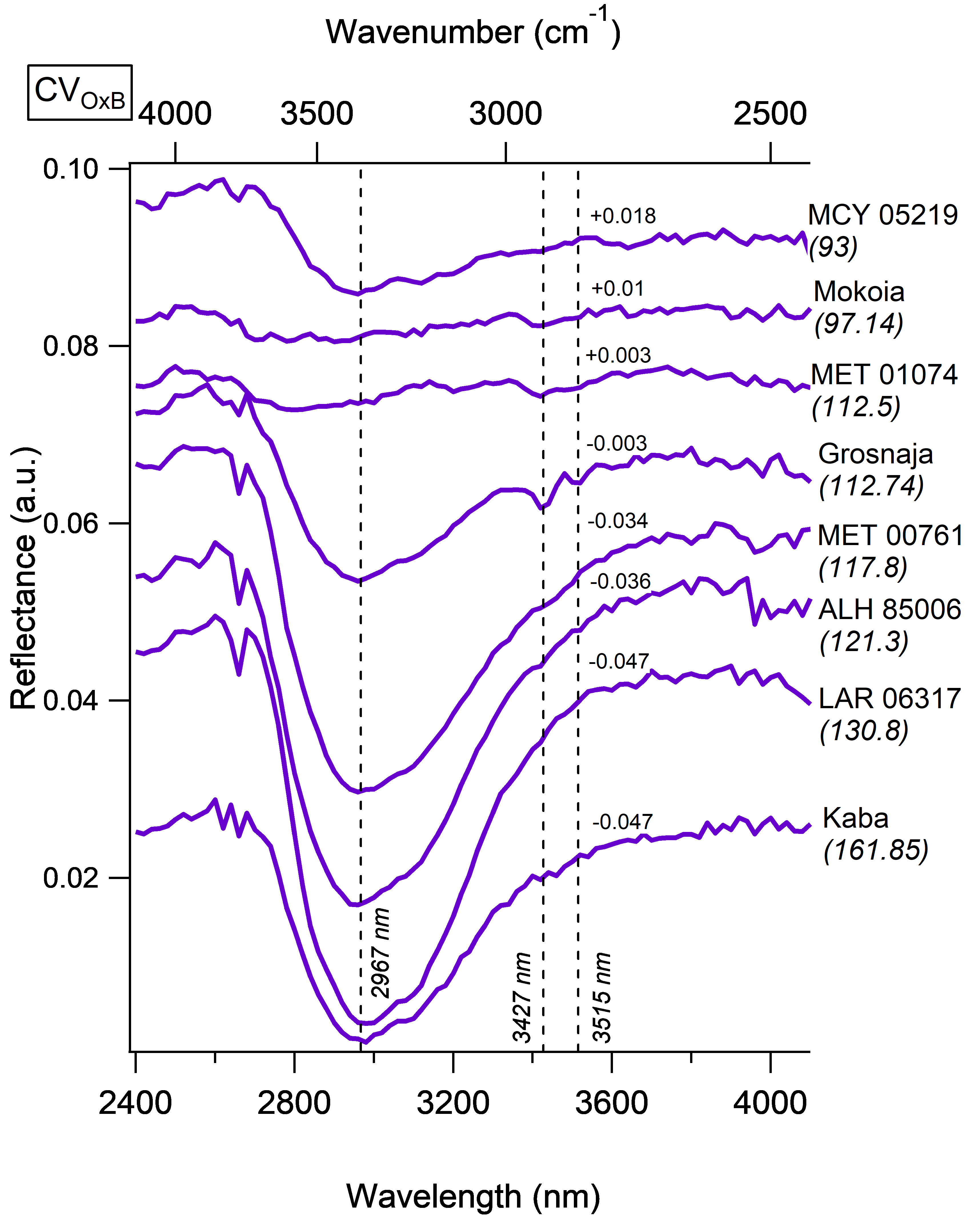}
\subcaption{}
\label{Fig:OxB_Hyd}
\end{subfigure}
\begin{subfigure}[c]{0.49\textwidth}
\includegraphics[width=\textwidth]{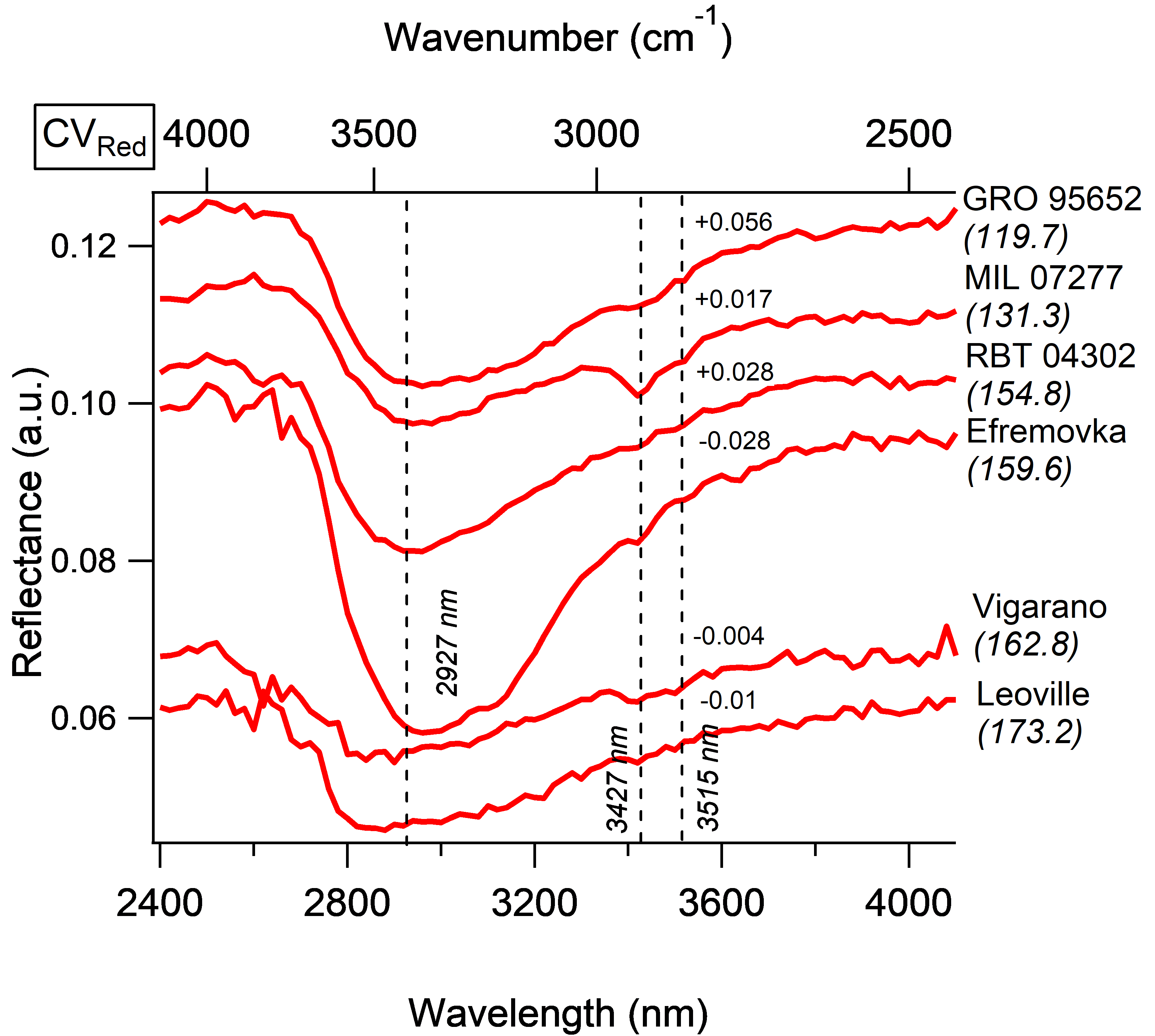}
\subcaption{}
\label{Fig:Red_Hyd}
\end{subfigure}
\caption{Reflectance spectra obtained for (a) CV$_\textrm{OxA}$ (b) CV$_\textrm{OxB}$ and (c) CV$_\textrm{Red}$. Spectra are shown between \SI{2600}{\nano\meter} and \SI{4000}{\nano\meter} and are plotted with an vertical offset, indicated by the ($+/-x)$ values, for better visibility. They are sorted by metamorphic grade with metamorphic grades increasing from bottom to top as indicated by the FWHM$_\textrm{D}$ (cm$^{-1}$) values \citep{Bonal2016} given in parenthesis behind each sample name. The dotted lines indicate the position of the \SI{3}{\micro\meter} band  minimum as well as the position of aliphatic bands.}
\label{Fig:Ox_Hyd}
\end{figure*}

\begin{figure*}[h]
\centering
\includegraphics[width=0.5\textwidth]{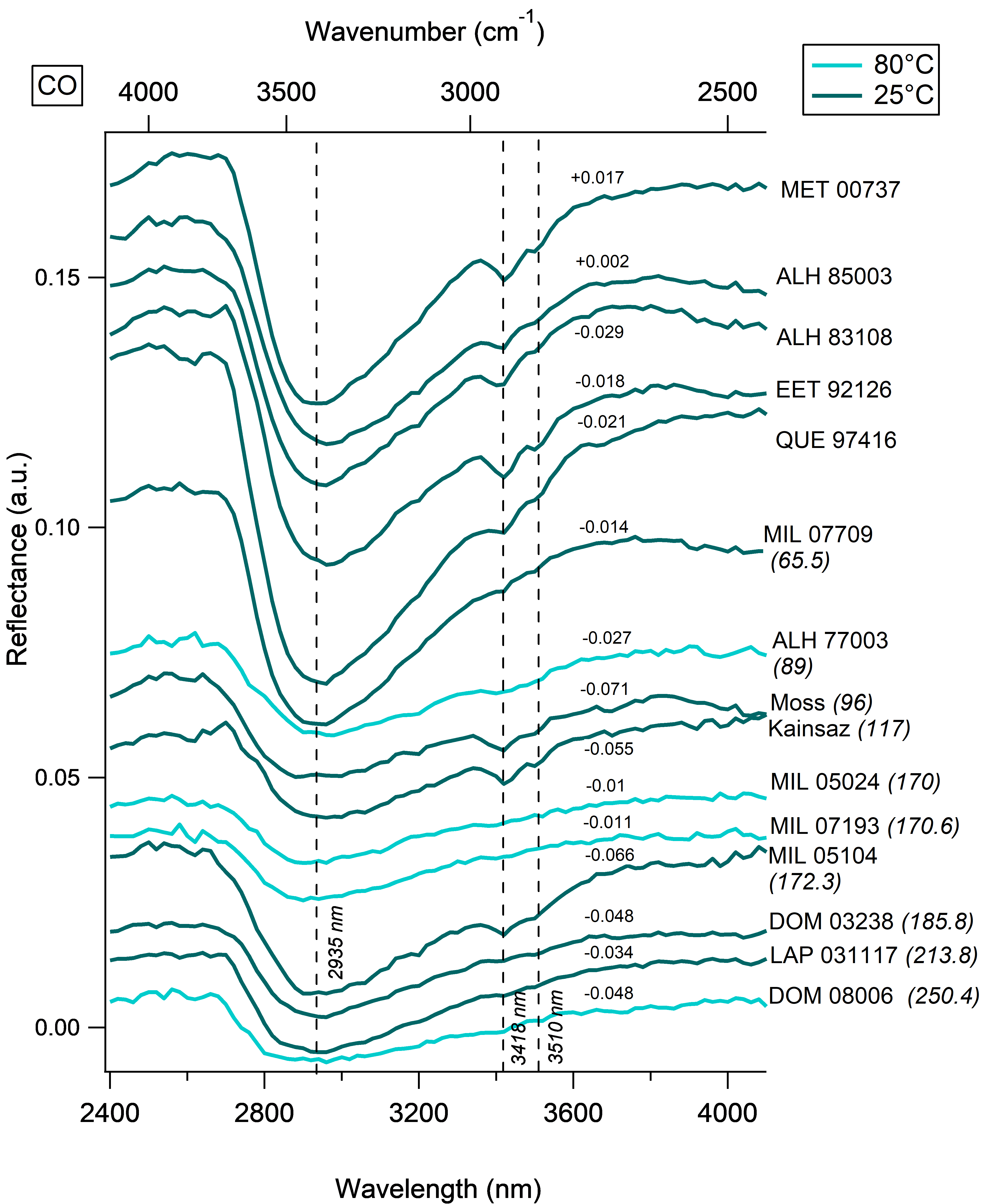}
\caption{Reflectance spectra obtained for COs. Spectra are shown between \SI{2600}{\nano\meter} and \SI{4000}{\nano\meter} and are plotted with an vertical offset, indicated by the ($+/-x)$ values, for better visibility. They are sorted by metamorphic grade with metamorphic grades increasing from bottom to top as indicated by the FWHM$_\textrm{D}$ (cm$^{-1}$) values \citep{Bonal2016} given in parenthesis behind each sample name. The dotted lines indicate the position of the \SI{3}{\micro\meter} band  minimum as well as the position of aliphatic bands.
Five of the samples were not provided with metamorphic grade values by \cite{Bonal2016}. Samples in dark blue were measured at \SI{80}{\celsius} while samples in light blue were measured at ambient temperatures.}
\label{Fig:CO_Hyd}
\end{figure*}

\begin{figure*}
\centering
\begin{subfigure}[c]{0.49\textwidth}
\includegraphics[width=\textwidth]{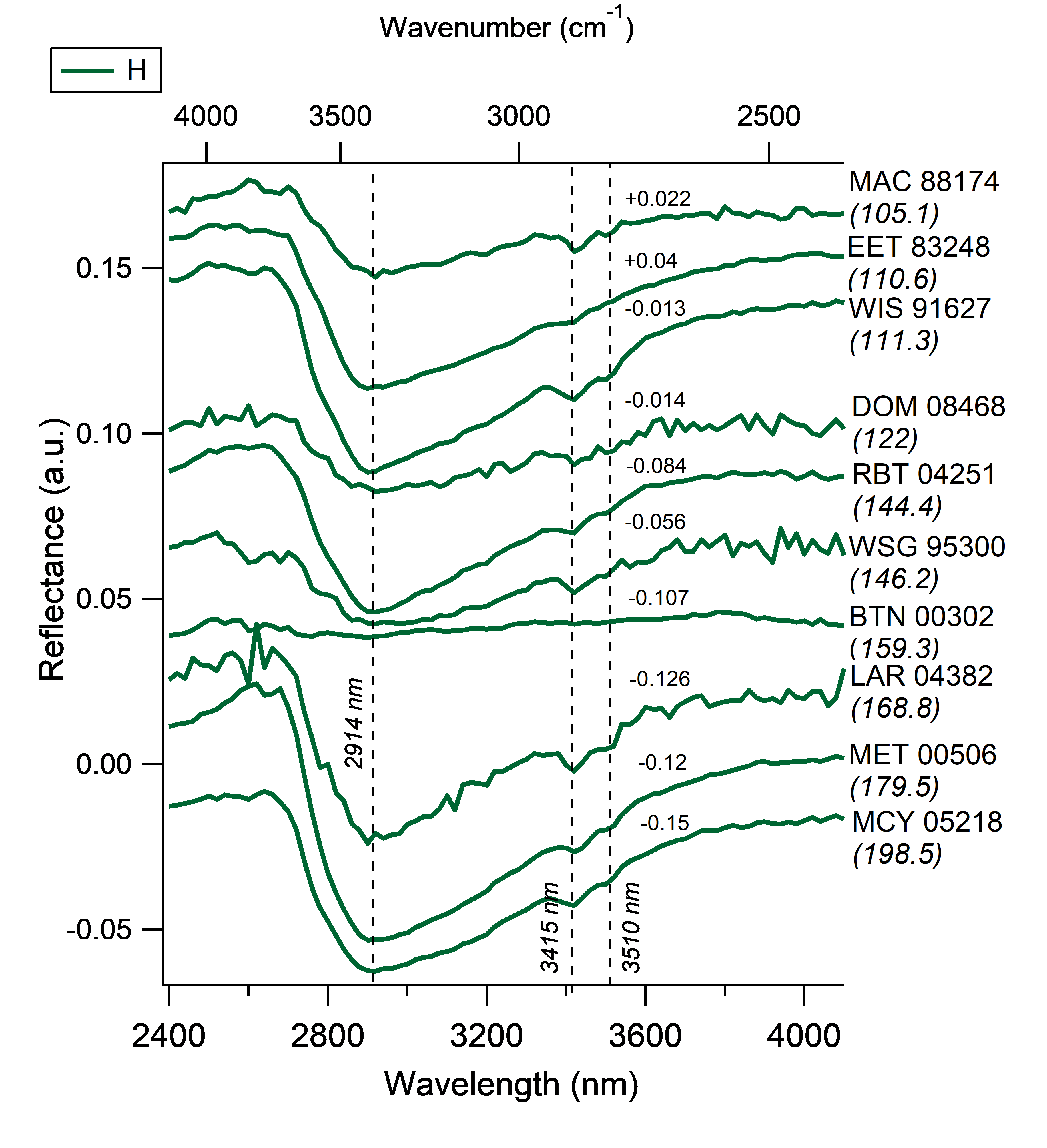}
\subcaption{}
\label{Fig:LL_H_Hyd}
\end{subfigure}
\begin{subfigure}[c]{0.49\textwidth}
\includegraphics[width=\textwidth]{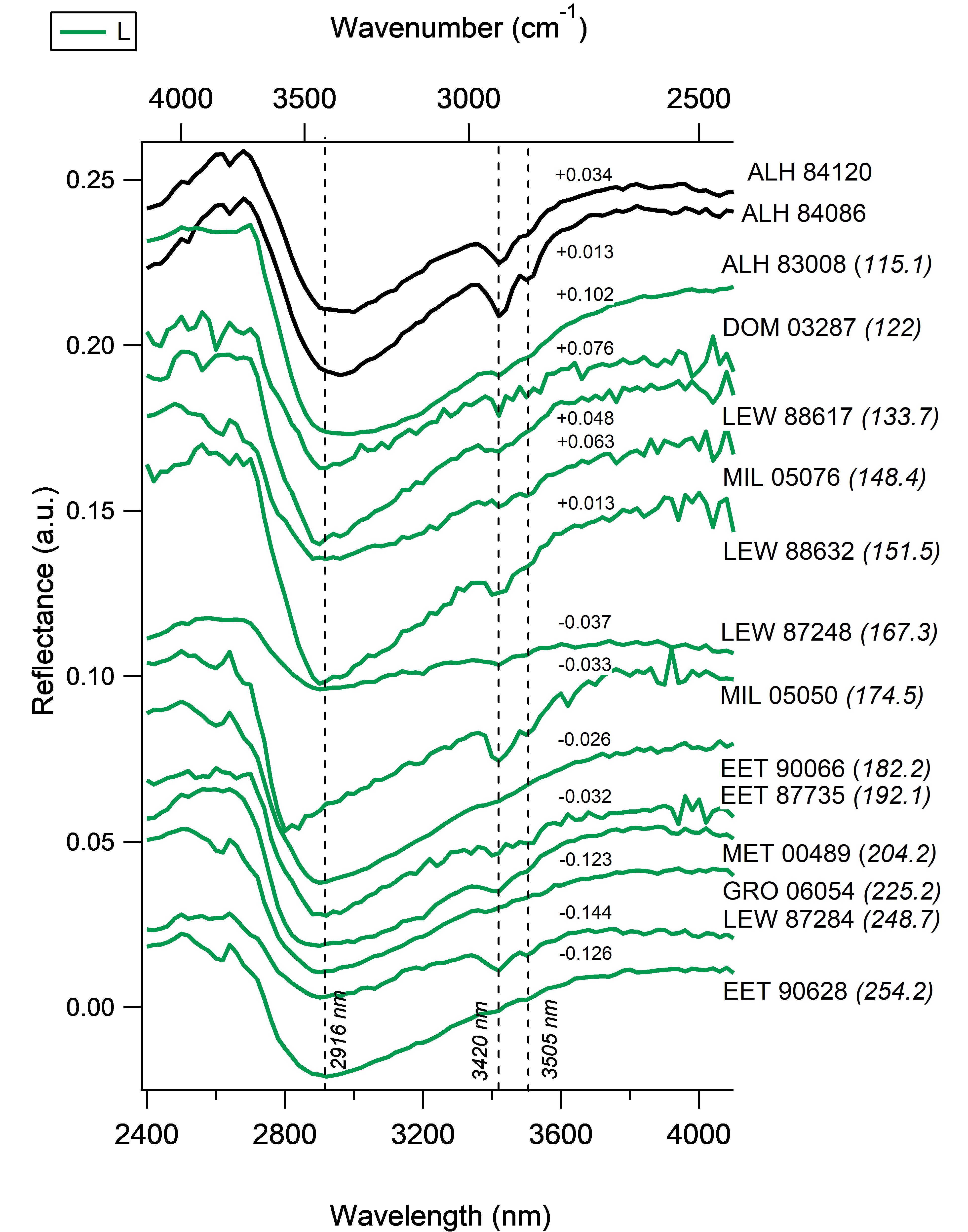}
\subcaption{}
\label{Fig:L_Hyd}
\end{subfigure}
\begin{subfigure}[c]{0.49\textwidth}
\includegraphics[width=\textwidth]{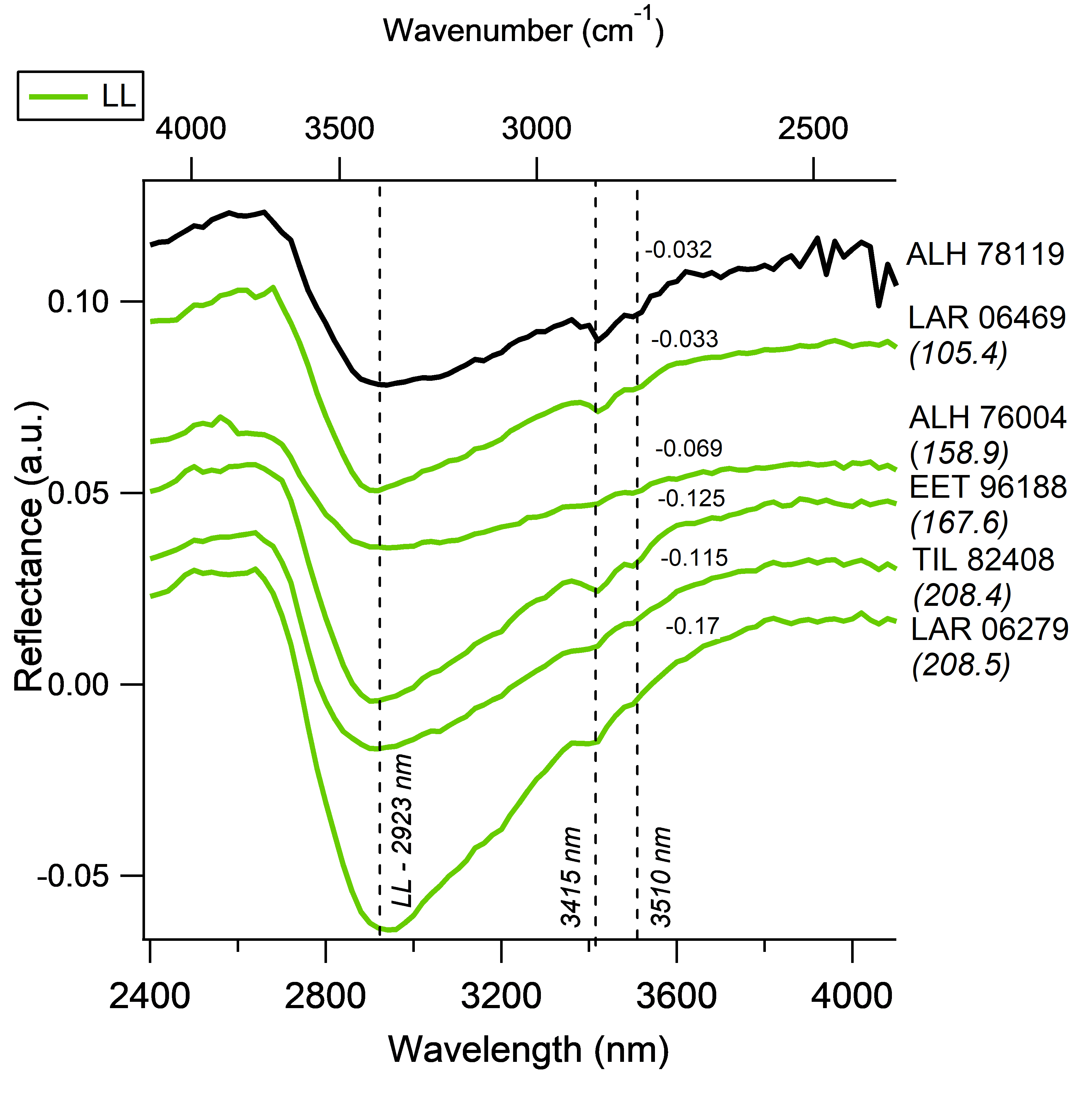}
\subcaption{}
\label{Fig:L_Hyd}
\end{subfigure}
\caption{Reflectance spectra obtained for (a) H (b) L and (c) LL UOCs. Spectra are shown between \SI{2600}{\nano\meter} and \SI{4000}{\nano\meter} and are plotted with an vertical offset, indicated by the ($+/-x)$ values, for better visibility. They are sorted by metamorphic grade with metamorphic grades increasing from bottom to top as indicated by the FWHM$_\textrm{D}$ (cm$^{-1}$) values \citep{Bonal2016} given in parenthesis behind each sample name. The dotted lines indicate the position of the \SI{3}{\micro\meter} band  minimum as well as the position of aliphatic bands.
Samples which were not provided with FWHM$_\textrm{D}$ ($cm^{-1}$) values by \cite{Bonal2016} are plotted in black.}
\label{Fig:Ord_Hyd}
\end{figure*}

\begin{figure*}[h]
\centering
\includegraphics[width=0.5\textwidth]{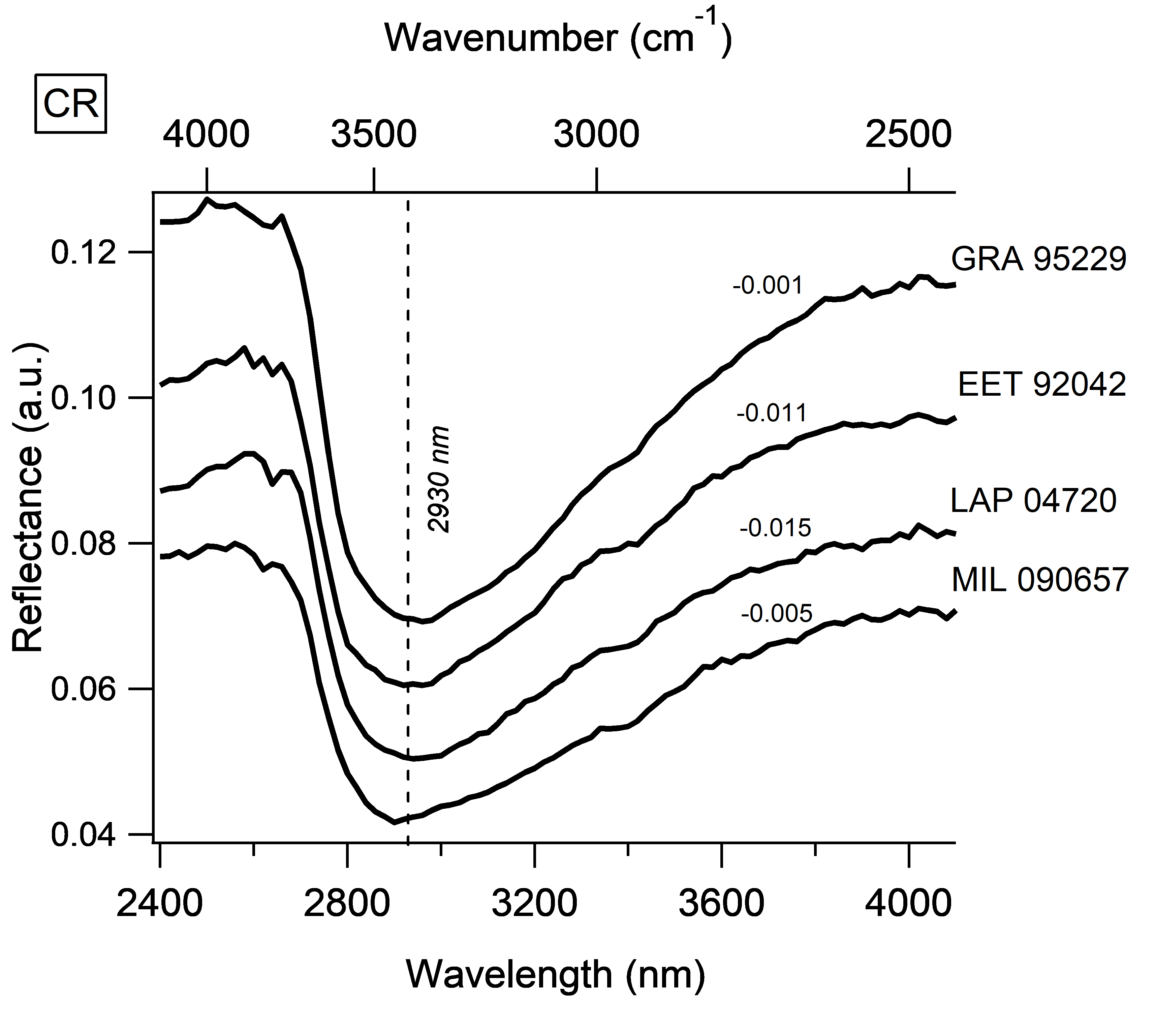}
\caption{Reflectance spectra obtained for CRs. Spectra are shown between \SI{2600}{\nano\meter} and \SI{4000}{\nano\meter} and are plotted with an vertical offset, indicated by the ($+/-x)$ values, for better visibility. The dotted line indicates the position of the \SI{3}{\micro\meter} band  minimum.}
\label{Fig:CR_Hyd}
\end{figure*}

\subsection{3~\si{\micro\meter} region absorption band}
Hydration of chondrites takes place on the asteroidal parent body through aqueous alteration and may occur through terrestrial weathering as well. These processes lead to previously non-hydrated minerals becoming hydrated as well as OH and H$_2$O being stored in the form of interlayer and adsorbed water.
In the reflectance spectra this hydration becomes visible as a convolution of three bands in the \SI{3}{\micro\meter} region. At $\sim \SI{2700}{\nano\meter}$ a band due the vibrational stretching modes of strongly bound OH is observed. Convoluted with this are two bands due to the vibrational stretching modes of adsorbed water at $\sim \SI{2900}{\nano\meter}$ and of strongly bound (interlayer) H$_2$O at $\sim\SI{3100}{\nano\meter}$ (\cite{Bishop_1994} and \cite{Frost_2000}). 
The shape, position and intensity of the \SI{3}{\micro\meter} band is dependent on the extent of hydration as well as the chemical composition of the chondrites (\cite{Bishop_1994} and \cite{Pommerol2008}). \\
To characterize the \SI{3}{\micro\meter} band, the position as well as the IBD$_\textrm{Hyd}$ are determined (Table \ref{Tab:ReflRes_Carb}, \ref{Tab:ReflRes_Ord}). The few samples that were measured at ambient temperature include a contribution due to water adsorption that takes place during the storage of the samples. The IBD$_\textrm{Hyd}$ of COs ALH 83108, EET 92126, MET 00737, MIL 05104, MIL 07687, MIL 07709, Moss, Kainsaz, LAP 031117, QUE 97416, ALH 85003 and DOM 03238 (Table \ref{Tab:ReflRes_Carb}) and  UOCs ALH 83008, ALH 84087 and ALH 84120 are, thus, less comparable to the rest of our sample set. \\
A pronounced hydration band is visible for most samples (Figs. \ref{Fig:Ox_Hyd} - \ref{Fig:Ord_Hyd}). CVs LAP 02206, Mokoia, MET 01074 and the UOC BTN 00302 are the exceptions showing only faint hydration bands. 
Generally, the shape of the bands are asymmetric with a minimum located between \SI{2900}{\nano\meter} and \SI{3000}{\nano\meter}.
Indeed, within the CV sub-groups, the band minimum is located at approximately \SI{2933}{\nano\meter} for CV$_\textrm{OxA}$, while for CV$_\textrm{OxB}$ and CV$_\textrm{Red}$ it is closer to \SI{2967}{\nano\meter} and \SI{2927}{\nano\meter}, respectively (Table \ref{Tab:ReflRes_Carb} and Fig. \ref{Fig:Ox_Hyd}). 
For the calculation of the average minimum position of CV$_\textrm{OxB}$, samples MET 01074 and Mokoia were excluded since their shallow hydration band made the determination of the minimum difficult.
For CO chondrites, the average hydration band minimum is located at approximately \SI{2935}{\nano\meter} (Table \ref{Tab:ReflRes_Carb} and Fig. \ref{Fig:CO_Hyd}).
For UOCs, it is located at $\sim \SI{2908}{\nano\meter}$ for H and is shifted to longer wavelength closer to \SI{2913}{\nano\meter} and \SI{2917}{\nano\meter} for L and LL, respectively (Table \ref{Tab:ReflRes_Ord} and Fig. \ref{Fig:Ord_Hyd}). Lastly, for CR chondrites the average minimum position is located closer to \SI{2930}{\nano\meter} (Table \ref{Tab:ReflRes_Carb} and Fig. \ref{Fig:CR_Hyd}).
On average, the \Area decreases from CV$_\textrm{Red}$ to CV$_\textrm{Ox}$ (CV$_\textrm{Red}$: 9.9 $\pm$ 1.3~\si{\percent},CV$_\textrm{OxB}$: 9.3 $\pm$ 2.0~\si{\percent}, CV$_\textrm{OxA}$: 9.7 $\pm$ 2.1~\si{\percent}). \\
Out of the 15 CO samples that were measured in this work, only 4 were heated to \SI{80}{\celsius}. This heating process will eliminate adsorbed water and can, therefore, lead to a change of the \Area value. In the following we only consider the four heated samples for the analysis of the hydration band area. The average hydration area of COs is 9.7 $\pm$ 1.3~\si{\percent} (Table \ref{Tab:ReflRes_Carb}) making them similarly hydrated to the the CVs considered in this work.\\
The common belief in the literature seems to be that UOCs are little to not hydrated with some exceptions (e.g. Semarkona and Inman) (e.g. \cite{NebProc}, \cite{Grossmann2000} and \cite{Quirico2003}). In contrast, UOCs measured in this work surprisingly all exhibit well distinct hydration bands (Fig. \ref{Fig:Ord_Hyd}). In fact, when comparing the average hydration area of UOCs (11.6 $\pm$ 2.1~\si{\percent} for H, 12.3 $\pm$ 1.3~\si{\percent} for L and 13.9 $\pm$ 1.1~\si{\percent} for LL) with those of the CV and CO chondrites, UOCs exhibit slightly higher IBD$_\textrm{Hyd}$ values. \\
For CR chondrites the \SI{3}{\micro\meter} absorption band is especially pronounced, exhibiting an average IBD$_\textrm{Hyd}$ value of 19.1 $\pm$ 1.0~\si{\percent}. This value exceeds that of all type 3 chondrites considered in this work, consistently with their post accretion history \citep{ClasMet}. Additionally, Fe oxyhydroxides due to terrestrial weathering might contribute to the \SI{3}{micro\meter} band \citep{Cloutis_CR}.

\subsection{Aliphatic bands}
Since organics are systematically present in primitive chondrites (\cite{Alexander2017}, \cite{ChondComp}), we expect to see two further absorption features positioned between $\sim \SI{3400}{\nano\meter}$ and $\sim \SI{3500}{\nano\meter}$. However, only CVs Axtell, MIL 091010, GRA 06101, Grosnaja, MET 01074, Mokoia, Efremovka and MIL 07277, COs MIL 05104, Kainsaz, Moss, QUE 97416, EET 92126, ALH 83108, ALH 85003 and MET 00737 and UOCs WIS 91627, RBT 04251, MCY 05218, EET 96188, LAR 06279, ALH 84120, ALH 84086, MET 00489 and LEW 87284 show well pronounced aliphatic bands. All other samples considered in this work exhibit either no or only very faint organic bands. 
There are some possible explanations for the varying abundances of organics in chondrites. For one, the abundance of organics decreases with increasing metamorphic grade. 
Furthermore, all organics are contained in the matrix material of the chondrites. Therefore, varying abundances of matrix material will lead to a variation in aliphatic band intensities. 
Nevertheless, no correlation between the presence of aliphatic bands and the metamorphic grade can be observed for the samples in this work (Fig. \ref{Fig:Ox_Hyd}, \ref{Fig:CO_Hyd}, \ref{Fig:Ord_Hyd} and \ref{Fig:CR_Hyd}). Furthermore, the abundance of matrix material for some of the CV chondrites considered in this work can be found in \cite{Bonal_TGA}. The aliphatic band intensity does not seem to be correlated to the matrix abundance either. It is, thus, not understood yet why organics are only present in some of the spectra. 


\section{Discussion}
\label{Sec:Discussion}
Our objectives are to (i) evaluate spectral variability between chondrite groups, (ii) determine whether post-accretion processes can be assessed based on reflectance spectra and (iii) investigate the link between chondrites and their possible parent asteroid.

\subsection{Reflectance spectra as a tool to investigate the hydration of chondrites}
\begin{figure*}[h]
\centering
\includegraphics[width=0.6\textwidth]{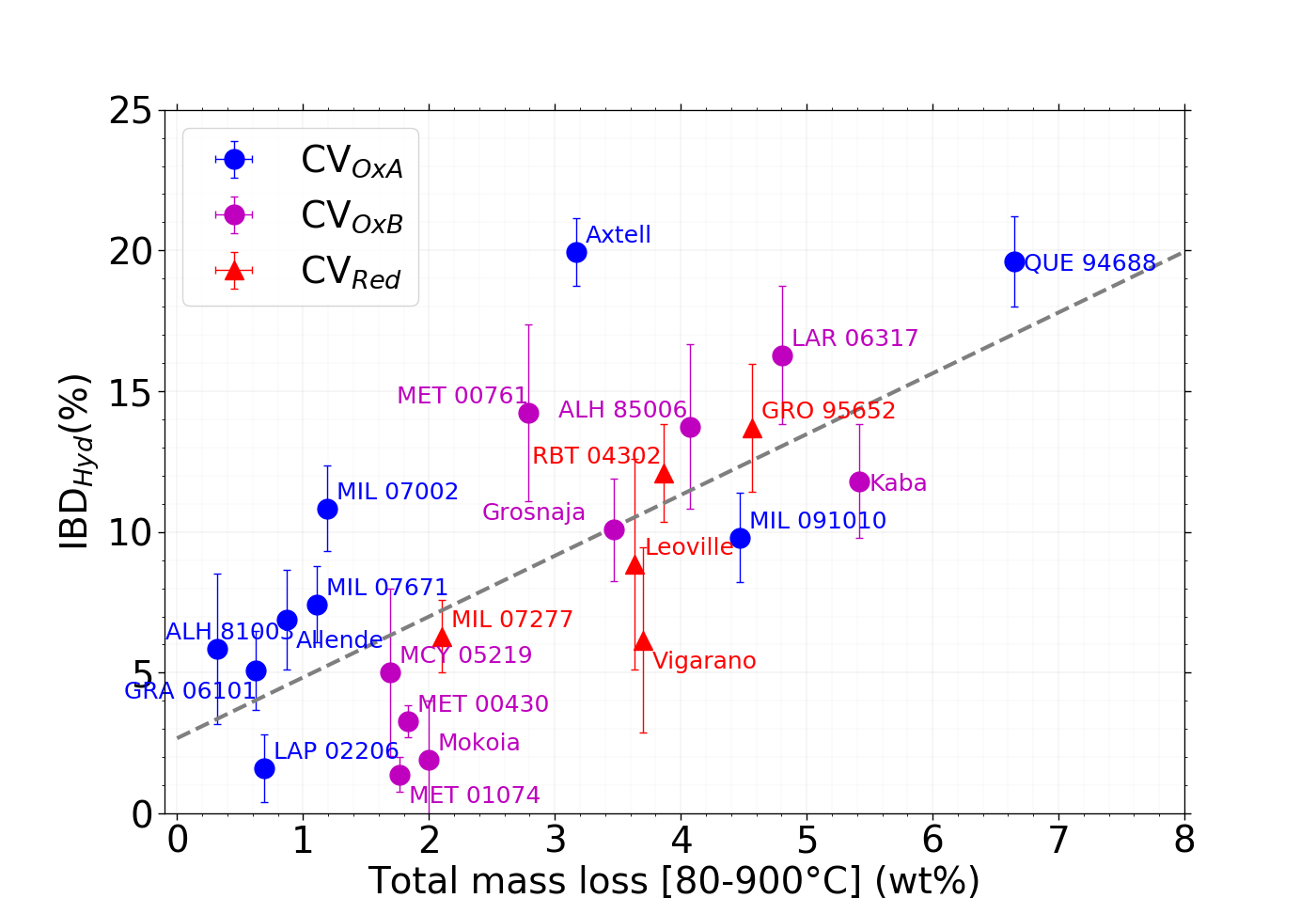}
\caption{IBD$_\textrm{Hyd}$ over the total mass loss measured by TGA. Reflectance spectra were measured on the bulk material at \SI{80}{\celsius} and under vacuum. TGA measurements were done in the \SI{80}{\celsius} to \SI{900}{\celsius} range.}
\label{Fig:Hyd_TGA}
\end{figure*}

As mentioned before, the \SI{3}{\micro\meter} band is related to the stretching and anti-stretching modes of metal-OH as well as bound H$_2$O and adsorbed water. We, thus, expect the IBD$_\textrm{Hyd}$ to reflect the hydration of the observed asteroids. To verify its suitability as a thin tracer for the hydration of asteroids, we consider the \Area series of CV chondrites, for which the hydration has been independently evaluated through Thermogravimetric Analysis (TGA) \citep{Bonal_TGA}.\\
A correlation can be observed ($r^2 = 0.5879$)(Fig. \ref{Fig:Hyd_TGA}) between the \Area and the total mass loss between \SI{80}{\celsius} and \SI{900}{\celsius}. Generally, the CV$_\textrm{OxA}$ considered in this work are less hydrated than CV$_\textrm{OxB}$ and CV$_\textrm{Red}$ with the exception of samples Axtell, QUE 94688 and MIL 091010 (Fig. \ref{Fig:Hyd_TGA}). These three samples were previously identified as weathered samples (\cite{Bonal_TGA} and references within) which could explain their increased hydration. Axtell shows a larger \Area than the trend would suggest. TGA and reflectance measurements were performed on different days and using different pieces of the same sample. Therefore, this could be an effect of sample heterogeneity. \\
The TGA measurements are based on the dehydration and dehydroxylation of (oxy)-hydroxides and phyllosilicates but also include contributions due to the decomposition of iron sulfides \citep{Garenne2016}. These iron sulfides cannot be distinguished from the hydrated minerals. The \Area value determined by reflectance spectroscopy on the other hand only includes contributions due to hydrated minerals. Therefore, we expect this to influence the correlation between the \Area and the TGA mass loss. 
We conclude that the IBD$_\textrm{Hyd}$ of the \SI{3}{\micro\meter} band is a qualitative tracer of the asteroids hydration.

\subsection{Thermal metamorphism effects on Reflectance spectra}
As discussed in Sections \ref{Sec:Introduction} and \ref{Ch:Analytical} the spectra of carbonaceous chondrites and UOCs exhibit two strong absorption features in the \SI{1}{\micro\meter} and \SI{2}{\micro\meter} region due to the presence of Fe-bearing silicates. 
The exact mineralogy of these silicates will vary between chondrite groups. 
Since chondrules are the dominant petrographic component in CVs, COs, CRs and UOCs, the spectra are most likely controlled by their chemistry. Furthermore, the mineralogy of the chondrites can vary within a chondrite group depending on the degree of secondary processes. Thus, by comparing these two spectral features with a quantitative parameter reflecting the thermal history of the samples \citep{Bonal2016} the objective is to find out if and in what way the spectral features reflect the asteroids thermal history.

\subsubsection{1 μm and 2 μm absorption bands}
\label{Sec:1000nmAnd2000nmDisc}
\begin{figure*}
\centering
\begin{subfigure}[c]{\textwidth}
\includegraphics[width=\textwidth]{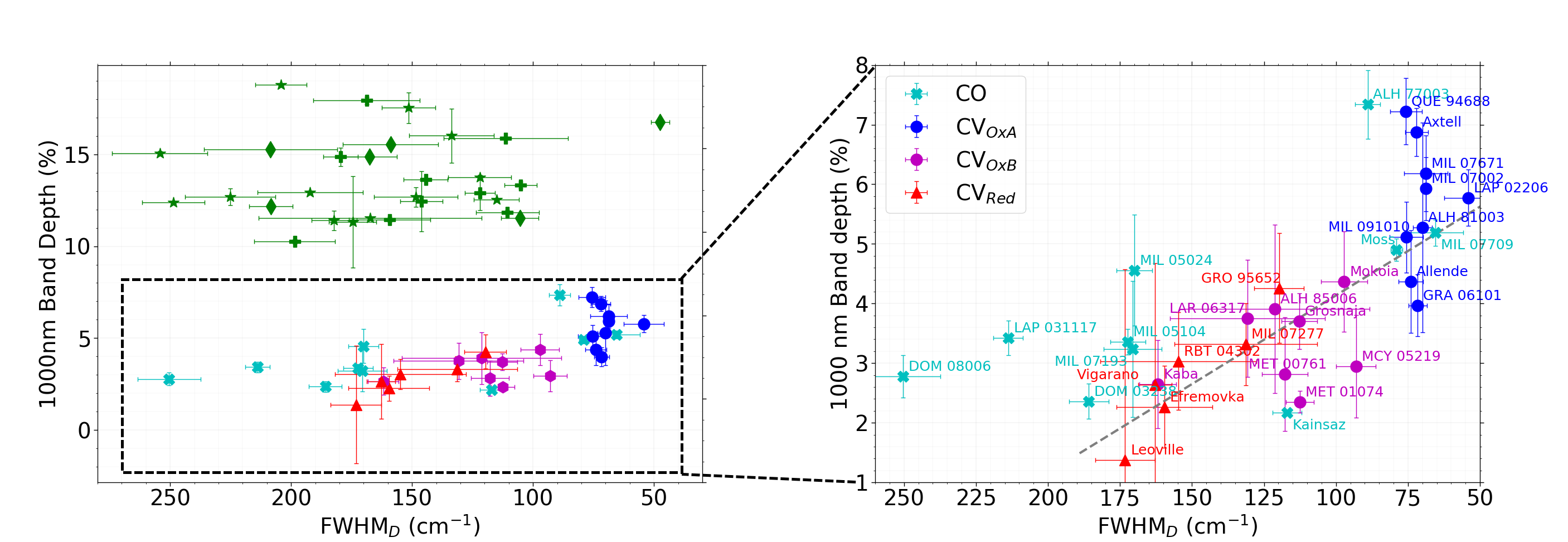}
\subcaption{}
\label{Fig:1000nmOverFWHMAll}
\end{subfigure}
\begin{subfigure}[c]{0.49\textwidth}
\includegraphics[width=\textwidth]{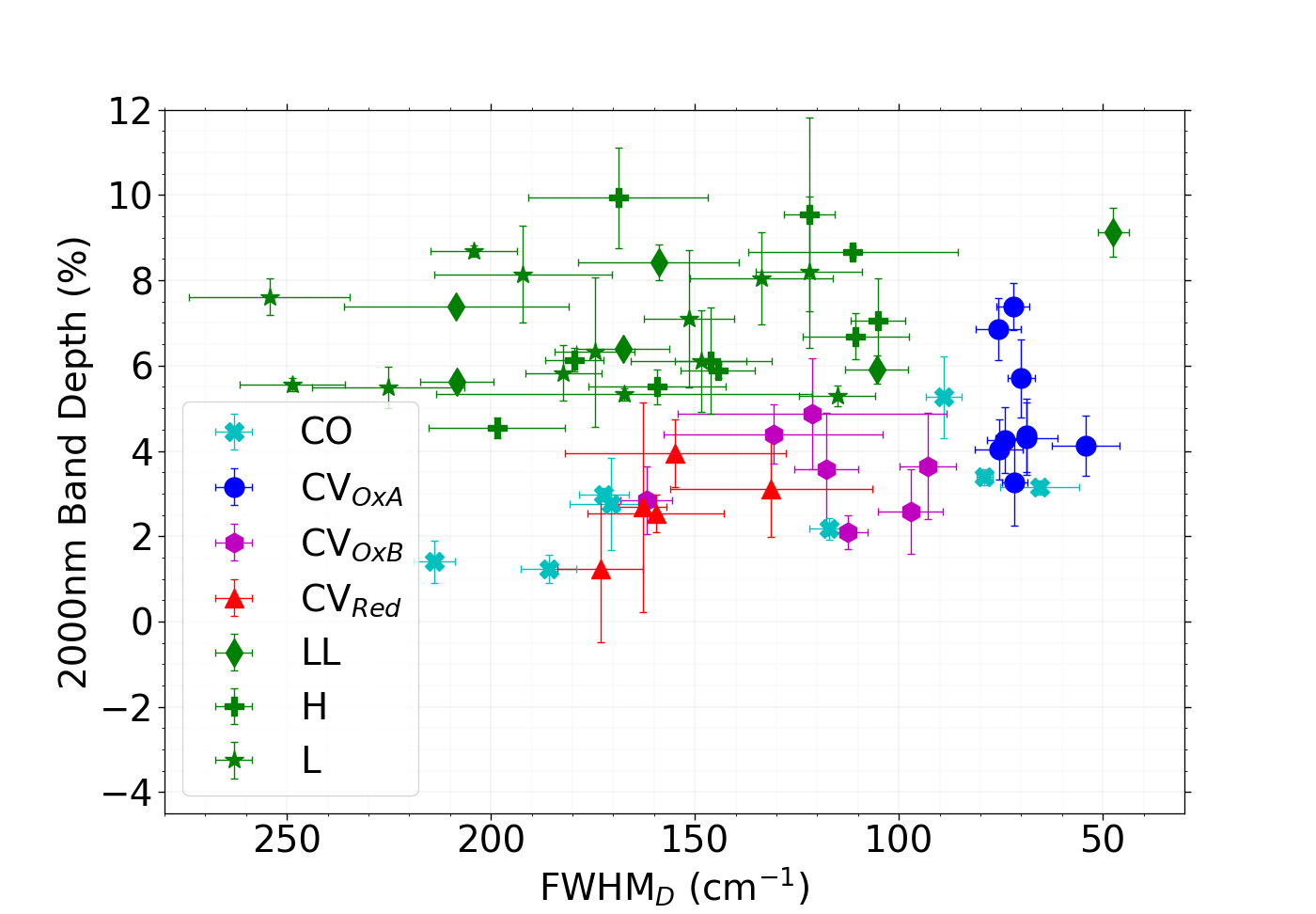}
\subcaption{}
\label{Fig:2000nmOverFWHMAll}
\end{subfigure}
\caption{Comparison between spectral values and the FWHM$_\textrm{D}$ value measured by Raman spectroscopy \citep{Bonal2016}. (a) \SI{1}{\micro\meter} band depth of CVs, COs and UOCs over the the FWHM$_\textrm{D}$ value (left) and the same for COs and CVs only (right); (b) \SI{2}{\micro\meter} band depth of CVs, COs and UOCs over the the FWHM$_\textrm{D}$ value.}
\label{Fig:Bands}
\end{figure*}

The \SI{1}{\micro\meter} band depth is correlated to the abundance of olivine in chondrites (see references in introduction). Furthermore, the abundance of iron in olivine increases with increasing metamorphic grade \citep{McSween1977_CO}. We, therefore, expect the \SI{1}{\micro\meter} band depth to increase with increasing metamorphic grade if the dominant silicate in this region is olivine (Fig. \ref{Fig:Bands}). Indeed, this can be observed for CV chondrites (Fig. \ref{Fig:1000nmOverFWHMAll}, $r^2 = 0.711$). Samples QUE 94688 and Axtell are the exceptions showing deeper \SI{1}{\micro\meter} bands than the trend would suggest. This might be related to the presence of oxy-hydroxides formed through terrestrial weathering. For UOCs and COs this trend is not seen (Fig. \ref{Fig:1000nmOverFWHMAll}).
The dominant silicate contributing to the \SI{1}{\micro\meter} region of UOCs is pyroxene \citep{Gaffey1976}. Thus, for UOCs the increase in olivine abundance does not affect the \SI{1}{\micro\meter} band depth as strongly as for carbonaceous chondrites. 
The absence of the trend within the considered CO
chondrites might be explained by their metamorphic
grades which are systematically lower than those of the considered CV chondrites (most COs considered here are indeed characterized by larger FWHM$_\textrm{D}$ values than CVs (Fig. \ref{Fig:1000nmOverFWHMAll})).  
Moreover, samples with the lowest metamorphic grades (DOM 08006, LAP 031117, DOM 03238, MIL 05104, MIL 07193 and MIL 05024) exhibit practically no absorption features (Fig. \ref{Fig:CO_Features}). 
Perhaps it is only above a given metamorphic temperature that sufficient chemical modification of olivine \citep{McSween1977} occurs resulting in significant spectral changes. Hence, low metamorphosed CO samples do not follow the trend defined by CV chondrites (Fig. \ref{Fig:1000nmOverFWHMAll}). \\

The \SI{2}{\micro\meter} band is dominated by contributions of pyroxene and spinel at around \SI{1900}{\nano\meter} (see references in introduction). Since spinel is mainly contained within the Calcium-Aluminium-Rich inclusions (CAIs) (e.g. \cite{CAIs}), we expect a variation of the spinel band depth as well as the spectral slope in the \SI{2}{\micro\meter} area with varying CAI abundances. The volume percent of CAIs contained in the different chondrite groups decreases in the order: CV chondrites (3 vol.$\%$), CO chondrites (1 vol.$\%$), CR chondrites (0.12 vol$\%$), H chondrites (0.01-0.2 vol.$\%$), L chondrites ($<$0.1 vol.$\%$) and LL chondrites ($<$0.1 vol.$\%$) \citep{ChondComp}.
Therefore, we expect that the \SI{2}{\micro\meter} region is dominated by pyroxene for CAI-poor groups such as CR, H, L and LL. For higher CAI abundances (i.e. CV and CO) we expect spinel to play a significant role in the \SI{2}{\micro\meter} band.
Moreover, the thermal metamorphism process leads to the chemical equilibration of the chondrite components. The transfer of chemical elements which are present in the matrix material, such as iron, to chondrite components, such as chondrules, results in an increase in band depths. 
However, no correlation between the \SI{2}{\micro\meter} band depth and the metamorphic grade can be observed for the chondrites measured in the present work (Fig. \ref{Fig:2000nmOverFWHMAll}). 
The UOCs and COs considered here show low metamorphic grades. Furthermore, pyroxene equilibrates at even higher temperatures than olivine \citep{ScottJones}. Perhaps the metamorphic temperature of the chondrites considered in this work was not sufficiently high for the equilibration of pyroxene to take place. 
Moreover, the high CAI abundance in COs and CVs could lead to a significant spinel band in the \SI{2}{\micro\meter} region overlaying with the pyroxene band. Taken all together, this could explain why there is no correlation observed between the \SI{2}{\micro\meter} band and the metamorphic grade.

\subsubsection{Spectral slopes}
\label{Sec:Slopes}
\begin{figure*}
\centering
\begin{subfigure}[c]{0.49\textwidth}
\includegraphics[width=\textwidth]{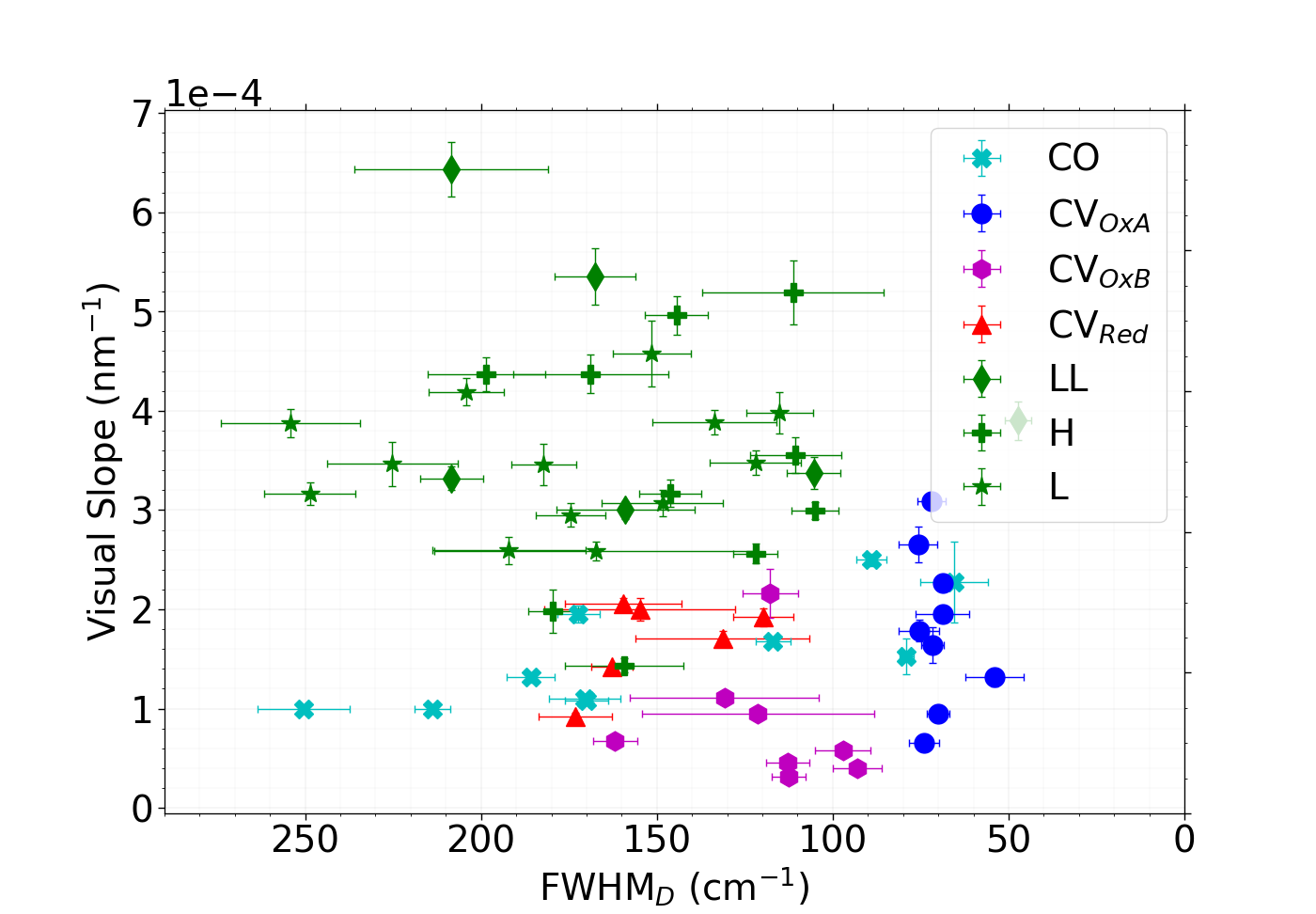}
\subcaption{}
\label{Fig:VisOverFWHMAll}
\end{subfigure}
\begin{subfigure}[c]{0.49\textwidth}
\includegraphics[width=\textwidth]{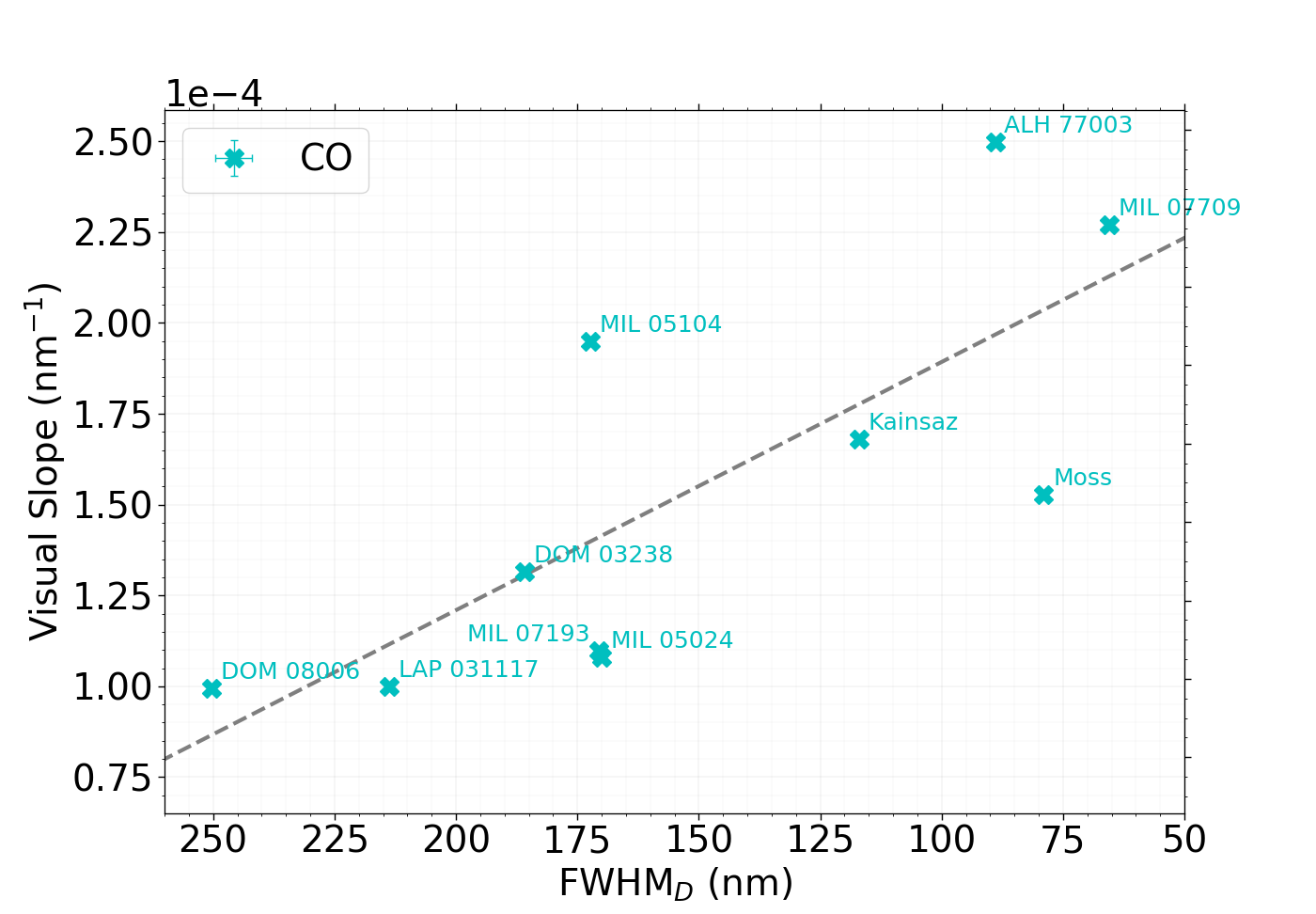}
\subcaption{}
\label{Fig:VisOverFWHMCO}
\end{subfigure}
\begin{subfigure}[c]{0.49\textwidth}
\includegraphics[width=\textwidth]{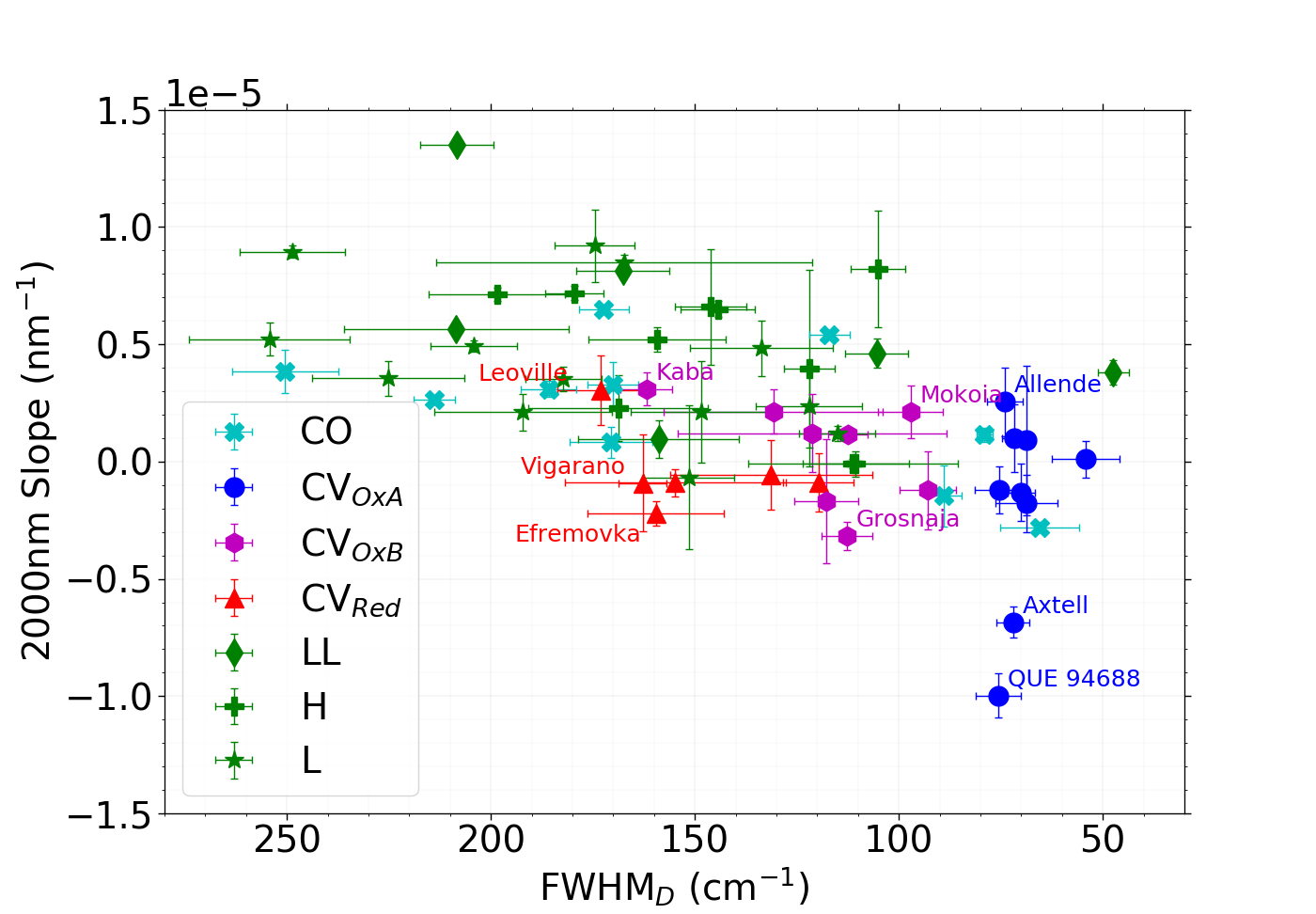}
\subcaption{}
\label{Fig:2000nmSlopeOverFWHM}
\end{subfigure}
\begin{subfigure}[c]{0.49\textwidth}
\includegraphics[width=\textwidth]{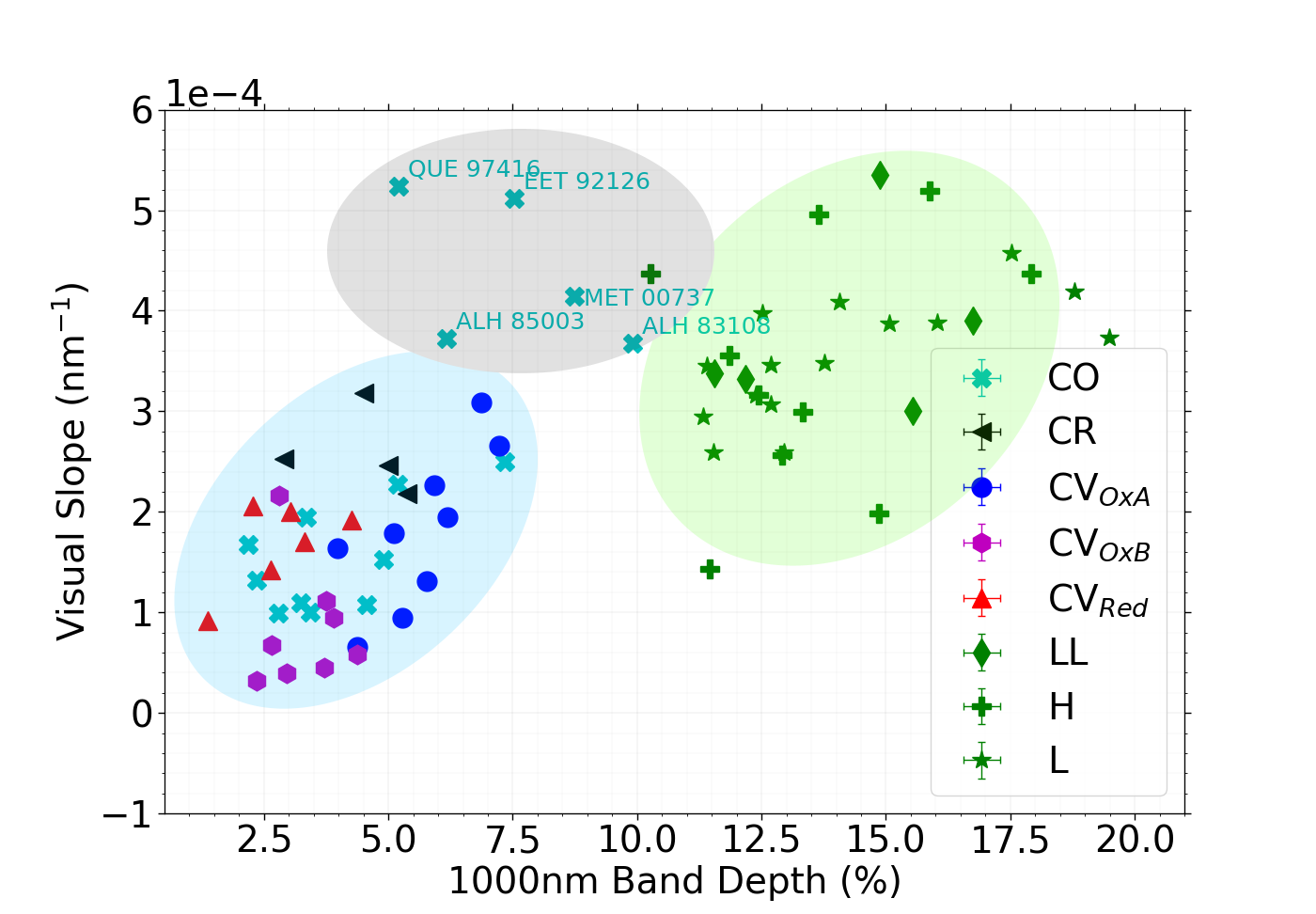}
\subcaption{}
\label{Fig:VisSlopeOver1000nm}
\end{subfigure}
\caption{Comparison between spectral values and the the FWHM$_\textrm{D}$ value measured by Raman spectroscopy \citep{Bonal2016}. (a) The visual slope of COs, CVs and UOCs over the FWHM$_\textrm{D}$ value; (b) The visual slope of COs over the FWHM$_\textrm{D}$ value; (c) The \SI{2000}{\nano\meter} slope of COs, CVs and UOCs over the the FWHM$_\textrm{D}$ value; (d) The visual slope of all considered chondrites over the \SI{1}{\micro\meter} band depth.}
\label{Fig:Slopes}
\end{figure*}

A negative trend between the \SI{2}{\micro\meter} spectral slope and the metamorphic grade of the chondrites is observed (Fig. \ref{Fig:2000nmSlopeOverFWHM}). The least metamorphosed UOCs and COs show positive to neutrally sloped spectra in the \SI{2}{\micro\meter} region (slopes between 0 and \SI{1.28e-5}{\per\nano\meter}). The higher metamorphosed CVs, on the other hand, show predominantly negative \SI{2}{\micro\meter} slopes (slopes varying from \SI{-0.99e-5}{\per\nano\meter} to \SI{0.31e-5}{\per\nano\meter}).
As explained above, we expect the \SI{2}{\micro\meter} spectral slope to become bluer with higher CAI content and thus with increased Fe content in spinel. 
This can be a further explanation for the differences in spectral slope between chondrite groups. As the abundance of CAIs decreases from CVs to COs to UOCs we expect the \SI{2}{\micro\meter} slope to become bluer from UOCs to COs to CVs.\\
It was suggested before that the overall spectral slope can be influenced by the terrestrial weathering experienced by the samples (\cite{Cloutis2012}, \cite{Salisbury1974}). This could explain why Axtell (CV$_\textrm{OxA}$) and QUE 94688 (CV$_\textrm{OxA}$) are falling off the trend, showing strong blue \SI{2}{\micro\meter} slopes (Fig. \ref{Fig:2000nmSlopeOverFWHM}). Furthermore, it was suggested that ``falls'' are redder sloped than ``finds'' since terrestrial weathering did not affect them as intensely \citep{Cloutis2012}.
The falls Allende, Mokoia, Kainsaz and Moss, indeed, show redder overall spectral slope in comparison to the ``finds'' (Fig. \ref{Fig:Ox_Features} and \ref{Fig:CO_Features}). However, this is not the case for falls Kaba, Grosnaja and Vigarano which stands in disagreement with the previous suggestions \citep{Cloutis2012}. \\
The UV absorption at wavelengths shorter than \SI{500}{\nano\meter}, due to iron in silicates and oxidized iron, are reflected in the visual slope. Indeed, steeper visual slopes indicate stronger absorption in the UV range, while more shallow visual slopes point towards more shallow UV absorption bands.
COs show a correlation between the visual slope and metamorphic grade of the samples (Fig. \ref{Fig:VisOverFWHMCO}, $r^2 = 0.586$). This indicates an increase of iron abundance in silicates and/or oxidized iron abundance with increasing metamorphic grade. On the other hand, for CVs and UOCs no trend can be observed (Fig. \ref{Fig:VisOverFWHMAll}).\\
As mentioned before, 5 of the 15 COs considered in this work (ALH 85003, ALH 83108, EET 92126, MET 00737 and QUE 97416) could previously not be provided with a FWHM$_\textrm{D}$ value \citep{Bonal2016}. Their deep spectral features point towards a strong metamorphic grade. To support this suggestion, the visual slope is compared with the \SI{1}{\micro\meter} band depth (Fig. \ref{Fig:VisSlopeOver1000nm}). Three main groupings can be observed. UOCs are clearly separated from carbonaceous chondrites by exhibiting deeper \SI{1}{\micro\meter} bands and slightly steeper visual slopes (Fig. \ref{Fig:VisSlopeOver1000nm}). Furthermore, a small group of CO chondrites can be observed which exhibit stronger visual slopes than the other carbonaceous chondrites. These samples are the 5 COs EET 92126, QUE 97416, ALH 85003, MET 00737 and ALH 83108 previously not provided with FWHM$_\textrm{D}$ values. Indeed, the strong iron abundances in silicates and/or oxidized iron abundances point towards high metamorphic grades in these 5 CO chondrites.

\subsection{Finding asteroidal parent bodies}

Several possible links between carbonaceous chondrites and asteroidal parent bodies have been suggested in the past which are not always coherent with each other.
\cite{Bell88} suggest K-type asteroids (Eos family) as a possible parent asteroid of CV and CO chondrites. CO, CK, CV and CR chondrites were suggested to be related to K-type asteroids \citep{Vernazza2015}. C-type asteroids were suggested as possible parent bodies for CK chondrites \citep{Gaffey1978} and CR chondrites \citep{Hiori1996}. This shows, that it is challenging to unambiguously identify the parent body of a meteorite based solely on remote spectroscopic data. 
 Since the return of asteroid samples with the Hayabusa 1 space mission \citep{Nakamura2011} the link between S-type asteroids and EOCs is confirmed. This information can be used to validate the method we chose for the asteroidal parent body search of carbonaceous chondrites. This will be further discussed in Section \ref{Sec:S-typeUOC} and \ref{Sec:CarbCondrLink}.

\subsubsection{Chosen spectral parameters for asteroid-chondrite comparison}
\label{Sec:SpecPara}
Spectral slopes and reflectance have been used in the past to compare meteorites and asteroids observations. However, the overall spectral slope is influenced by a lot of external factors such as measuring geometry, grain size or weathering degree. Indeed, \cite{Salisbury1974} showed that when artificially weathering a L6 chondrite, the spectral slope becomes steeper while the \SI{1}{\micro\meter} region remains unchanged. 
Because they are exposed to the space environment, asteroid surfaces can experience specific processes described as space weathering (SW). 
For silicate-rich (S-type) asteroids SW leads to the reddening and darkening of the spectra \citep{Marchi2005}. For carbonaceous asteroid (C-Type) large advances have been made in the last few years to understand SW effects (e.g. \cite{Brunetto2015}, \cite{Brunetto2020}, \cite{Lantz2013}, \cite{Matsuoka_2020}), in particular in the framework of the Hayabusa-2 and Osiris-REx sample-return missions.
From ion- or laser-irradiation experiments, spectral reddening was observed for some CV, CO and CM samples (\cite{Brunetto2014}, \cite{Lazzarin2006} and \cite{Moroz1996}). For Tagish Lake, an ungrouped CC, spectral flattening was observed \citep{Vernazza2013}. For Murchison (CM), spectral flattening, darkening and decrease of the \SI{0.7}{\micro\meter} and \SI{3}{\micro\meter} was observed \citep{Matsuoka_2020}.
In their study, \cite{Lantz2017} showed complex behaviors for carbonaceous chondrites, with dark samples revealing a bluing, and brighter sample showing reddening with increasing irradiation dose. There are still several open questions regarding the effects of SW on the spectra of dark objects. How does it effect other CC-groups (e.g. CK or CR)? Is there a connection between different petrographic types of CV, CO or CR and the amount of SW experienced on their asteroidal parent bodies?
What is the dependency of SW signatures on petrographic type of CV, CO or CR? 
Until these questions can be answered, we restrict the asteroid-meteorite comparison on the \SI{1}{\micro\meter} and \SI{2}{\micro\meter} absorption features, since they seem to be the least affected by SW.
Lastly, temperature differences between the measurements done on meteorite and asteroid exist but have a negligible influence on the band positions of the spectra \citep{Dunn2013}, typically below the spectral resolution of the measurement from this study.\\
As these results show, the \SI{1}{\micro\meter} and \SI{2}{\micro\meter} absorption features appear to be mostly unaffected by processes such as SW, grain size and temperature, making them a promising spectral feature for the identification of parent bodies. 
\cite{Devogele2018}, for example, suggests a link between L-type asteroids and CAI rich chondrites such as CV and CO chondrites based on a strong spinel absorption feature observed in the asteroid spectra. Similar arguments can be found in \cite{Burbine1992} and \cite{Sunshine}.
Therefore, in this work we base our spectral comparison on the \SI{1}{\micro\meter} and \SI{2}{\micro\meter} absorption features and not on the spectral slopes. The comparisons between asteroid and chondrite features are made in Figure \ref{Fig:AstComp}.

\subsubsection{Spectral differences between chondrite groups}
\label{Sec:SpecDiff}
Our results show that carbonaceous chondrites can generally be distinguished from UOCs by exhibiting much less pronounced spectral features (Figs. \ref{Fig:Ox_Features}, \ref{Fig:CO_Features} and \ref{Fig:Ord_Features}). Between carbonaceous chondrite groups this separation is not as clear. This can make a clear identification of asteroidal parent bodies difficult, since different chondrite groups overlap. This is especially true for CO and CV chondrites. Both groups exhibit similar \SI{1}{\micro\meter} and \SI{2}{\micro\meter} absorption band depths as well as band positions and spectral slopes (Fig. \ref{Fig:2000nmOver1000nm}, \ref{Fig:1000PosOver1000} and \ref{Fig:2000PosOver1000}). 
CR chondrites can be clearly separated from CO and CV by showing much lower \SI{1}{\micro\meter} band positions (Fig. \ref{Fig:1000PosOver1000}). 
CK chondrites can be distinguished from CO, CV and CR by having deeper \SI{1}{\micro\meter} absorption band depths, similar to those of UOCs. They exhibit less deep \SI{2}{\micro\meter} absorption bands and their band positions are located at longer wavelengths than UOCs (Fig. \ref{Fig:2000nmOver1000nm}, \ref{Fig:1000PosOver1000} and \ref{Fig:2000PosOver1000}).

\subsubsection{S-type asteroids and UOCs}
\label{Sec:S-typeUOC}
Ordinary chondrites (OCs) are pieces of S-type asteroids as first proposed through reflectance spectroscopy (e.g. \cite{Chapman1973}) and confirmed by the return of the Hayabusa-1 space mission in 2010 \citep{Nakamura2011}.
More specifically, the connection was found between EOCs (LL4 to LL6) and S-Type asteroids.
In terms of reflectance spectral features, EOCs exhibit stronger absorption features and stronger peak reflectance values in the visible range than UOCs. The \SI{2}{\micro\meter} band positions of UOCs might be shifted slightly towards shorter wavelengths due to lower amounts of Fe in pyroxene.\\
Our results show, the \SI{1}{\micro\meter} and \SI{2}{\micro\meter} band depth of the S-type asteroid spectrum match those of UOCs (see Fig. \ref{Fig:2000nmOver1000nm}). 
Furthermore, the \SI{1}{\micro\meter} band position is in good agreement with the UOC values (Fig. \ref{Fig:1000PosOver1000}). 
As expected, the \SI{2}{\micro\meter} band position of the UOCs is shifted towards lower wavelengths ($\sim$ \SI{1840}{\nano\meter}) in comparison to the S-type asteroid ($\sim$ \SI{1950}{\nano\meter})(Fig. \ref{Fig:2000PosOver1000}).   \\
In conclusion, it is possible to identify S-type asteroids as the parent body of UOCs based on the reflectance spectral parameters. This validates our approach. 

\subsubsection{Carbonaceous chondrite asteroid links}
\label{Sec:CarbCondrLink}
\begin{figure*}
\centering
\begin{subfigure}[c]{0.49\textwidth}
\includegraphics[width=\textwidth]{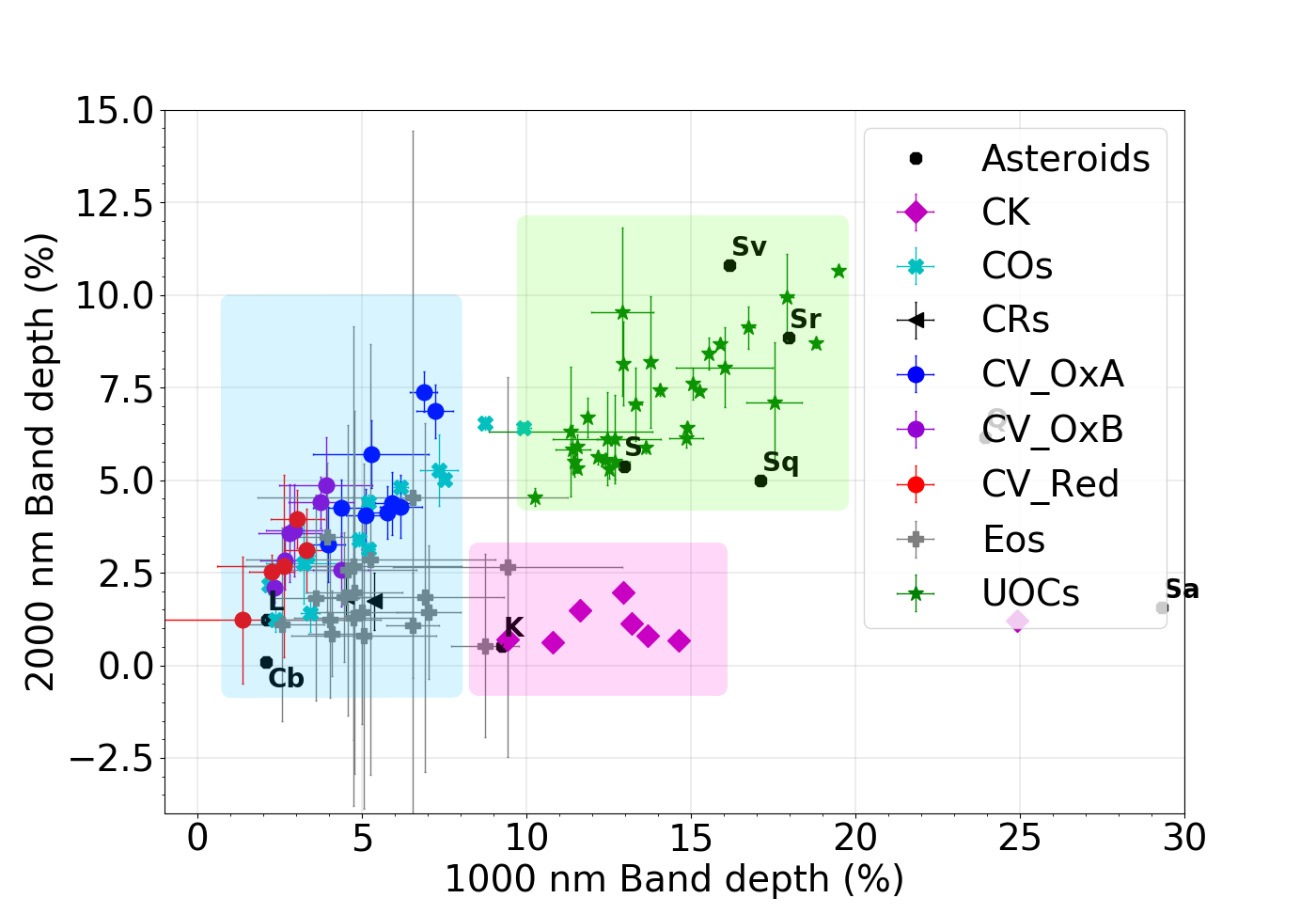}
\subcaption{}
\label{Fig:2000nmOver1000nm}
\end{subfigure}
\begin{subfigure}[c]{0.49\textwidth}
\includegraphics[width=\textwidth]{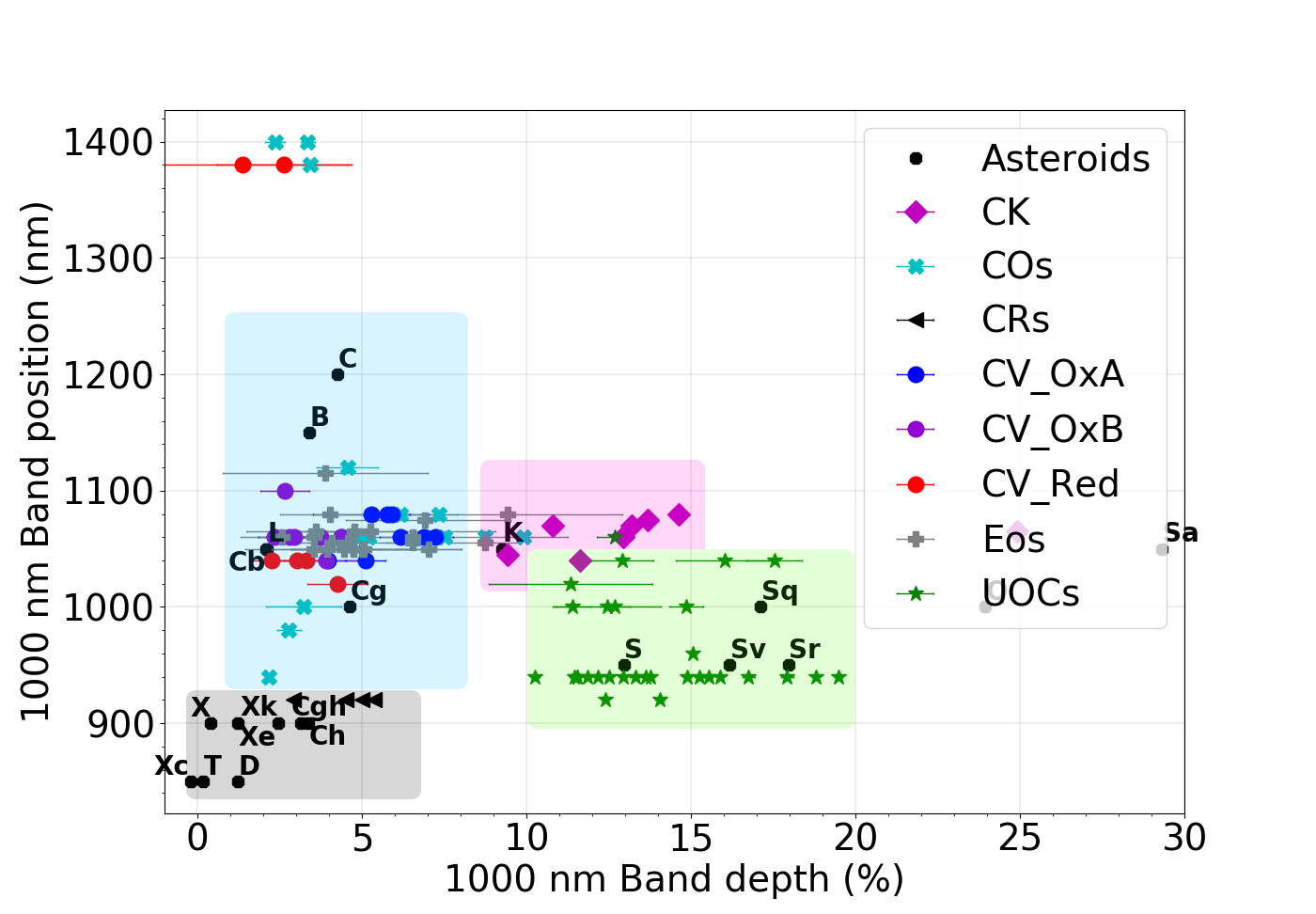}
\subcaption{}
\label{Fig:1000PosOver1000}
\end{subfigure}
\begin{subfigure}[c]{0.49\textwidth}
\includegraphics[width=\textwidth]{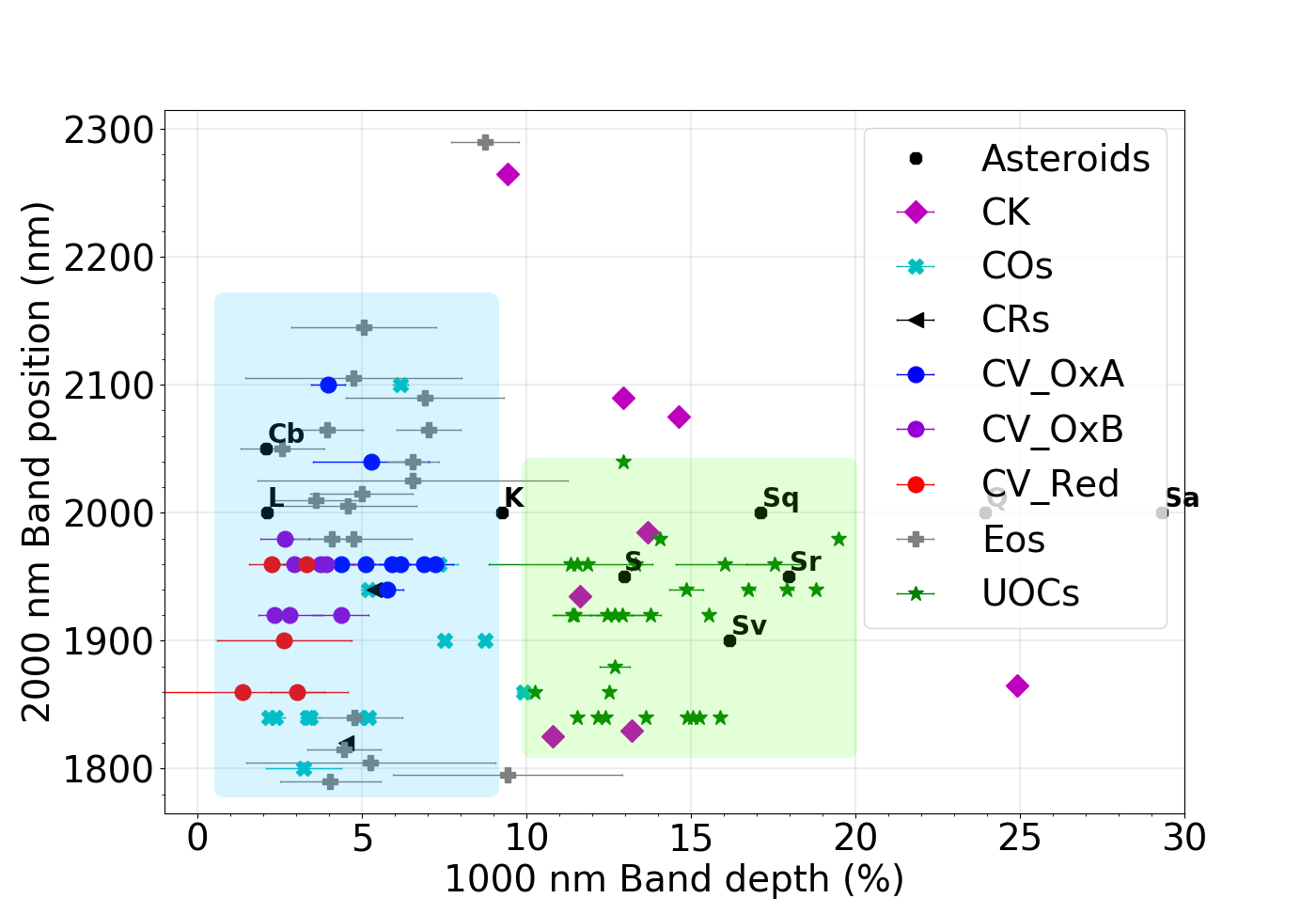}
\subcaption{}
\label{Fig:2000PosOver1000}
\end{subfigure}
\caption{Comparison between spectral values of asteroids (in black symbols) and chondrites (in colored symbols). (a) Comparison of the \SI{2}{\micro\meter} band depth with the \SI{1}{\micro\meter} band depth of chondrites and asteroids; (b) \SI{1}{\micro\meter} band position over the \SI{1}{\micro\meter} band depth; (c) \SI{2}{\micro\meter} band position over the \SI{1}{\micro\meter} band depth.}
\label{Fig:AstComp}
\end{figure*}

Taking the difficulties discussed in sections \ref{Sec:SpecPara} and \ref{Sec:SpecDiff} into account, similarities between asteroid and CC spectra can be analyzed (Fig. \ref{Fig:2000nmOver1000nm}, \ref{Fig:1000PosOver1000} and \ref{Fig:2000PosOver1000}).
Good matches of band depths and positions can be found between L-Type and Cb-Type asteroids and COs and CVs (Fig. \ref{Fig:AstComp}). The same can be said when comparing Eos spectra with COs and CVs. C, Cg and B-Type asteroids match COs and CVs well in the \SI{1}{\micro\meter} range but do not show any \SI{2}{\micro\meter} feature that could be included in the comparison.
As for CVs and COs, CRs show a similar \SI{1}{\micro\meter} and \SI{2}{\micro\meter} band depths as L, Cb and Eos asteroids. The band positions, however, are closer to X, Xe, Xk, Cgh or Ch-Type asteroids (Fig. \ref{Fig:1000PosOver1000} and \ref{Fig:2000PosOver1000}). However, none of these asteroid matches show any \SI{2}{\micro\meter} feature that could be included in the comparison. It should also be mentioned here that other criteria such as albedo were not considered in this comparison but can be used to further narrow down possible links between asteroids and meteorites. The albedo of CR chondrites (p$_v$ = 0.15 \citep{Piironen1998}) is generally lower than that of high albedo Xe-asteroids (p$_v$ $>$ 0.3 \citep{Fornasier2011}) making this affiliation unlikely. A larger set of CR chondrites should be considered as well to increase the statistics of the results. 
The four CRs analyzed here are finds from Antarctica making the presence of terrestrial weathering products probable. Since CR chondrites are rich in metal and sulfides \citep{ChondComp} these weathering products can have a larger influence on the slope as well as the \SI{1}{\micro\meter} band of the reflectance spectra. For future works it would be interesting to leach these samples, ridding them from the weathering products and repeating the asteroid-meteroid comparisons.
The \SI{1}{\micro\meter} and \SI{2}{\micro\meter} band depths and positions of CK chondrites match those of K-type asteroids (Fig. \ref{Fig:AstComp}). Due to the faint \SI{2}{\micro\meter} features of the CKs analyzed here, the \SI{2}{\micro\meter} band position values vary over a broad range (Fig. \ref{Fig:2000PosOver1000}) making matches less clear. Additionally, Eos family asteroids show a similar \SI{1}{\micro\meter} band positions and \SI{2}{\micro\meter} band depths and positions as CKs. The difference in \SI{1}{\micro\meter} band depth might be due to space weathering effects on Eos family asteroids.\\
We conclude, L-type, Cb-Type and Eos family asteroids match the spectral values of CV and COs in compliance with previous suggestions by \cite{Bell88} and \cite{Devogele2018}. CR chondrites match the band positions of X, Xe, Xk, Cgh and Ch-type asteroids.
Lastly, K-type and/or Eos family asteroids match CK chondrites in compliance with \cite{Vernazza2015}.

\subsection{\SI{3}{\micro\meter} band for asteroid-chondrite comparison}
\begin{figure*}[h]
\centering
\includegraphics[width=0.5\textwidth]{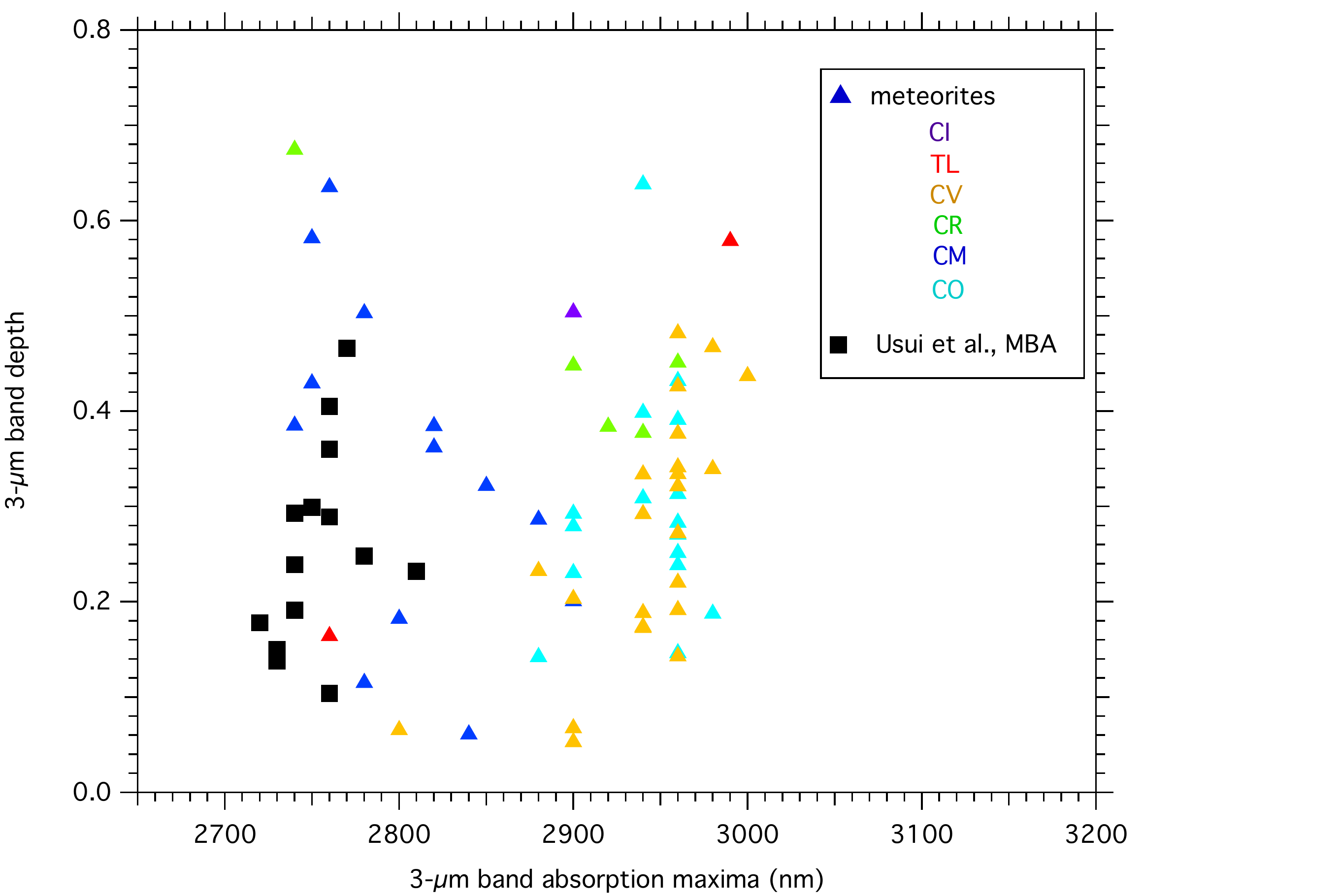}
\caption{The \SI{3}{\micro\meter} band depth over the \SI{3}{\micro\meter} band position for carbonaceous chondrites and C-Type main belt asteroids (MBA). All CVs, COs and CRs were taken from this work. The CIs, CMs and Tagish Lake were taken from \cite{Potin2020}. Lastly, the C-Type MBA come from \cite{Usui2018}.
}
\label{Fig:3microOverPos}
\end{figure*}

As mentioned before, the \SI{3}{\micro\meter} hydration band is not present in the asteroid spectra used in this work. 
Shown in Figure \ref{Fig:3microOverPos} is the \SI{3}{\micro\meter} band depth over the \SI{3}{\micro\meter} band position for the CC measured in this work as well as some CM, CI and Tagish Lake samples from \cite{Potin2020}. Additionally, some C-type main belt asteroid (MBA) spectra taken from \cite{Usui2018} are shown. The \SI{3}{\micro\meter} band shows strong spectral variations between and within different CC groups. Furthermore, a good match between the \SI{3}{\micro\meter} band of the C-type asteroids and type 2 CMs as well as possibly Tagish Lake is seen. No match is seen with type 3 CC. 
To match our type 3 CC spectra we expect asteroid spectra which show a strong \SI{3}{\micro\meter} band at longer wavelengths than those matching CMs. No such asteroid spectra are available yet. 
We, thus, conclude that the \SI{3}{\micro\meter} band has the potential to be a good tracer for meteorite parent bodies. 
To further investigate the link between chondrites and their parent asteroids using reflectance spectroscopy we therefore suggest to extend the asteroid spectra, typically acquired in the 450 - 2450~\si{\nano\meter} spectral range to the 450 - 4000~\si{\nano\meter} spectral range, thus including the \SI{3}{\micro\meter} band.

\section{Conclusion}
To increase our understanding of asteroid reflectance spectral features we acquired reflectance spectra (from \SI{340}{\nano\meter} to \SI{4200}{\nano\meter}) of 23 CV, 15 CO, 4 CR and 31 UOCs. Clear differences in the reflectance spectra of different chondrite groups as well as a large variability between chondrites of the same group were observed. The parameters used for comparing the spectra were the band depth, position and slope in the \SI{1}{\micro\meter} and \SI{2}{\micro\meter} region, the visual slope and peak reflectance in the visible wavelength range and the IBD$_\textrm{Hyd}$ as well as location of the \SI{3}{\micro\meter} band. Furthermore, we looked at the overall spectral slope of all the spectra.
In comparison to carbonaceous chondrites, UOCs systematically exhibit deeper absorption features which are located at lower wavelengths. 
Type 2 CR chondrites exhibit absorption features in the \SI{1}{\micro\meter} region at even lower wavelengths than UOCs, thus clearly separating them from type 3 chondrites. 
On the other hand, none of the considered spectral parameters allow to separate CV and CO chondrites. This is not surprising due to their comparable mineralogical composition \citep{ClasMet}. \\
Secondly, we investigated the link between these reflectance spectral features and the thermal history of the chondrites. 
Several spectral features appear to be controlled by
the metamorphic grade of the samples. 
The depth of the \SI{1}{\micro\meter} absorption band becomes deeper with increasing metamorphic grade along the considered series of CV chondrites. 
This was not observed previously \citep{Cloutis2012} which might be related to the significantly greater number of CV chondrites considered in the present work. To be noted, this correlation was not observed for the other chondrite groups.
The absence of the trend within the considered CO chondrites might be explained by systematically lower metamorphic grades than those of the considered CV chondrites. 
Perhaps it is only above a given metamorphic temperature that sufficient chemical modification of olivine \citep{McSween1977} occurs resulting in significant spectral changes. 
No correlation between the \SI{2}{\micro\meter} absorption feature and the metamorphic grade was observed. 
For CO chondrites the visual slope becomes steeper with increasing metamorphic grade. This indicated an increase in iron in silicates and/or oxidized iron in the samples with increasing metamorphic grade. 
The spectral slope in the \SI{2}{\micro\meter} region seems to be negatively correlated to the metamorphic grade for all samples. 
This indicates a decrease in pyroxene content with increasing peak metamorphic temperature (\cite{Cloutis2012}, \cite{Bonal2016}).\\
Lastly, we investigated the genetic link between asteroids and chondrites based on the reflectance spectral features.
The comparison of spectral features of asteroids
and chondrites led to a good match between the following types: S-type asteroids have similar \SI{1}{\micro\meter} and \SI{2}{\micro\meter} absorption band depths as UOCs. This is expected \citep{Nakamura2011}, and legitimates our approach. The parent bodies of CK and CV/CO chondrites have previously been suspected to be K-type
asteroids, particularly Eos family members \citep{Clark2009}.
This genetic affiliation is confirmed by the data we looked at for CK chondrites which exhibit similar \SI{1}{\micro\meter} and \SI{2}{\micro\meter}
absorption features as K-type asteroids and similar \SI{1}{\micro\meter} band positions and \SI{2}{\micro\meter} band depths and positions as Eos family members.
For CO/CV chondrites the absorption band depths and \SI{1}{\micro\meter} absorption band position match those of the Eos family. However, a match with L-type and Cb-Type asteroids exists as well. 
This is consistent with previous work (\cite{Sunshine}, \cite{Devogele2018}) suggesting a link between CAI-rich chondrites and L-type asteroids.\\
Lastly, we underline the potential of the \SI{3}{\micro\meter} band to constrain asteroid-meteorite links.\\

\textit{Acknowledgments.} This work has been funded by the Centre National d’Etudes Spatiales (CNES–France) and by the ERC grant SOLARYS ERC-CoG2017-771691


%
%

  \bibliographystyle{elsarticle-harv} 
  \bibliography{thesis}







\end{document}